\documentclass[12pt,preprint]{aastex}

\usepackage{rotating}

\newcommand{\msol}{M{$_{\odot}$}}

\newcommand{\kms}{km~s{$^{-1}$}}

\slugcomment{DRAFT: \today}
\shorttitle{Galactic Center}
\shortauthors{Bally et al.}

\begin{document}

\title{The Bolocam Galactic Plane Survey: 
         $\lambda$ = 1.1 and 0.35 mm  Dust Continuum 
         Emission in the Galactic Center Region}

\author{John Bally\altaffilmark{1}, 
	     James Aguirre\altaffilmark{2},
	     Cara Battersby\altaffilmark{3},
	     Eric Todd Bradley\altaffilmark{4},
	     Claudia  Cyganowski\altaffilmark{5},
	     Darren Dowell\altaffilmark{6},
	     Meredith Drosback\altaffilmark{7},
	     Miranda K Dunham  \altaffilmark{8},
	     Neal J. Evans II \altaffilmark{9}, 
	     Adam Ginsburg\altaffilmark{10},
              Jason Glenn\altaffilmark{11},
              Paul Harvey\altaffilmark{12},
              Elisabeth Mills \altaffilmark{13},
              Manuel Merello \altaffilmark{14}     
              Erik Rosolowsky \altaffilmark{15},
              Wayne Schlingman \altaffilmark{16},
              Yancy L. Shirley \altaffilmark{17},
              Guy S. Stringfellow\altaffilmark{18},
              Josh Walawender\altaffilmark{19}, and 
              Jonathan Williams\altaffilmark{20}
}
\affil{{$^1$}{\it{CASA, University of Colorado, UCB 389, 
      Boulder, CO 80309},        \email{John.Bally@casa.colorado.edu}}  }
\affil{{$^2$}{\it{Department of Physics and Astronomy, University of Pennsylvania, 
      Philadelphia, PA          } ,  \email{jaguirre@sas.upenn.edu}}}   
\affil{{$^3$}{\it{CASA, University of Colorado, UCB 389, 
      Boulder, CO 80309},        \email{Cara.Battersby@colorado.edu}}  }     
\affil{{$^4$}{\it{Department of Physics and Astronomy, University of Central Florida,           
                                              } ,   \email{tbradley@physics.ucf.edu}} }  
\affil{{$^5$}{\it{Department of Astronomy, University of Wisconsin, 
       Madison, WI  53706} ,      \email{ccyganow@astro.wisc.edu}}}  
\affil{{$^6$}{\it{Jet Propulsion Laboratory, California Institute of Technology, 4800 Oak Grove Dr. , 
      Pasadena, CA  91104},    \email{cdd@submm.caltech.edu}}}
\affil{{$^7$}{\it{Department of Astronomy, University of Virginia, P. O. Box 400325,
      Charlotesville, VA 22904},     \email{drosback@virginia.edu}}}
\affil{{$^8$}{\it{Department of Astronomy, University of Texas,   1 University Station C1400, 
      Austin, TX 78712},            \email{nordhaus@astro.as.utexas.edu}}}      
\affil{{$^9$}{\it{Department of Astronomy, University of Texas,   1 University Station C1400, 
      Austin, TX 78712},            \email{nje@astro.as.utexas.edu}}}
\affil{{$^{10}$}{\it{CASA, University of Colorado, UCB 389 
       Boulder, CO 80309},       \email{Adam.Ginsburg@colorado.edu}}}       
\affil{{$^{11}$}{\it{CASA, University of Colorado, UCB 389, 
      Boulder, CO 80309},        \email{Jason.Glenn@colorado.edu}}}
\affil{{$^{12}$}{\it{CASA, University of Colorado, UCB 389, 
      Boulder, CO 80309},        \email{pmh@astro.as.utexas.edu}}}
\affil{{$^{13}$}{\it{Department of Physics Astronomy, University of California, Los Angeles,
      Los Angeles, CA  90095},  \email{millsb@astro.ucla.edu}}}      
\affil{{$^{14}$}{\it{Department of Astronomy, University of Texas,   1 University Station C1400, 
      Austin, TX 78712},            \email{manuel@astro.as.utexas.edu}}}
\affil{{$^{15}$}{\it{Department of Physics and Astronomy, University of British Columbia, 
       Okanagan,   3333 University Way, Kelowna BC V1V 1V7 Canada} ,              
                                                    \email{erik.rosolowsky@ubc.ca}}}      
\affil{{$^{16}$}{\it{Steward Observatory, University of Arizona,  933 North Cherry Ave.,
      Tucson, AZ 85721 },         \email{wschlingman@as.arizona.edu }}}
\affil{{$^{17}$}{\it{Steward Observatory, University of Arizona,  933 North Cherry Ave.,
      Tucson, AZ 85721 },        \email{yshirley@as.arizona.edu}}}  
\affil{{$^{18}$}{\it{CASA, University of Colorado, UCB 389, 
      Boulder, CO 80309},       \email{Guy.Stringfellow@colorado.edu}}}
\affil{{$^{19}$}{\it{Institute for Astronomy (IfA), University of Hawaii, 640 N. Aohoku Pl., 
       Hilo, HI 96720},               \email{joshw@ifa.hawaii.edu}}}
\affil{{$^{20}$}{\it{Institute for Astronomy (IfA), University of Hawaii,  2680 Woodlawn Dr.,
       Honolulu, HI 96822},     \email{jpw@ifa.hawaii.edu}}}

\begin{abstract}

The Bolocam Galactic Plane Survey (BGPS) data for a six square 
degree region of the Galactic plane containing the Galactic  center 
is analyzed and compared to infrared and radio continuum data.  The BGPS 
1.1 mm emission consists of clumps interconnected by a network of 
fainter filaments surrounding cavities,  a few of which are filled with 
diffuse near-IR emission indicating the presence of warm dust  or with radio 
continuum characteristic of  HII regions or supernova  remnants.   
New 350 $\mu$m images of  the environments of the two brightest regions, 
Sgr A and B, are  presented.   Sgr B2 is the brightest mm-emitting clump
in the Central Molecular Zone and may be forming the closest analog to a 
super star cluster in the Galaxy. The Central Molecular Zone (CMZ) contains 
the  highest concentration of mm and  sub-mm emitting dense  clumps in the 
Galaxy.     Most 1.1 mm features at positive longitudes are seen in silhouette  
against  the 3.6 to 24  $\mu$m background observed by the Spitzer Space 
Telescope.    However,  only a few  clumps at negative longitudes  are 
seen in absorption, confirming the hypothesis that positive  longitude clumps 
in the CMZ  tend to be on the  near-side of the Galactic center,  consistent with 
the   suspected orientation  of the central bar in our Galaxy.   Some 1.1 mm 
cloud surfaces are seen in  emission  at 8 $\mu$m, presumably due to  
polycyclic aromatic hydrocarbons  (PAHs).     A  $\sim$0.2\arcdeg\  ($\sim$30 pc) 
diameter cavity and infrared  bubble between $l \approx$ 0.0\arcdeg\   and 
0.2\arcdeg\ surrounds the Arches and Quintuplet clusters and Sgr A.   The 
bubble contains several  clumpy dust filaments that point toward Sgr A$^*$;   
its potential role  in their  formation  is explored.    
 
Bania's Clump 2,  a feature  near $l$ = 3\arcdeg\  to 3.5\arcdeg\ which  exhibits  
extremely broad molecular  emission  lines  ($\Delta V >$ 150  km~s$^{-1}$),  
contains dozens of 1.1 mm clumps.  These clumps are  deficient in near- and 
mid-infrared emission in the Spitzer images when  compared to  both the inner 
Galactic plane and the  Central Molecular Zone.   Thus, Bania's Clump 2 is 
either  inefficient in forming stars or is in a pre-stellar phase of clump evolution.

The Bolocat catalog of 1.1 mm clumps contains 1428 entries in the Galactic
center between $l$ = 358.5\arcdeg\ to $l$ = 4.5\arcdeg\ 
of which about 80\% are likely to be  within about 500 pc of the center.   The
mass-spectrum  above about 80 M$_{\odot}$ can be described by
a power-law $\Delta N / \Delta M =  N_0 M^{-2.14 (+0.1,-0.4)}$.   The power-law
index is somewhat sensitive to systematic grain temperature variations, 
may be highly biased by source confusion,  and is very sensitive to the spatial 
filtering inherent in the data acquisition and reduction.

\end{abstract}


\section{Introduction}

The formation of massive stars and star clusters remains one of the
outstanding problems in star formation research \citep{stahler00, bally05}.  
The Central Molecular Zone (CMZ) of our Galaxy,  located within about
500 pc of the nucleus, contains about 5 to 10\% of the molecular gas mass
and the largest concentration of massive stars and star clusters in the
Milky Way \citep{MorrisSerabyn1996, Ferriere2007}.   The CMZ hosts 
two of the most massive, densest, and remarkable star forming complexes 
in the Milky Way, Sgr A and Sgr B2 \citep{Yusef-Zadeh2008,Yusef-Zadeh2009}.    
Giant molecular clouds (GMCs)  in the CMZ are one to two orders of 
magnitude denser and have over an  order of magnitude larger line-widths 
than typical GMCs in the Galactic plane beyond 3 kpc from the nucleus 
\citep{bally87,bally88,oka98,tsuboi99, oka01}.
The study of the interstellar medium in the CMZ and the immediate
vicinity of the central black hole is an essential step towards understanding
star and cluster formation in extreme environments such as starburst
galaxies and active galactic nuclei.  The CMZ may be a Galactic analog of 
the nuclear star-forming rings observed in the centers of star forming barred 
galaxies  \citep{KormendyKennicutt2004,KormendyCornell2004, Liszt09}.

The Galactic plane just outside the CMZ between  $l$ = 1.3\arcdeg\ to 
5\arcdeg\  contains several
remarkable, localized cloud complexes with unusually large velocity
extents.    Longitude-velocity diagrams in species such as HI, CO,  and 
high-density  gas tracers such as CS and HCO$^+$ show that near 
$l$ = 355\arcdeg ,  1.3\arcdeg , 3\arcdeg , and 5\arcdeg ,  these cloud 
complexes have velocity extents of 100 to 200 km~s$^{-1}$ in regions 
less than 0.5\arcdeg\ ($\sim$ 75 pc) in diameter 
\citep{dame2001,Liszt06,Liszt08}.   These remarkable cloud complexes 
may trace the locations where gas is entering the dust lanes at the
leading edges of the bar in the center of our Galaxy and passing through
a series of shocks \citep{Liszt06}, or dust lanes along the bar's leading 
edge seen nearly end-on.

A crucial step in the observational study of massive star and cluster 
formation is the identification and characterization of clumps that
will soon form or are actively forming stars,  without biases introduced 
by targeting sources with already known signposts of star formation 
such as masers, IR sources, or HII regions.  Massive stars and
clusters form from cool high-density clumps with very large column
densities and extinctions that can sometimes exceed $A_V = $ 
100 magnitudes.  Such clumps
are best investigated at millimeter and sub-millimeter wavelengths.  
Spectral lines provide excellent diagnostics of line-of-sight motions in a
cloud.    However, the interpretation of various gas tracers that produce 
emission  lines in this portion of the spectrum can be very difficult.    
Variations in tracer abundances caused by depletions 
and complex chemical processing, uncertainties in excitation conditions,
and the impacts of radiation fields and shocks make the derivation of
column densities, masses, and other physical properties of clumps
highly uncertain.  

The continuum emission from  warm dust provides a somewhat more 
reliable tracer of the column  density and clump masses.     At  1.1 mm,  
$h \nu / k ~\approx$ 13 K,  and most sources are expected to be optically thin.   
Thus,  the column  density is roughly proportional to the temperature and 
optical depth of  the emitting medium.   
Shorter  wavelength observations  can be more difficult to obtain due  
to lower  atmospheric transmission and  greater variations in sky  emission 
(``atmospheric"  or  ``sky noise'').  Furthermore,  the stronger  dependence 
of column density on   temperature when  $h \nu / k ~>$ T$_{dust}$ makes
the derivation of physical parameters  more difficult.

Contamination of the dust continuum emission in the Bolocam 1.1 mm 
filter by spectral lines and free-free emission is likely to be small, less than a
few percent in almost all cases.  As discussed in detail in \citet{Aguirre2010},
the Bolocam 1.1 mm filter excludes the bright 1.3 mm CO lines.   In the
worst case, that of the hot cores in Sgr B2, contamination by other spectral
lines can account for at most 20\% of the detected 1.1 mm flux.  Such spectral
line emission is much weaker in the other clumps.  Contamination by free-free
emission  would only occur  from ultra-compact
HII regions.  However, such objects tend to be only a few arc-seconds
in diameter,   and will be  beam diluted and
completely overwhelmed by the larger-angular scale dust continuum.
Comparison of the 20 cm radio continuum emission (see discussion of  Figure 11 
below) shows that 20 cm radio and 1.1 mm dust continuum emission are not 
correlated, supporting   the argument that there is little contamination of the 1.1 
mm fluxes by   free-free emission.

This paper presents two new data sets obtained with the 10.4 meter Caltech
Submillimeter Observatory;  1.1 mm continuum observations  obtained with Bolocam, 
and 350 $\mu$m data obtained with SHARC-II.   These data are compared 
with a variety of previously published infrared and radio continuum data. 
The new data presented here are part of a 1.1 mm continuum  Bolocam Galactic 
Plane Survey  (BGPS) \citep{Aguirre2010,Rosolowsky2010,Dunham2010} 
that has mapped about 170 square degrees of the northern Galactic Plane.  
The BGPS data were released into the public domain 
on 24 June  2009  through the Infrared Processing and Analysis Center (IPAC) 
at the  California Institute for Technology 
(http://irsa.ipac.caltech.edu/data/BOLOCAM$\_$GPS/).

This paper presents an analysis of BGPS clumps in the region between  
$l$ = 358.5 and +4.5\arcdeg\ which contains the Galactic nucleus,  the 
Central Molecular Zone (CMZ) of dense, turbulent,  and warm molecular 
clouds  \citep{MorrisSerabyn1996}, and the most  prominent 
high-velocity-dispersion cloud complex,  Bania's Clump 2  \citep{Bania_Clump2} 
located  at $l$ = 3.2\arcdeg .    Additionally,  350 $\mu$m images are presented 
of  the  two most luminous star forming complexes in the CMZ, Sgr A  and its  
neighboring cloud complexes, and the  Sgr B2  region.   
Previous (sub)millimeter studies of the Galactic  center   
\citep{lis91,lis94,liscarlstrom94,lis98,lis01} tend to be based on maps  of  small 
fields.    The wide-field BGPS data are presented and compared to  images 
obtained with the Spitzer Space Telescope at 3.6 to 8.0 $\mu$m  
\citep{Arendt2008} and  Spitzer 24 $\mu$m images now in the public domain
\citep{Yusef-Zadeh2009}. 
BGPS is also compared with the SCUBA / JCMT 450 and 850 $\mu$m dust 
continuum maps \citep{pierce-price00} 
from an approximately 2\arcdeg\  by 0.5\arcdeg\ region containing the CMZ.  
Finally, the 1.1 mm data are compared with  the 20 cm radio continuum 
\citep{yusef-zadeh2004}  to constrain recent and on-going 
massive star formation activity in the CMZ  and Bania's Clump 2.   

This paper is organized as follows:  
Section 2 describes the
observations and analysis methods, including a discussion of mass estimation
(Section 2.1), and the determination of dust properties for the two fields where
350 $\mu$m observations are presented (Section 2.2).    
Section 3 presents the observational results.  
Section 3.1 
presents a discussion of clump mass spectra extracted from the Bolocat catalog
\citep{Rosolowsky2010}.   
Section 3.2
describes the  emission from the Central Molecular Zone containing the Sgr A, 
B1, B2,  and C complexes and its relationship to objects detected at other 
wavelengths.    
Section 3.3  presents a more detailed discussion of the environment of Sgr A.    
Section 3.4 presents a discussion of the  emission along  the inner Galactic 
plane and Bania's Clump 2.
Section 3.5 presents a short discussion of features likely to be located 
in front  of or behind the center of the Galaxy in the Galactic plane.   
Section 4.1 explores the energetics of the cavity in the Sgr A region.
Section 4.2 discusses the clumps that may be fueling star formation in Sgr A.
Section 4.3 discusses the low star formation rate in Bania's Clump 2.   
Conclusions  are  presented in Section 5.  Detailed descriptions of some of 
the brighter or  otherwise more noteworthy sources are given in an Appendix.
A  tabulation of Bolocat clumps located within the field of view 
covered by Figure \ref{fig1} and \ref{fig2}  along with their derived 
properties is  given in Table 3 in the electronic version of this paper.

\section{Observations}

The observations reported here were obtained with Bolocam\footnote
{http://www.cso.caltech.edu/bolocam} between  June 2005 and July 2007
on the Caltech Submillimeter Observatory (CSO) 10-meter diameter telescope.
SHARC-II observations were obtained during the best weather on these
observing runs and in April 2008.     The observations used in this paper were 
obtained  on the dates listed in  Table~1. 

Bolocam \citep{glenn03} is a 144-element bolometer array with between 
110 and 115 detectors working during the observations reported here.
The instrument consists of a monolithic wafer of silicon nitride
micromesh AC-biased bolometers cooled to 260 mK.  All data were
obtained with a 45 GHz bandwidth filter centered at 268 GHz ($\lambda$
= 1.1 mm) which excludes the bright 230 GHz J=2-1 CO line.  However, as
discussed in \citet{Aguirre2010},  the effective central 
frequency is 271.1 GHz.   The individual bolometers are arranged on a 
uniform hexagonal grid.  Each bolometer has an effective Gaussian 
beam with a FWHM diameter of 31\arcsec , but the maps presented here 
have an effective resolution of 33\arcsec\  due to beam-smearing by the
scanning strategy, data sampling rate, and data reduction as discussed 
by \citet{Aguirre2010}. At an assumed distance of 8.5 kpc to the Galactic 
center, the 33\arcsec\  diameter Bolocam effective resolution corresponds to a 
length-scale of  1.36 pc and 1\arcdeg\ corresponds to  about 148.7 pc.  
The instantaneous field-of-view (FOV) of the focal-plane array is 7\arcmin .5
but the effective filter function limits spatial frequency sensitivity to about 3\arcmin\
to 5\arcmin .   Bolocam observations  were only obtained when atmospheric 
conditions  were clear with a 225 GHz  zenith opacity ranging from 0.06 to 
0.15.  During better observing conditions, 350 $\mu$m SHARC-II 
observations were  obtained as discussed below.

Data were obtained by raster-scanning Bolocam on the sky at a rate
of 120\arcsec\  per second.  Each `observation' consisted of three pairs
of orthogonal raster-scans each covering a full square-degree or
a 3\arcdeg\ by 1\arcdeg\  field 
in Galactic coordinates with scan-lines separated
by 162\arcsec\ on the sky.  The second and third pairs of orthogonal
raster-scans were offset from the first pair by $\pm$ 42\arcsec .  
Four to over a dozen `observations'  taken on different
days were averaged to obtain the maps presented here.

The Bolocam raw data were reduced using the methods described in detail 
in \citet{Aguirre2010} based on the earlier techniques described in  
\citet{enoch06} and  \citet{laurent05}.    The new reduction pipeline handles 
the large dynamic range and complex structure characteristic of Galactic star 
forming regions better.   In summary, real-time pointing data were merged with 
the time-series data from each bolometer whose exact location in the CSO focal-plane 
is known.  Principal component analysis (PCA) is used to remove signals 
common to all  bolometer  channels,  such as the DC level from atmospheric 
loading and sky brightness fluctuations.     Simulations have shown that for data 
containing faint sources which  are sparsely distributed,  the removal of the 
first three  orthogonal  PCA components suffices to correct for atmospheric 
effects produced by sky noise  (see Enoch et al. 2006 for details).   However, 
in the Galactic plane the presence of bright ($>$ few Jansky) sources and 
crowding required the removal of 13 PCA components.  Following removal of
signals attributed to the atmosphere, the maps are then ``iteratively mapped"
to restore astronomical structure on angular scales comparable to the size 
of the Bolocam array \citep{enoch06,Aguirre2010}.  
The final data were sampled onto a uniform 7\arcsec.2  grid.
The data were flux calibrated using observations of the planets Mars, Uranus,
and Neptune.    In the Galactic center fields 
presented here, the 3 $\sigma$ r.m.s. noise is about 40 mJy beam$^{-1}$.

The V1.0 data released at the IPCA site use 7\arcsec.2 pixels.  The data value
in each pixel corresponds to the flux measured at that location on the sky by the
33\arcsec\ effective beam and is in units of Jy/beam.
The area under a 33\arcsec\ gaussian beam is the same as the area under a top-hat
beam with a radius of $r_B = FWHM / (4 ln 2)^{1/2}$ = 19.8\arcsec\ or an aperture
with a  diameter of $\approx$ 40\arcsec .    There are $\pi r_B^2 / 7.2^2$ = 23.8 pixels
in such an aperture. Thus, to convert the data in the V1.0 images from Jy/beam to 
Jy/pixel where each pixel is a 7\arcsec.2 square, divide the data by 23.8.   To measure
the total flux over some aperture, sum the pixels values 
in the aperture (after converting to Jy/pixel units).  

Only flux produced by structure smaller  than about one-half of the Bolocam 
array-size of  approximately 7.5\arcmin\  can be measured reliably.    
Larger-scale emission  smoothly distributed in 2 dimensions  is  under-represented.    
However,  long filaments  or structure in which  at least one dimension is compact 
compared to the Bolocam array size can be recovered.   A more detailed discussion 
of the spatial  transfer-function of the BGPS is  given in \citet{Aguirre2010}. 

Comparison between Bolocam and 1.2 mm data acquired with two other
instruments, MAMBO on the IRAM 30 meter \citep{Motte2003,Motte2007}
and SIMBA on the SEST \citep{Matthews2009} 15 meter, are presented
in \citet{Aguirre2010}.    \citep{Aguirre2010} find that  the BGPS Version 
1.0  Bolocat and image data release nead a mean correction factor of 1.5 
to be applied to align BGPS fluxes with these previous measurements and 
to yield dust spectral indices when compared to data at other wavelengths.

During the very best weather conditions at the CSO when the zenith sky opacity at
225 GHz was less than about 0.06, observations of selected fields were obtained
at 350 $\mu$m with the 384 element SHARC-II focal plane array
\citep{Dowell2003}.  
SHARC-II  consists of a 12 by 32 pixel array of  ``pop-up" bolometers providing
an effective resolution (beam diameter) of 
about 9\arcsec\  on the CSO.    The SHARC-II  350 $\mu$m filter is matched to the 
atmospheric window in this part of the spectrum and has an effective
passband of $\Delta \lambda / \lambda$ =  0.13 (111 GHz) centered at $\lambda$
350 $\mu$m.   

Observations of five slightly  overlapping  fields in the Sgr A region  were obtained in 
June 2005 using   10\arcmin\  by 10\arcmin\  BOX$\_$SCANs.  A second set of
observations were obtained on 6 April 2008 using a 30\arcmin\ by 30\arcmin\
BOX$\_$SCAN covering the fields near Sgr A using a scan rate of 60\arcsec\  per second
for a total integration time of 40 minutes.     The Sgr B2 observations were obtained
on  31 March 2004  with a 13\arcmin\ $\times$ 15\arcmin\  BOX$\_$SCAN with a scan-rate
of 30\arcsec\ per second.   The total  integration time was 20 minutes.  
While the June 2005 data were reduced with the facility software package, 
CRUSH \citep{Kovacs2008}, the 350 $\mu$m images shown  in the figures
were processed with the code SHARCSOLVE written by co-I Dowell.    Minimal filtering
was used;  many iterations fit only an offset and gain to each bolometer time-stream, 
a spatially flat atmospheric signal at each time,  ending with a few iterations of fitting 
an atmospheric gradient over the detector array.

Due to their  better uniformity, only the March 2004 and April 2008 images 
are used in the figures.   The June 2005 observations were used as an independent check
on flux calibration and image fidelity.     Although there are slight differences in the
background intensities, the overall structure of the emission during the two epochs
agrees extremely well.  The flux calibration on compact objects such as G359.984,-0.0088]
agrees to within 10\% (29  vs 32 Jy/beam) once an annular background is subtracted.   
The 3$\sigma$ sensitivity  of these observations is about  1Jy/beam at 350 $\mu$m.

\subsection{Mass and Column Density Estimation}

Estimation  of the dust column densities and masses from the 1.1 mm emission 
requires  knowledge of the dust properties such as emissivity, spectral index,  
and temperature, which requires  input from additional data.   Column densities,
densities,  and masses  are estimated for a subset of the brightest clumps  based  
on dust  parameters appropriate for  the Solar neighborhood  (Table~2).    
Masses, column densities, and densities estimated for all Bolocat entries 
are  tabulated in Table 3 in the electronic version of this paper.     For all mass
estimation given in this paper, the images and fluxes reported in the V1.0 data 
release have been multiplied by an empirically determined scaling factor of 1.5
as discussed above.

The mass within the  33\arcsec\  BGPS effective beam
(or equivalently, in a 40\arcsec\  diameter circular aperture) is given by
$$
M  = {{1.0 \times 10^{-23} ~{ S_{1.1} (Jy) } D^2} \over  {\kappa _{1.1} B_{1.1}(T) }}
$$
where $ B_{1.1}(T)$ is the Planck function for dust temperature $T$,
and
$\kappa _{1.1}$ is the dust opacity at $\lambda$ = 1.1 mm 
in units of cm$^2$~$g^{-1}$.  As a function of wavelength,
$\kappa _{\lambda} = \kappa _{0} ( \lambda _{0} / \lambda )^{\beta}$
where $\beta$ is the dust emissivity.  \cite{hildebrand83}
calibrated $\kappa$ at $\lambda$ = 0.25 mm, giving a value
$\kappa _{0.25}$ = 0.1 cm$^{2}$~g$^{-1}$ for interstellar dust
and a gas-to-dust ratio of 100.  For $\beta$ = 1.5, the Hildebrand opacity
implies $\kappa _{1.1}$ = $1.08 \times 10^{-2}$~cm$^{2}$~g$^{-1}$.
\cite{ossenkopf94}  (OH94) calculated the opacity of grains with
ice mantles that have coagulated for $10^5$ years at a gas density
of $10^6$ cm$^{-3}$.  \citet{enoch06}  interpolated logarithmically the 
OH94 opacities to obtain
$\kappa _{1.1}$ = $1.14 \times 10^{-2}$~cm$^{2}$~g$^{-1}$, 
close to the estimate based on the Hildebrand opacity. 
As discussed in \citet{Aguirre2010} the effective central frequency of
the Bolocam 1.1 mm filter is 271.1 GHz.
Log-interpolating the OH94 opacity at 1.00 mm  to this frequency implies 
$\kappa _{1.1}$ = $1.14 \times 10^{-2}$~cm$^{2}$~g$^{-1}$.  
Adopting this value  gives
$$
M =  14.26  (e^{13.01/T(K)} - 1)  S(Jy)  D^2_{kpc} ~~~~~~~(M_{\odot})
$$
where $D^2_{kpc}$ is in units of 1.0 kpc and $T$ is in Kelvin.     

The greater metallicity of the Galactic center may result in a larger 
dust opacity, and possibly a different dust emissivity law.   This effect has not
been taken into account in the present analysis.   The spatial filtering
inherent in BGPS, and large uncertainties inherent in the estimation
of masses using spectral lines, it is not at present possible to 
determine clump masses with sufficient precision to detect a difference
in the dust properties. Use of a larger dust opacity would decrease the 
estimated masses and column densities given here.    

The dust temperature in the Central Molecular Zone has been studied
by \citet{Sodroski1994}  who used COBE/DIRBE 140 to 240 $\mu$m data to 
conclude that 15 to 30\% of the emission arises from molecular clouds
with temperatures of about 19 K, 70 to 75\% arises from the HI phase
with a dust temperature of 17 to 22 K and less than 10\%  comes from
$\sim$ 29 K dust in the extended HII phase.
\citet{RodriguezFernandez04} used the Infrared Space Observatory
(ISO) to study a representative sample of 18 molecular clouds not associated 
with prominent thermal radio continuum emission.  They found typical dust 
temperatures, measured from $\lambda$ = 40 to 190 $\mu$m,  required a 
combination of a  cold, 15 K component and a warm component with
a temperature ranging from 27 to 42 K.  Thus, in the analysis presented 
here, the dust column densities and masses are parameterized in terms of
a fiducial dust temperature of 20 K.     For T =  20 K, 
$$
M = 13.07 ~ S(Jy)  D^2_{kpc}     ~~~~~~~(M_{\odot})
$$
which at the 8.5 kpc distance to the Galactic center gives
$M = 944 ~S(Jy)  ~ M_{\odot}$.    This mass can be converted to a column
density averaged over the beam,
$$
N(H_2) = {{M} \over { \pi \mu_{H2}  m_H  r_B^2}} ~~~~~~~(cm^{-2})
$$
where $M$ is in grams, $\mu_{H2}$ = 2.8 is the mean molecular 
weight per hydrogen molecule \citep{Kauffmann2008}, 
$r_{B} = D_{GC} \times FWHM / ( 205265 \times  \sqrt{4~ ln{2}})$ is the effective 
beam radius in centimeters at the Galactic center,  and 
$D_{GC}$ = 8.5 kpc.   
Following \citet{Kauffmann2008},  the mean molecular weight per 
particle,  $\mu_p \approx 2.37$ should be used when estimating quantities 
such as collision rate or pressure.   For BGPS, $FWHM  \approx  33$\arcsec ,
giving effective beam radius or  $R_B = 19.8$\arcsec  ,
which at a distance of 1 kpc corresponds to $r_B(kpc) = 2.97 \times 10^{17}$ cm
and at the distance to the Galactic center is $r_B = 2.52 \times 10^{18}$ cm.
The resulting column density per beam is 
$$
 N(H_2) =  2.19 \times 10^{22} [e^{13.0 / T(K)} - 1] ~S(Jy) 
 \approx 2.0 \times 10^{22} ~ S(Jy)  ~~~~~~~~(cm^{-2})
$$
where the right side was evaluated for T = 20 K.   This column density 
corresponds to a visual extinction of 
$A_V \approx 10.7$ magnitudes assuming the usual Bohlin \& Savage
conversion of $N(H) = 1.9 \times 10^{20} ~cm^{-2}$ per visual magnitude 
in the standard Johnson / Cousins V filter centered at 0.55 $\mu$m .    These
column densities can be converted into surface densities;
$\Sigma (g~cm^{-2}) = 4.65 \times 10^{-24} N(H_2)$ or 
$\Sigma (M_{\odot}~pc^{-2}) = 0.048  A_V$. 

For extended sources,  masses are given by 
$M = \mu m_H N(H_2) A_S$ where $A_S$ is the area subtended by
the aperture used to estimate the flux.   At  the Galactic center,  the mass is given by 
$M_{tot} = 944 ~ \langle S_{1.1} (Jy) \rangle ~  [ \Omega _S / \Omega_B]  $~M$_{\odot}$  
where $< S_{1.1} (Jy) >$  is the flux averaged over the
solid angle $\Omega _S$  subtended by the measurement aperture
and $\Omega_B = \pi r_B^2 \approx 2.9 \times 10^{-8}$  is the BGPS beam
solid angle in steradians.  
A  temperature of $T$ = 20 K is assumed to evaluate the numerical 
coefficient. 

An estimate of the volume density can be made by assuming that
each clump is spherical and uniformly filled with emitting matter.   Table~2, 
columns 9 and 10 give the densities estimated from the fluxes measured
in 40\arcsec\ and 300\arcsec\ diameter apertures under the assumption that 
each clump is either 40\arcsec\ or 300\arcsec\ in diameter.  Interferometric
measurement  of clumps in the Solar vicinity show that much of the
millimeter continuum emission arises from regions having diameters of order 
$10^3$  AU or less.    Thus, BGPS clumps are likely to consist of clusters of
unresolved objects much smaller than the beam.   The BGPS-based
density  estimates are volume averaged quantities.    Inside cores, the
densities are likely to be much higher while in-between cores, densities
may be lower.   

Table 2 lists the locations, fluxes,  masses and densities for a sample of 
regions, many of which are marked in Figure \ref{fig9}. $S_{40}$ gives 
the flux measured with the FWHM = 33\arcsec\ Bolocam 
effective resolution at the location of the brightest emission in each source. 
The uncertainties in the listed quantity range from about 20\% to over 30\% for 
bright sources due to a combination of calibration errors, and incomplete flux 
recovery  in the iterative mapping phase of data reduction.    The 1.1 mm 
images have also been smoothed with a ``tophat"  function having a 20 pixel 
radius to provide an estimate of the total flux produced
by extended  objects in a 300\arcsec\ diameter, circular region.  At the
distance of the Galactic center (8.5 kpc), these  masses  are 
given by  $M(300) =  5.0 \times 10^4 ~S_{300}(Jy) $~M$_{\odot}$ for 20 K dust
where $S_{300}(Jy) $ is the average flux density in the 300\arcsec\ 
aperture.   (Note that
for several high-latitude regions suspected to be in the foreground,   
a distance of 3.9 kpc is assumed in estimating masses as indicated in
the comments column.)  These mass estimates are lower bounds because
they only refer to the flux in small-scale structure to which Bolocam is most 
sensitive.     Because BGPS data starts to lose sensitivity to structure
on scales ranging from 3\arcmin\  to 6\arcmin ,  300\arcsec\  is the
largest aperture for which reliable photometry can be obtained.   As discussed
below,  comparison of mass determinations in the 300\arcsec\  aperture  
to previously published measurements for regions such as Sgr A and Sgr B2 
agree to within the estimated uncertainties.

\subsection{Dust Properties From 350 $\mu$m and 1.1 mm data}

Two fields around Sgr A and Sgr B2 were observed at 350 $\mu$m.  The
350 $\mu$m to 1100 $\mu$m flux ratio can be  used to constrain dust 
temperatures or emissivities.   Three  corrections must be applied to the 
SHARC-II data prior to comparisons with BGPS;  matching of angular resolution, 
correction for signals-picked  up in the sidelobes, and matching of the spatial 
frequency response functions.   No correction has been made for contamination 
by spectral lines.

The SHARC-II 350 $\mu$m images were obtained with a 9\arcsec\  FWHM beam.   
Thus, they have to be convolved with a gaussian function with 
$\sigma$ =  13.5\arcsec\ to match the 33\arcsec\ effective BGPS beam.   Because
the data values in both the 350 and 1100 $\mu$m images are in units of 
Janskys per beam,  the pixel values in the convolved SHARC-II images must be 
multiplied by the ratio of beam areas (13.4 when scaling from 9\arcsec\ to
a 33\arcsec\ beam).       

Inspection  of the 350 $\mu$m beam delivered by the CSO on bright point sources 
such as planets shows that in addition to the main 9\arcsec\ beam, there is a low-level error 
pattern several arc-minute diameter  with a roughly hexagonal shape.   The 
main-beam efficiency of the CSO at 350 $\mu$m is only about  $0.4 \pm 0.15$.
In a complex field such as the Galactic center with an abundance 
of bright, extended  emission,   fluxes will be over-estaimated when compared 
with a point source calibrator in the main beam due to the signal entering the sidelobes.    
Because bright emission 
near both Sgr A and Sgr B2 is extended on scales of many arc-minutes, we assume that 
about 30\% of the signal at a typical location in the maps arises from signal picked
up in the sidelobes.   As an approximate correction for this effect, we scale the 
convolved 350 $\mu$m images by a factor  $0.7$ to correct for clumps contributing
emission through the side-lobes.

Comparison of the two 350 $\mu$m maps of the Sgr A region discussed above 
shows that the rapidly-scanned SHARC-II images recover flux on larger
angular scales than the 1.1 mm BGPS images.  Thus, the SHARC-II images
are spatially filtered with an "unsharp mask".   The mask in constructed by 
convolving  the 350 $\mu$m with a gaussian function having a kernel 
$\sigma _k$, and subtracting a fraction, $f$ of the mask  from the original image.  
The optimal values of $\sigma _k$ and $f$ were chosen by finding the closest
match between the spatial structure of the  resulting "unsharp-masked" SHARC-II
image with  the corresponding BGPS image.  For the Sgr A field, the best choice
is  $\sigma _k$ = 90 \arcsec\  (FWHM =  212\arcsec ) and $f$ = 0.8; for Sgr B2, 
$\sigma _k$ = 68\arcsec\  (FWHM = 160\arcsec ) and $f$ = 0.5.  
The beam-matched, spatially filtered  350 $\mu$m images 
were divided by  the corresponding 1.1 mm BGPS images to form  flux-ratio 
maps.

The observed intensity of continuum emission at each location in a map
is given by 
$S_{\nu} = \Omega_{beam} [1 - e^{- \tau_{\nu d}}] B_{\nu}(T) ~~(Jy)$
where $\Omega_{beam}  \approx 2.9 \times 10^{-8} sr $ is the beam solid 
angle corresponding to the 33\arcsec\ effective resolution of the 1.1 mm
data and the convolved, beam-matched,  350 $\mu$m observations,
$\tau_{\nu}$ is the optical depth of the emitting dust, and
$B_{\nu}(T)$ is the Planck function at each frequency.  The optical depth in
the sub-mm to mm regime scales with frequency as $\nu ^{\beta}$ where
$\beta$ is the emissivity power-law index.  For small grains, $\beta$ is
close to 2 in the mm to sub-mm regime but can be smaller if the grains
have experienced substantial growth or are coated by ice mantles
\citep{ossenkopf94}.    In the optically thin limit, the 
flux-ratio at each point  in a ratio map is given by 
$$
R = S_{350 \mu m} / S_{1.1 mm} 
    =   {   { \nu _{350}^{3 + \beta} [exp (h \nu _{1.1} / k T_d) -1]}  \over
              { \nu _{1.1}^{3 + \beta} [exp (h \nu _{350} / k T_d) -1]}           }
$$
which can not be solved analytically for $T_d$.   A look-up table is generated 
for each choice of $\beta$ and $T_d$ and used to analyze points in the ratio 
maps.   For hot $\beta$=2.0 dust where both 350 $\mu$m and 1100 $\mu$m are
in the Rayleigh-Jeans limit, the observed intensity ratio should be
around $R$ = 100.  Larger ratios require $\beta > $2.
Where the pixels have good signal at both 350 $\mu$m and 1.1mm, 
flux ratios have values  $R < $ 70, corresponding to dust temperatures less than
70 K for $\beta$ = 2 dust.     The interpretation of these results will be discussed
below.   

\section{Results}

\begin{figure}
\epsscale{1.0}
\center{\includegraphics[width=1.1\textwidth,angle=90]{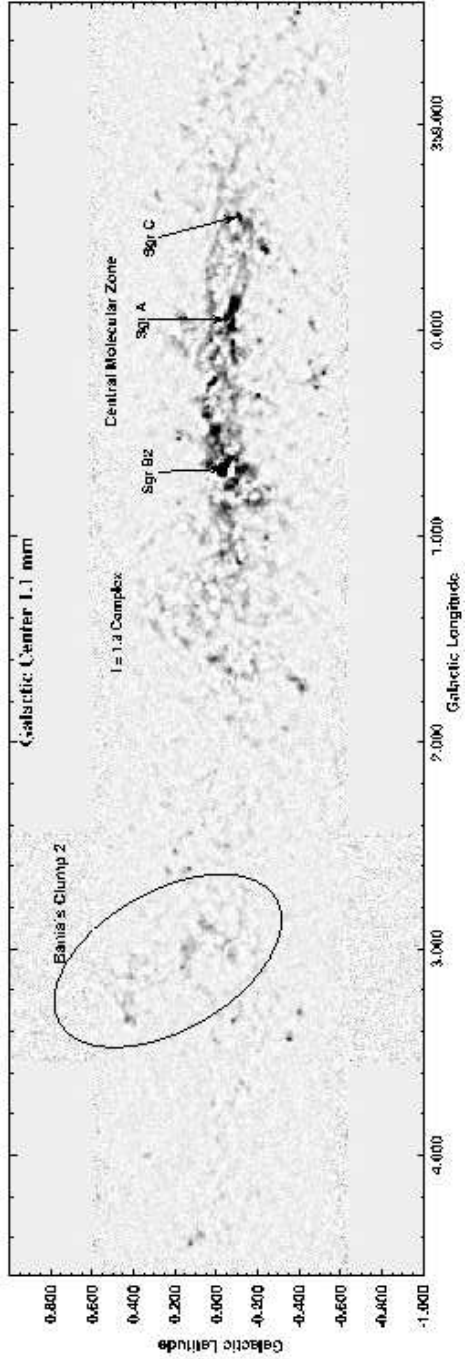}}
\caption{A 1.1 mm BGPS continuum image showing the 
central parts of the Galactic Plane from $l$ = 358.5\arcdeg\ 
to 4.5\arcdeg .   This image has been processed with 13 PCA
components and 50 iterations of the iterative mapper.  
While this level of processing restores much of the faint 
structure, it removes flux from bright sources ($>$ about 5 Jy) 
and creates significant negative bowls around them.   
The display presented here uses a logarithmic scale
from -0.5 to 100 Jy/beam.}
\label{fig1}
\end{figure}

\begin{figure}
\epsscale{1.0}
\center{\includegraphics[width=1.1\textwidth,angle=90]{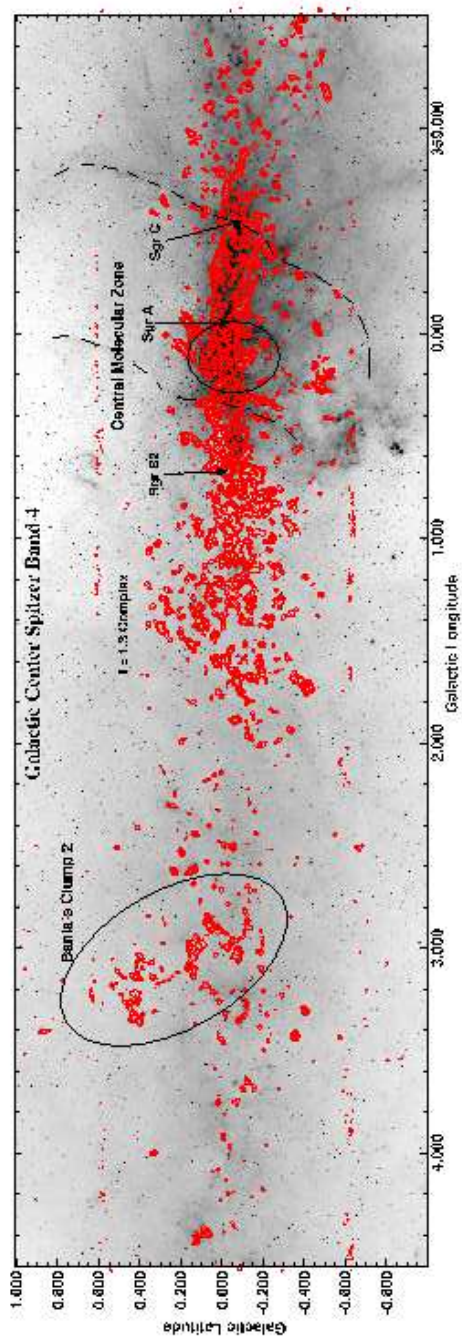}}
\caption{
BGPS 1.1 mm contours superimposed on 
the Spitzer Space Telescope 8 $\mu$m image shown on a logarithmic
intensity scale and at the same spatial scale as Figure 1.  Various regions
discussed in the text are marked.    Bania's Clump 2 is shown on the left and the $l$ =
1.3\arcdeg\ complex is located in the middle.
The oval to the left of Sgr A shows the approximate outline of the
infrared cavity that contains the Arches and Quintuplet clusters and is rimmed by
mid-IR and radio continuum emission.   The dashed line-segments
illustrate the apparent walls of a larger cavity bounded by Spitzer 8 and 24
$\mu$m emission.  The interior of this region is filled with 20 cm radio continuum.
See text for discussion.  The 1.1 mm contour levels are at
0.10, 0.20, 0.37, 0.73, 1.4, 2.7, 5.3, 10.3, and 20 Jy/beam.}
\label{fig2}
\end{figure}

The inner two degrees of the Galactic plane (Figures \ref{fig1} and \ref{fig2}) 
contain the brightest millimeter-wave continuum emission in the 
sky.   The Central Molecular Zone (CMZ, which consists of the high-surface
brightness region between $l ~ \approx$ 359\arcdeg\ to 1.8\arcdeg ) 
has a radius of about 200 to 300 pc and 
contains the highest concentration of dense gas and dust in the Galaxy
\citep{MorrisSerabyn1996}.    

\begin{figure}
\epsscale{1.0}
\center{\includegraphics[width=1.2\textwidth,angle=90]{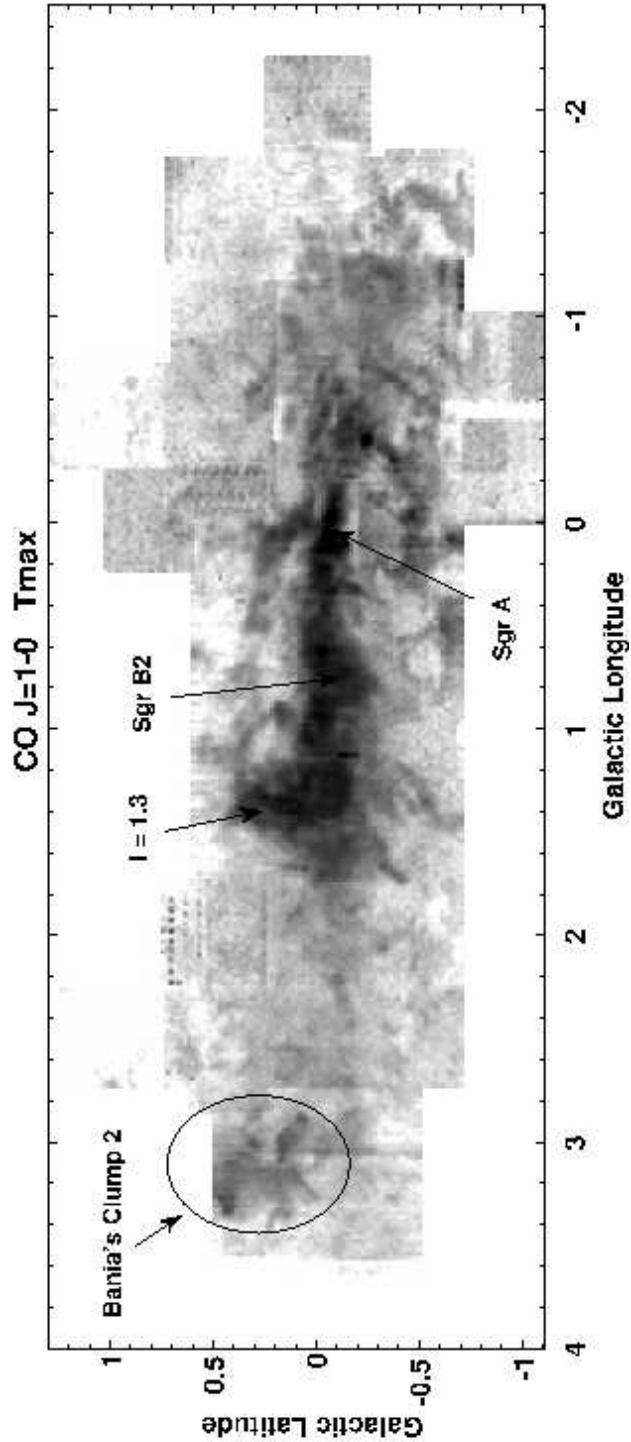}}
\caption{The Central Molecular Zone in the the Galactic center as
traced by the peak antenna temperature, T$_A^*$ of the  J =1--0 
$^{12}$CO  transition.    The location of Bania's Clump 2 is indicated 
along with some of the other prominent Galactic center features.
Taken from the AT\&T Bell Laboratories  data presented by 
\citet{bally87,bally88}. }
\label{fig3}
\end{figure}

Figure \ref{fig3} shows an image of the peak antenna temperature of the 
$^{12}$CO J = 1 -- 0 line taken from the data set presented by \citet{bally87,bally88}.
Comparison of Figures  \ref{fig1} through \ref{fig3} and maps of dense gas 
tracers such as CS \citep{MiyazakiTsuboi2000} shows that the brightest 1.1 mm 
continuum emission is closely associated  with the brightest CO and CS emission.    

\begin{figure}
\epsscale{1.0}
\center{\includegraphics[width=1.2\textwidth,angle=90]{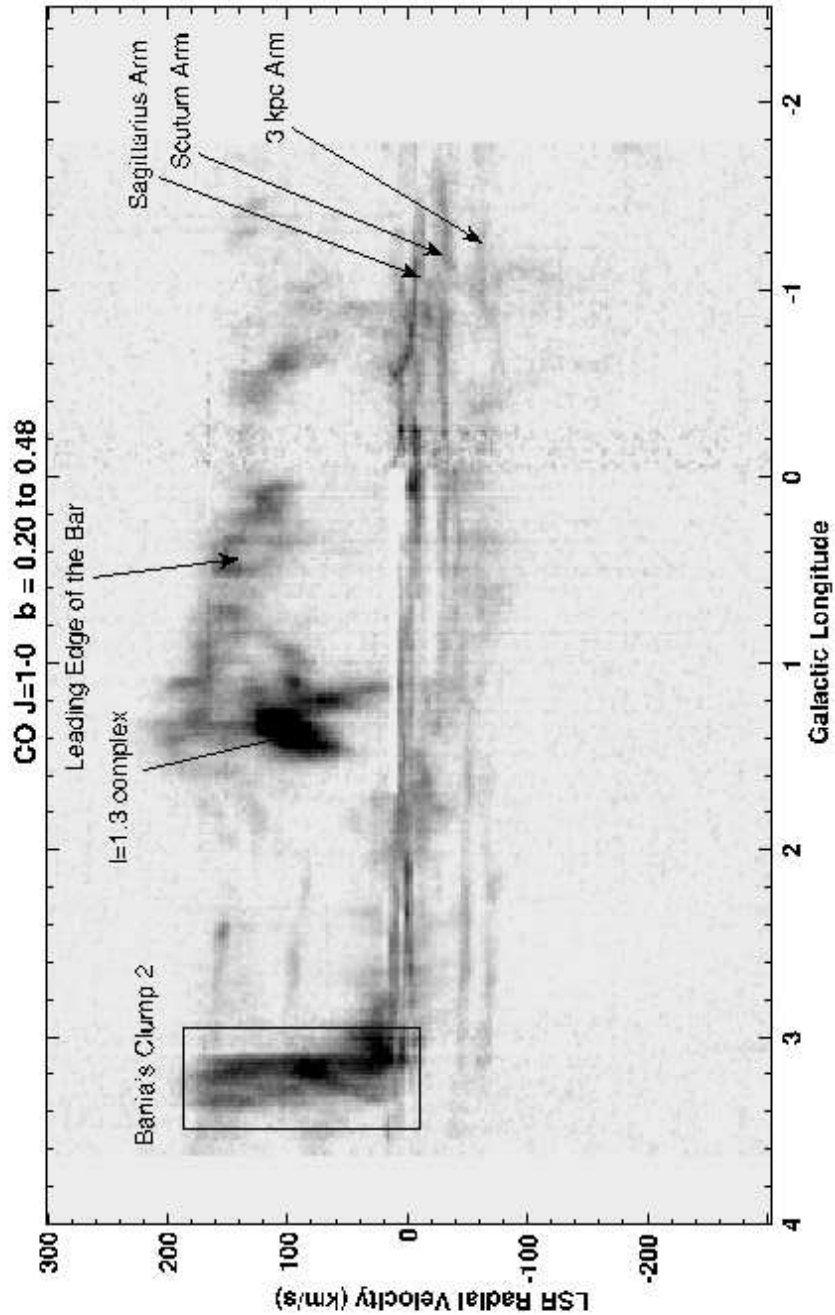}}
\caption{A longitude-velocity diagram showing the Central Molecular Zone 
in the Galactic center as traced by the peak antenna temperature, 
T$_A^*$ of the  J =1--0  $^{12}$CO  transition averaged over the Galactic 
latitudes from $b$ = 0.20\arcdeg\ to 0.48\arcdeg .  The location of Bania's Clump 2
is indicated along with some of the other prominent Galactic center features.
Taken from the AT\&T Bell  Laboratories  data presented by \citet{bally87,bally88}. }
\label{fig4}
\end{figure}

Figure \ref{fig4} shows the spatial-velocity behavior of
CO-emitting gas at positive latitudes integrated from $b$ = 0.20 to 0.48\arcdeg .
Emission from the Galactic disk is confined to velocities
between $V_{LSR}$ = -- 70 and +10 km~s$^{-1}$  and can be distinguished from
the Galactic center gas by its relatively narrow line-widths of order $\Delta V \sim$
few km~sec$^{-1}$.     Foreground CO emission is mostly 
associated with the nearby Sagittarius Arm (near $V_{LSR}$ = 0 km~s$^{-1}$), 
the Scutum-Centaurus Arm (near $V_{LSR}$ = -35 km~s$^{-1}$), and
the 3 kilo-parsec Arm (near $V_{LSR}$ = -60 km~s$^{-1}$).  The systematic
progression in (negative) radial velocity as one approaches the Galactic center
provides evidence for the presence of a central bar \citep{binney1991}.
The high velocity CO feature (near $V_{LSR}$ = 100 to 200  km~s$^{-1}$)
between $l$ = --1.5\arcdeg\ and +2\arcdeg\   has been called the  ``expanding 
molecular  ring"  (EMR) which in the barred models of the Galaxy is
interpreted as gas in $x_1$ orbits located mostly along the leading edge of the 
central bar (labeled as ``leading edge of the bar''  in Figure \ref{fig4})
whose  positive longitude side is thought to be closer to us
\citep{MorrisSerabyn1996}.     The  major axis of the bar is thought to have an 
angle with respect to our line-of-sight  between 20\arcdeg\ to 
45\arcdeg \citep{binney1991, Bissantz03, Ferriere2007, Pohl08}.  

Figures \ref{fig3} and \ref{fig4} show the remarkable molecular  
feature  known as Bania's Clump 2 \citep{Bania_Clump2}.  This feature is localized
to a roughly 1/2 degree diameter ($\sim$70 pc) region in the spatial direction.
But, as illustrated by Figure \ref{fig4}, it extends over a velocity range of order
200 km~s$^{-1}$.   Figures \ref{fig1} and \ref{fig2} show that Bania's Clump 2 
is prominent at 1.1 mm, consistent with the previous detection of emission by 
high density gas tracers.  Throughout the Galactic plane,  bright dust continuum 
sources generally indicate the presence of embedded massive stars or clusters.   
But, as discussed  below,  there is virtually no massive star or cluster formation 
in this region  as indicated by the lack of infrared and radio  continuum emission. 

The 1.1 mm data presented here are qualitatively similar to the recent ATLASGAL
survey of the Galactic plane conducted with the LABOCA array on the APEX 
12 meter diameter telescope in Chile at a wavelength of 870 $\mu$m 
\citep{ATLASGAL2009}.    Although the angular resolution of ATLASGAL is nearly
a factor of two better than BGPS,  the images presented in  \citet{ATLASGAL2009} 
show similar structures to what is seen in BGPS.  Specifically, the CMZ and 
Bania's Clump 2 dominate the emission in the region between $l$ = 358.5\arcdeg\
and 4.5\arcdeg .   A quantitative comparison of fluxes in ATLASGAL, BGPS, and
eventually Herschel Space Observatory images, will enable the measurement
of dust temperatures and emissivities.

\subsection{Clump Mass Spectra in the Galactic center}

The Bolocat clump catalog contains  the highest concentration of  clumps in 
the sky towards  the inner few degrees of the Galaxy, and  the 1.1 mm emission 
is  about  5  times  brighter than adjacent parts of Galactic plane.  
The region between $l$ = 0\arcdeg\ to 1.5\arcdeg\ contains 542 Bolocat
entries,  or 361 objects per square degree.  In contrast, the Galactic disk away 
from the Galactic center between $l$ = 5.5\arcdeg\ and 11.5\arcdeg\ contains 
511  entries in 6 square degrees or 85 objects per square degree on average.  
This field is chosen for comparison since it likely has a similar surface 
density of foreground BGPS clumps as the Galactic center region.   The
symmetrically located field in the 4-th quadrant was not observed to comparable
depth due to its lower elevation as seen from Mauna Kea.   The
surface density of clumps is somewhat larger around $l$ = 20\arcdeg\ to 35\arcdeg\ 
where the line-of-sight passes through the tangent points in the Molecular Ring.
Thus, the surface density of clumps in the CMZ is about 4.3 times higher than 
in the plane.  When Bolocat entries located at higher latitudes above and 
below the CMZ are excluded, the surface density of Bolocat clumps located
close to the Galactic center is more than a factor of 5 higher.   This is a lower limit 
since confusion is likely to make Bolocat have lower fidelity in the Galactic center.   
Thus, at least  80\% of  Bolocat entries  in the inner few degrees of the Galaxy 
are likely to be within  500 pc of the nucleus, located at approximately 8.5 kpc from 
the Sun.   Less  than 20\% are likely to be located in  the molecular ring at 
distances of 2 to 6 or 11 to 14 kpc from the Sun.  

\begin{figure}
\epsscale{1.0} 
\center{\includegraphics[width=0.7\textwidth,angle=0]{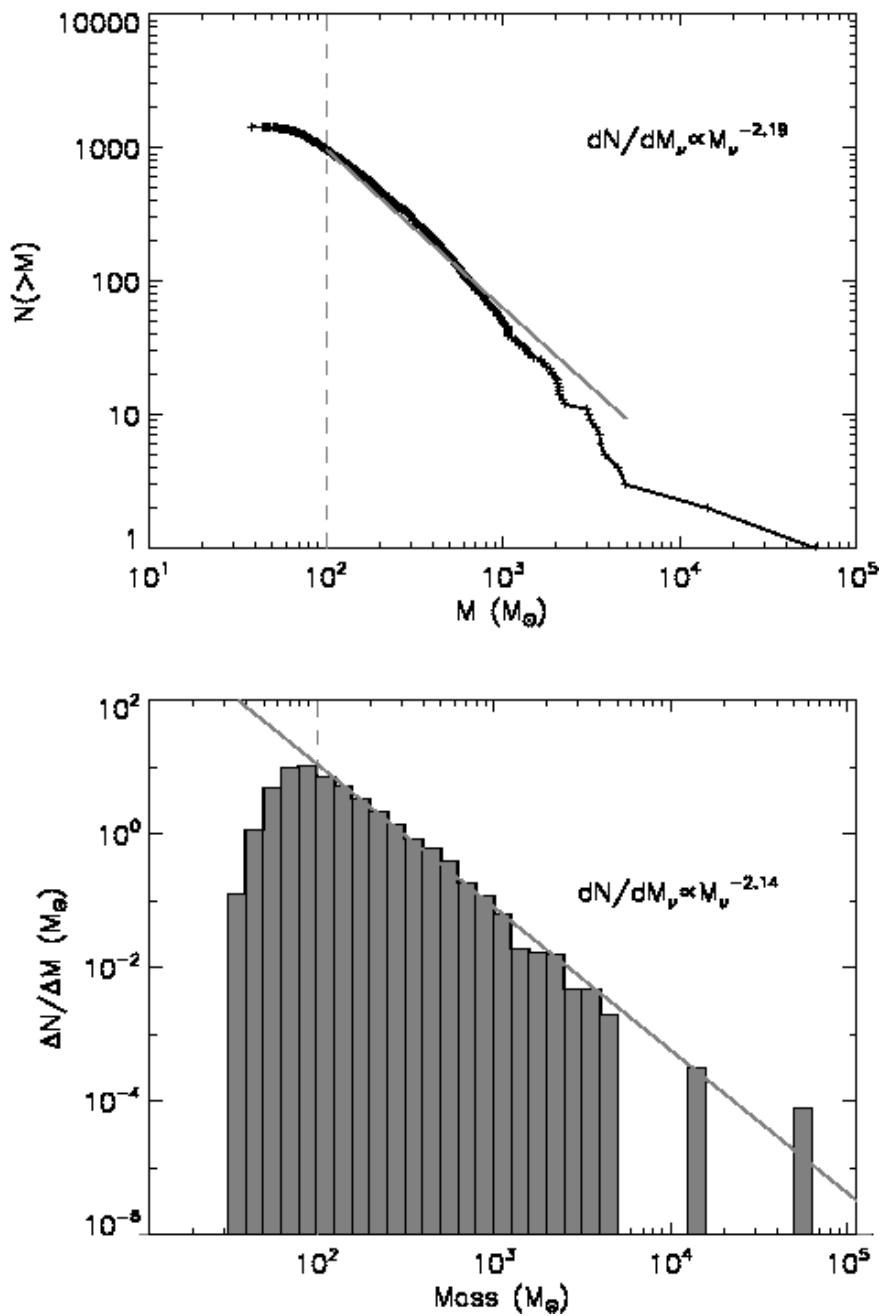}}
\caption{The mass spectrum of  Bolocat  clumps
derived from the fluxes measured in a 40\arcsec\ diameter aperture
under the assumption that all Bolocat clumps are located at
D = 8.5 kpc and have a temperature of T = 20 K.  The top panel shows 
the cumulative distribution along with a line representing a differential 
mass distribution with the indicated slope.  The bottom panel shows the 
differential mass distribution with counts in bins whose width increased
logarithmically along with a best fit slope to the differential mass distribution
for  constant width mass-bins.} 
\label{fig5}
\end{figure}

The concentration of clumps in the CMZ permits the generation of a 
clump mass spectrum for a large sample of objects located at a common 
distance between about 8 and 9 kpc.    The formulae in section 2.1 are used 
to estimate individual clump masses from the peak flux per beam in a 
40\arcsec\  diameter (1.64 pc at D = 8.5 kpc) aperture tabulated in column 
13 in Bolocat.   Figure \ref{fig5} shows the mass spectrum using the assumption 
that all clumps have the same temperature of $T_{dust}$ = 20 K.   For masses 
above about  100 M$_{\odot}$,   the mass spectrum can be fit by a power-law,  
$dN/dM = k M^{\alpha}$ with an index $\alpha  \approx$  --2.14 $\pm$ 0.1.  
On this scale, the Salpeter Initial Mass Function (IMF) of stars has a slope of
--2.35.   Between about $400$ and $10^4$ M$_{\odot}$, the slope is slightly 
steeper with  $\alpha  \approx$  --2.58 $\pm$ 0.1, but there are several very 
massive clumps such as Sgr B2 which lie well above the extrapolated curve.
Figures 6 to 8  show mass-spectra under different assumptions about the dust
temperature and its variation with either Galacto-centric distance or the peak
1.1 mm flux in each clump.   These spectra are based on the fluxes reported
in the V1.0 data release and have not been scaled by the factor of 1.5. 

Great caution must be exercised in generating mass-spectra for diffuse
objects and in the comparison of such spectra  with those obtained 
with different telescopes, at different wavelengths, and using different 
tracers.   Mass-spectra will be biased by the properties of the data acquisition 
strategy, the properties of reduction pipelines, the definitions and algorithms 
used to define `objects'.    Nevertheless, mass-spectra are frequently presented
in the literature on molecular clouds and their condensations.   Despite 
their potential pitfalls,  mass-spectra were computed because BGPS contains 
one of the largest samples of dense clumps ever obtained. 

\citet{MiyazakiTsuboi2000} present a mass spectrum for molecular clouds in the 
Galactic center, finding that for masses greater than $10^4$ M$_{\odot}$, the slope of
the mass function is $\alpha$ = --1.59 $\pm$ 0.07, considerably shallower than the
slope of BGPS clumps.   Inspection of their J = 1 -- 0 CS images shows that the
emission is continuous, area-filling, and dominated by extended structure in both
the $l -b$ and $l-$V maps;  it is far less `lumpy' than the BGPS maps which are 
dominated by small-scale strucutre.   Thus, it is not surprising that the overlap
in mass between the CS and BGPS mass-spectra is small or that the majority of
CS features are considerably more massive than the BGPS clumps.  
The slope of the 1.1 mm BGPS mass  spectrum above about 100 M$_{\odot}$ is 
similar to  the  mass spectrum of   low-mass cloud cores  in the  Solar vicinity which 
has a slope  of around --2.3,   similar to the stellar  initial mass function 
\citep{Motte1998,Motte2001,Andre2007,Enoch2008};   it is considerably steeper than  
found for CO-emitting  GMCs in the Galactic plane.   

Because at the distance of the Galactic center, most BGPS clumps will form clusters 
rather than individual (or multiple) stars, it may be more appropriate to compare 
the BGPS mass spectra with the mass spectra of star clusters.  Observations of
Galactic star clusters \citep{Battinelli1994} and OB associations 
\citep{McKee_Williams1997}  reveal steep mass spectra 
scaling as $M^{-2}$.   Massive young star clusters in the merging 
galaxy pair known as the `Antennae'  also have $dN/dM = N_0 M^{-2}$
\citep{Zhang_Fall1999} in the mass range $10^{3.5}$ to $10^{6.5}$  M$_{\odot}$.
Similar results have been found for clusters in M51 \citep{Bik2003} and
older clusters in the Large Magellanic Cloud \citep{deGrijs_Anders2006}.   
Nearby low-mass star-forming cores,  more distant and more massive BGPS 
clumps in the Galactic center, and Galactic and extra-galactic  star clusters 
share the  common trait of having steep mass spectra.   That all three classes of
object are gravitationally bound may have something to do with this trait.

The  BGPS data acquisition and reduction pipeline results in images with highly 
attenuated response to extended structure, effectively acting as a high-pass
filter that attenuates low-spatial frequencies \citep{Aguirre2010}.   Furthermore, 
the Bolocat watershed algorithm \citep{Rosolowsky2010} tends to subdivide 
sources with complex structure into  multiple clumps.  The Sgr B2 complex 
provides a good example.   While single dish maps in tracers such as CO or 
other high-dipole moment  molecules such as CS show that this region consists 
of a single cloud  complex  about 0.2\arcdeg\ by 0.3\arcdeg\ in extent (e.g. 
Figure \ref{fig3}),  the BGPS images such as Figures \ref{fig9}  highlight local 
maxima  and interconnecting ridges.  Bolocat  subdivides the Sgr B2 complex 
into  several dozen clumps.  
Attenuation of low spatial frequencies combined with the watershed algorithm 
of Bolocat may remove objects from the high-mass, large-extent portion of the 
mass-spectrum and re-distribute the flux into a larger number of lower-mass 
entries.   This has the effect of steepening the power-law index of the derived 
mass-spectrum.  Detailed simulations to quantify this effect are needed and will
be the subject of a future paper.   In complex fields such as the
Galactic plane,  line-of-sight blending of physically unrelated sources may lead to
further confusion which can alter the slope of the mass function.   We urge the
reader to exercise great caution in the interpretation of mass spectra of diffuse
objects.

Bolocat \citep{Rosolowsky2010} tabulates fluxes in four different ways.  In 
addition to the flux in a  40\arcsec\ aperture,  fluxes are also measured in 80\arcsec\  
and 120\arcsec\ apertures.  The catalog also tabulates the total flux in the area
assigned to each clump along with a beam-deconvolved   ``effective''  radius of 
the clump (the radius of a circle that has the same area as the clump). 
Mass-spectra based on fluxes measured in larger  apertures have
shallower slopes by about 0.1 to 0.2.    

\begin{figure}
\epsscale{1.0} 
\center{\includegraphics[width=0.7\textwidth,angle=0]{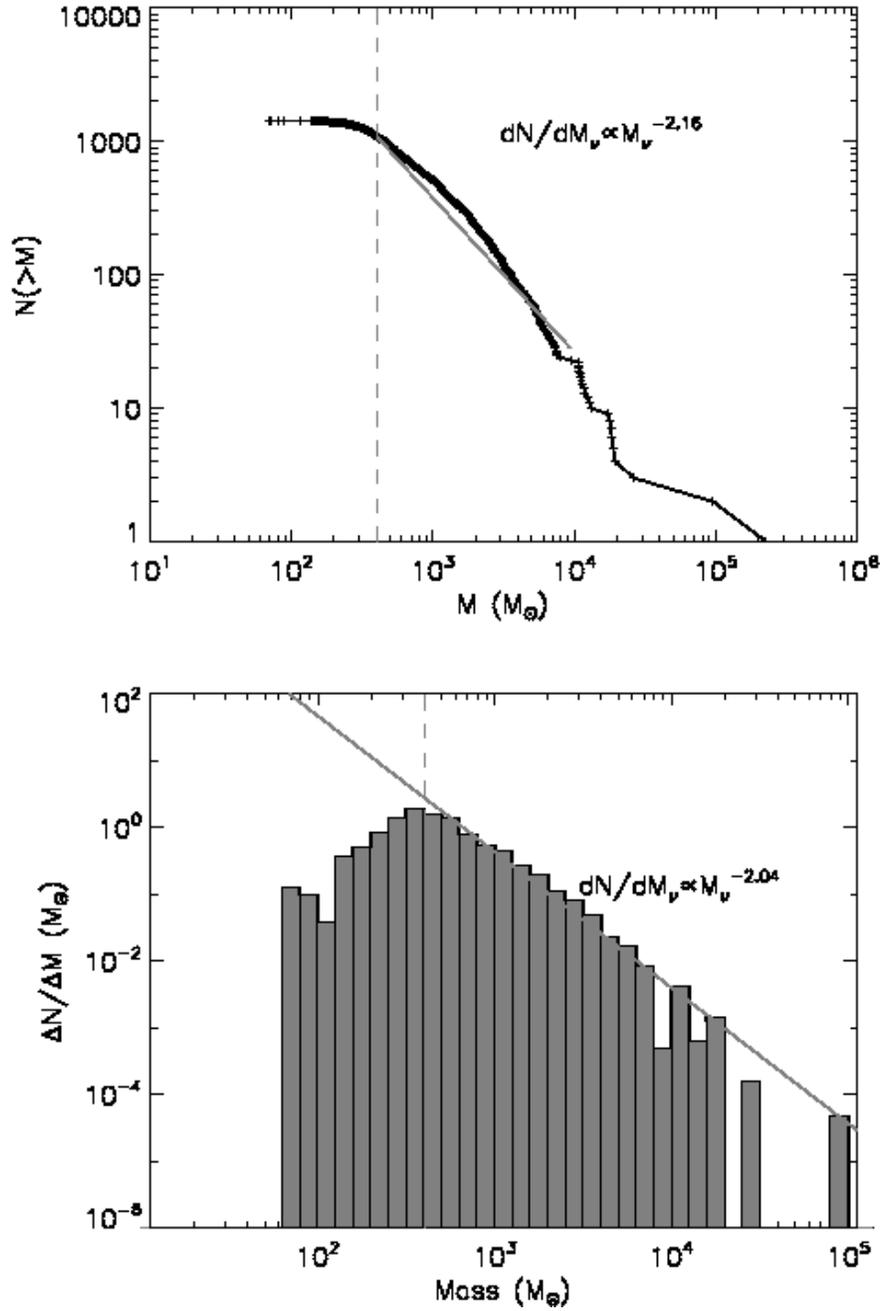}}
\caption{Same as Figure 5 but using fluxes measured in a 120\arcsec\
diameter aperture. } 
\label{fig6}
\end{figure}

The mass-spectrum derived from the Bolocat fluxes measured
in a 120\arcsec\ diameter aperture using a constant dust temperature of 20 K
has a power-law index $\alpha$ = --2.04 (Figure \ref{fig6}), slightly shallower than 
the index based on measurements in a 40\arcsec\ aperture.   This difference may 
be due to increased source confusion within the larger aperture.  The relatively
small slope change may suggest that source blending and confusion already 
affects the slope based on fluxes in the smaller aperture.   The slope also 
depends slightly on the widths of the bins used to generate the histograms. 

A curious puzzle about the Galactic center is that the dust temperature appears
to be lower than the gas temperature despite the high average density.   
Both the COBE and ISO satellites  determined that the average dust temperature 
in the CMZ ranges from 15 to 22 K \citep{Reach95,lis01}.  Thus, the use of 
$T_{dust}$ = 20 K in mass estimation is justified.    However, the dense gas 
associated with this dust is considerably warmer with 80\% of the ammonia 
having temperatures between 20 and 80 K and 18\% warmer than 
80 K \citep{Nagayama07}.  

Is  the slope of the derived mass spectrum sensitive to variations in $T_{dust}$?
Two types of dust temperature variations are considered to answer this question:
First,   the dust temperature of a clump is assumed to decrease with increasing 
projected  distance from Sgr A as a power law.  Second, the dust temperature of 
each clump is assigned a  value depending on the observed flux;   a 100 Jy source such as 
Sgr B2 is assumed to have dust with T = 50 K while a 1 Jy source is assumed to have
T = 20 K.  This corresponds to a dust temperature varying with flux, $F$, as $F^{0.2}$.

Given a source of  luminosity $L$ located at a distance $r$ from a grain,  
the mean grain temperature depends on the luminosity $L$ and distance $r$  as 
$$
T(r) = T_0 L^{1 / \gamma _g} r^{- 2 / \gamma _g}
$$ 
\citep{ScovilleKwan1976} where $T_0$ is a normalization constant.   For blackbodies 
larger than the emitted and 
absorbed wavelengths,   $\gamma _g = 4$.  For interstellar grains, $\gamma _g$ 
ranges between $5$ and $6$ for  grain emissivity power-law indices ranging from 1 to 2.   
Assume that the heating luminosity $L$ is generated by stars in Galactic disk and bulge,
that the luminosity-to-mass ratio is constant, and  grains at a given distance $r$  from
the Galactic center are heated  mostly by the stars within a sphere of radius $r$ centered
on the Galactic center.   Thus, the enclosed luminosity,  $L(r)$,  is a function of $r$ which
can be  estimated by assuming that the density of stars decreases with increasing distance 
from the Galactic center as $\rho(r') = \rho _0 r'^{- \delta}$.   For $\delta$ = 2  
(the singular isothermal sphere), integrating the mass distribution from   $r'$ = 0 to 
$r$ implies that the mass enclosed within a radius $r$ is proportional to $r$.  
Thus, the luminosity $L(r)$ is also proportional to $r$.   Inserting this in the
above equation for $T(r)$ gives a simple approximation for the radial variation of the grain
temperature
$$
T(r) = T_0  r^{- 1 / \gamma _g}
$$
where $T_0$ is  normalized to a temperature at some radius.   This estimate applies
to grains which are located mostly at cloud surfaces or otherwise are exposed 
to the visual and near-IR component of starlight.   As a concrete example,
choose $\beta$ = 1 and $\gamma_g$ = 5 so that $T(d) = T_0 r^{-0.2}$ because it 
yields reasonable cloud surface temperatures over a wide range of Galactocentric 
distances.   Taking $T_0$ = 50 K at $r$ = 10 pc implies $T$ = 27.5 K at $r$ = 200 pc,  
15.5 K   at $r$ = 3 kpc, and 12.5 K at 8.5 kpc.  This parametrization  provides 
reasonable base temperatures at the surfaces of clouds lacking internal or 
close-by heating sources.    It is used to test the sensitivity of the mass spectrum 
to a systematic decrease of grain temperature with increasing galacto-centric distance.

\begin{figure}
\epsscale{1.0} 
\center{\includegraphics[width=0.7\textwidth,angle=0]{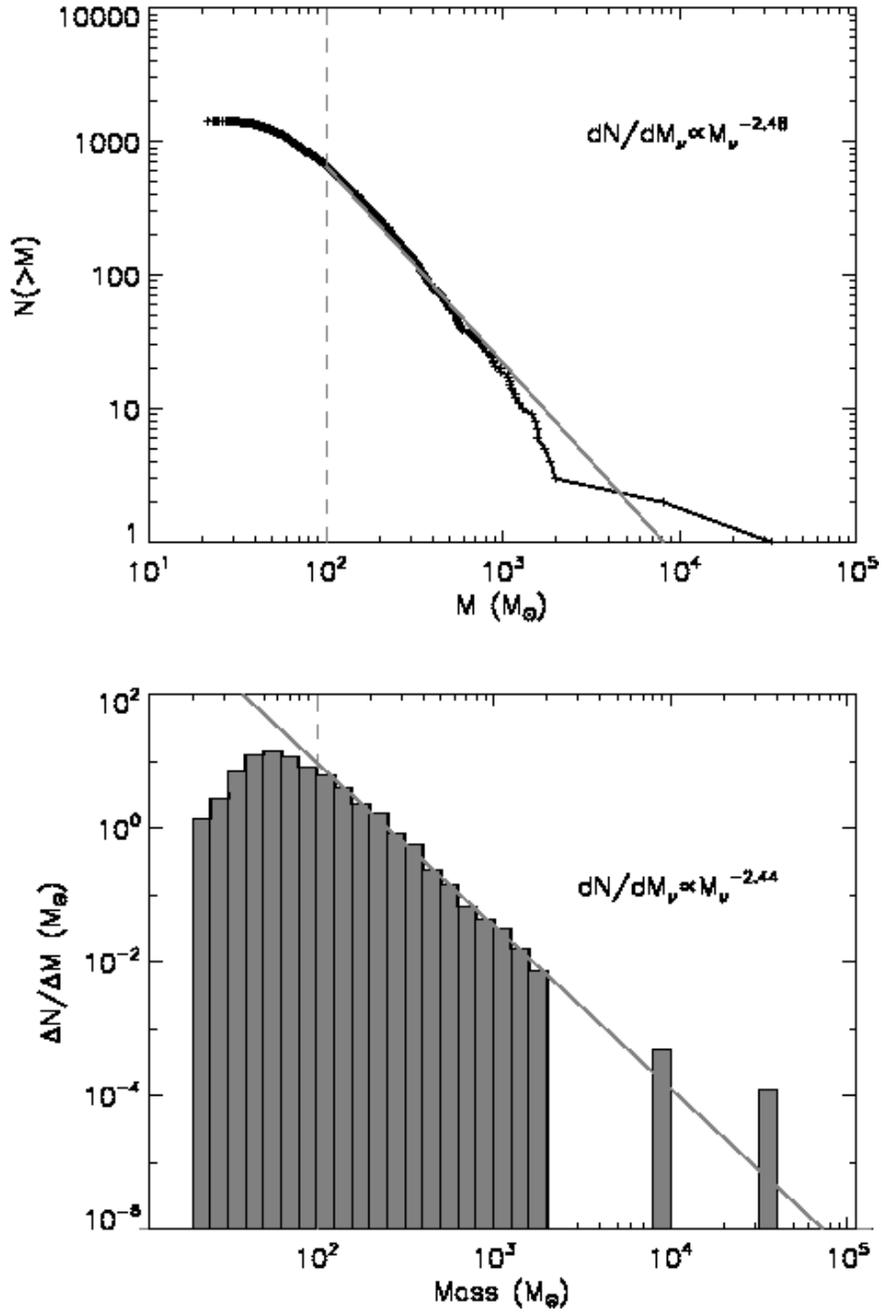}}
\caption{Same as Figure 5 using fluxes measured in a 40\arcsec\
diameter aperture but assuming that the dust temperature declines
with increasing distance,  $d$,   from Sgr A as a power-law with 
$T =T_0 d^{-0.2}$.  See text for details.} 
\label{fig7}
\end{figure}

Systematic  temperature variations do not have a large impact on the derived 
slope of the mass spectrum.    However, the elevated temperature does decrease the
estimates of total mass.  The mass-spectrum for  $\gamma _g$ = 5 corresponding 
to a radial temperature gradient expected for  a grain emissivity that decreases with 
wavelength  as a power-law with index  $\beta$ = 1 has a slope of   $\alpha$ = --2.44
(Figure \ref{fig7}).   

\begin{figure}
\epsscale{1.0} 
\center{\includegraphics[width=0.7\textwidth,angle=0]{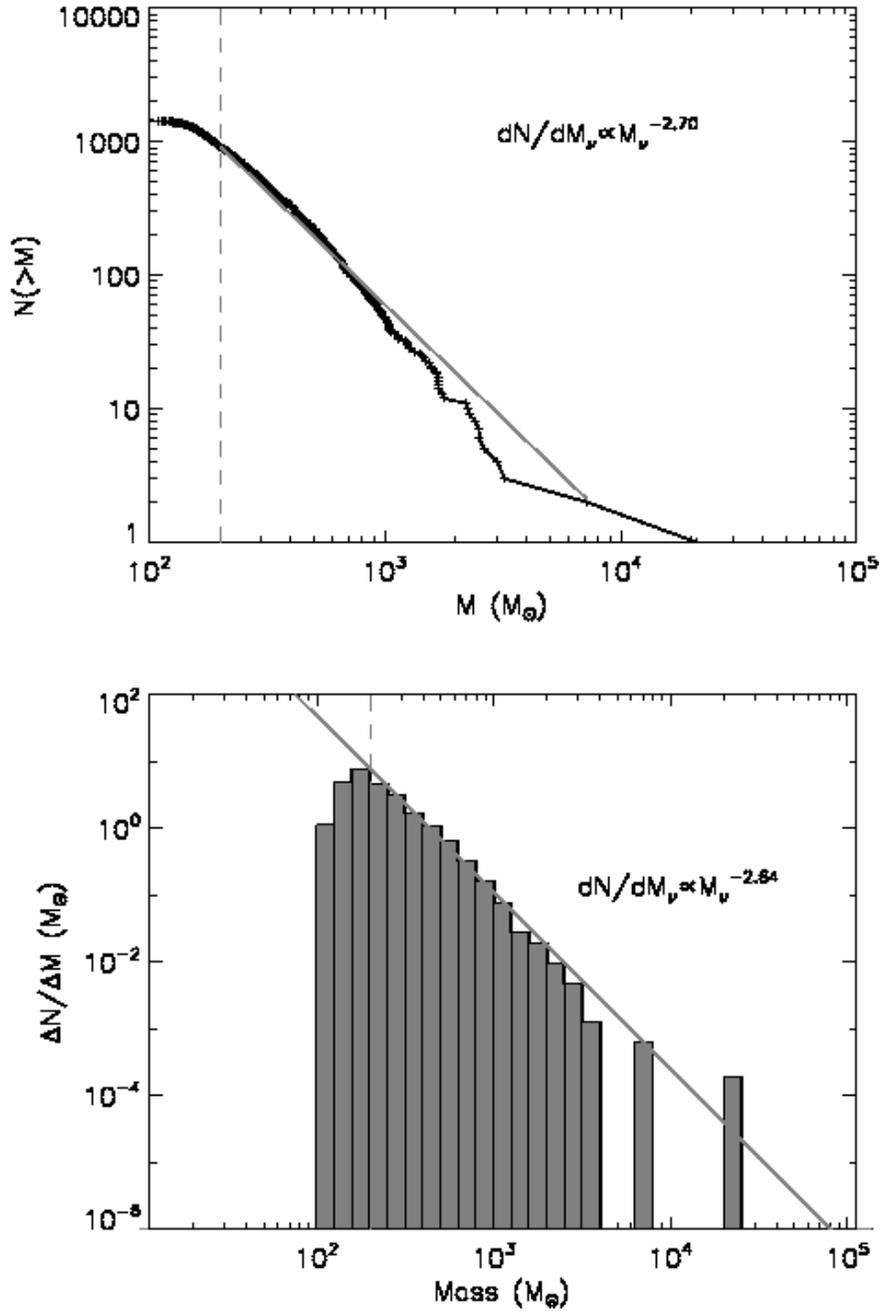}}
\caption{Same as Figure 5 using fluxes measured in a 40\arcsec\
diameter aperture but assuming that the dust temperature increases
with increasing 1.1 mm flux, $F$, as a power-law with $T = T_0 F^{0.2}$.  
See text for details.} 
\label{fig8}
\end{figure}

Finally, it is possible that internal heating by massive stars results in an average
dust temperature that is a function of 1.1 mm flux.  To determine the consequences 
of such a model on the derived mass-spectrum, we assume that the dust temperature
varies as $T = T_0 F_{1.1}^{0.2}$ where the normalization constant $T_0$ is
chosen to give a dust temperature of 20 K for a 1.1 mm flux density per beam of 1 Jy.
This model  results in a steeper mass spectrum shown in Figure \ref{fig8} with
a slope of $\alpha$ = -2.64.  

These results show that the mass spectrum of mm clumps having masses above
about 70 M$_{\odot}$ is represented by a steep power-law comparable to  low-mass star
forming cores with masses below about 10 M$_{\odot}$ 
\citep{Motte1998,Motte2001,Andre2007}.    The derived power-law index
is relatively insensitive to assumptions about the dust temperature.   In summary, the power-law
index of Bolocat clumps in the Galactic center is 2.14 (+0.4,--0.1).  The mass-spectrum in the 
Galactic center is steeper than GMC mass-spectra in the inner galaxy and LMC and 
comparable to the outer  Galaxy and M33 \citep{Rosolowsky2005}.   However, great
caution must be exercised in the interpretation of these mass spectra owing to the
data acquisition and reduction methodology.

The total H$_2$ mass  of the 1428 Bolocat  clumps between  $l$ = --1.5\arcdeg\ to 4.5\arcdeg , 
measured in a 40\arcsec\ aperture, assuming that all are located at a distance of 8.5 kpc,  
have a dust temperature of 20 K, and multiplied by the empirically determined scaling factor of 1.5 needed to bring BGPS fluxes into agreement with MAMBO (Motte  et al. 2003, 2007) and SIMBA 
(Matthews et al. 2009) gives $M(H_2)  \sim 6.6  \pm 1 \times ~10^5$~M$_{\odot}$. 
Assuming a temperature gradient  that decreases with distance from Sgr A as proposed 
above gives the lowest total mass of order $\sim 5.7 \pm 1.2 \times 10^5$~M$_{\odot}$.
Assuming a dust temperature that scales with detected 1.1 mm flux as described above 
gives a total mass of $\sim 1.1 \pm 0.23 \times 10^6$~M$_{\odot}$ in a 40\arcsec\ aperture.
In the 120\arcsec\ aperture,  (assuming a constant dust temperature of 20 K) this mass grows to 
$\sim 5.5 \pm 1.2  \times 10^6$~M$_{\odot}$.   Using the total integrated flux (Bolocat 
column 19) within the effective radius  (Bolocat column 12) results in a total  mass  
$\sim 6.2 \pm 1.8 \times 10^6$~M$_{\odot}$.  

Summing all BGPS flux in this field above 100 mJy
gives a mass  of about $\sim 6.3 \pm 1.8 \times 10^6$~M$_{\odot}$.
The total mass of molecular gas within 
500 pc of the center has been estimated to be between $3$ to $8 \times 10^7$ ~M$_{\odot}$
\citep{Sodroski1994,Dahmen1998,tsuboi99}.  Thus, the total 1.1 mm emission from clumps 
in a 40\arcsec\ aperture represents about  0.7\% to 2.2\% of the total mass in molecular gas
while the mass in a 120\arcsec\ aperture is about 7\% to 18\% of the total.   Summing the  
flux above 100 mJy correponds to 8\% to 21\% of the total molecular gas mass.
This is  about a factor of several larger than what was found in  the Gem OB1 association 
toward the Galactic  anticenter  \citep{Dunham2010}, indicating that a larger fraction
of the molecular gas near the Galactic center is in compact clumps.  

It is important to recall that BGPS  filters out extended emission on scales larger than about 
5\arcmin\ and that the 1.1 mm data is only sensitive to dust in compact structure.     
Such spatial-filtering is not present present in the single-dish spectral line 
data which therefore can trace gas in extended regions.   Thus, the small mass fraction 
detected by BGPS may be a consequence  of the spatial filtering or diffuse, extended emission.  
Spatial filtering of the spectral line data to match BGPS could be used to constrain the
relative abundances of dust and the tracers producing the lines.

\subsection{The Central Molecular Zone}

\begin{figure}
\epsscale{1.0}
\center{\includegraphics[width=1.0\textwidth,angle=0]{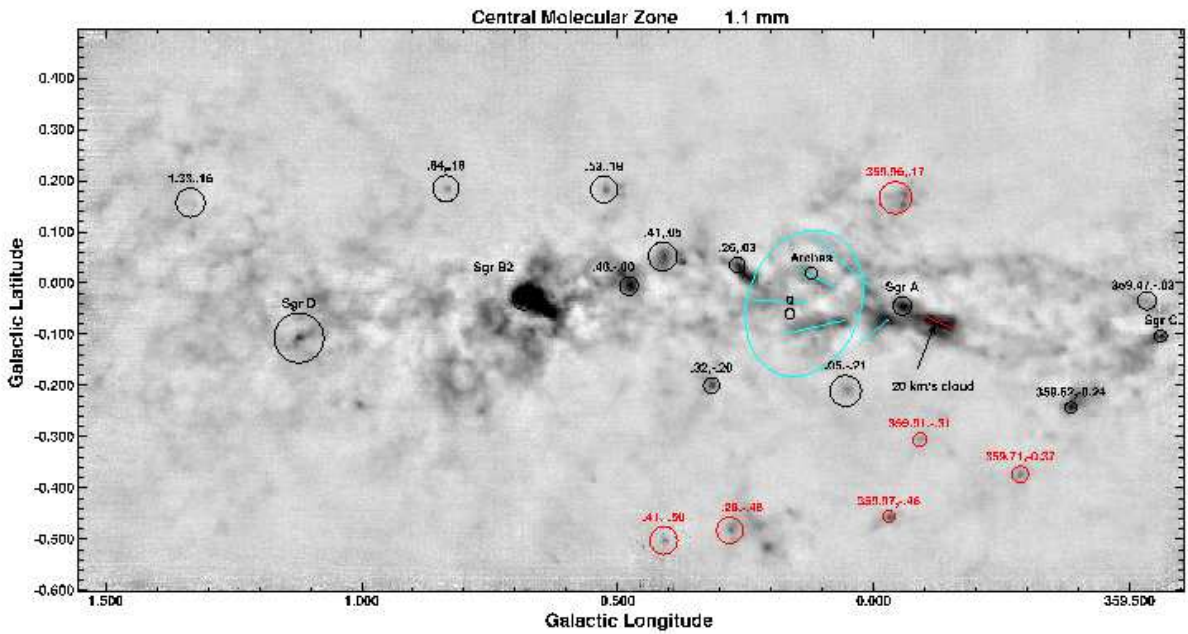}}
\caption{The 1.1 mm image of the Central Molecular Zone.
The red lines show the orientations and locations of cometary clouds
and chains of clumps that face Sgr A.  These features are located in the 
interior of the cavity surrounding Sgr A that is discussed in the text
and form the walls of its multiple chambers.  The features marked in
red are likely to be foreground clouds at an assumed distance of
3.9 kpc as indicated in Table 1.}
\label{fig9}
\end{figure}

Figure \ref{fig1}   shows an image of the 1.1 mm  emission from 
the Galactic center region.   Figure \ref{fig2} shows this image as contours
superimposed on the Spitzer Space Telescope IRAC 8$\mu$m image.
As at most long IR, sub-mm, and radio wavelengths, the 
area-integrated 1.1 mm emission is about 5 times brighter
than in adjacent portions of the inner Galactic plane.   While most of
the Galactic plane mm-wave continuum emission is dominated by
discrete clumps \citep{Aguirre2010,Rosolowsky2010,Dunham2010},
the CMZ contains bright  arcminute-scale extended 
structures  interconnected by a lacy network of filaments.  Figure
\ref{fig9} shows the inner 2 square degrees of the Galactic plane
at 1.1 mm.   Figure \ref{fig10} shows the same field at 8.0 $\mu$m as observed
by the Spitzer Space Telescope \citep{Arendt2008} with contours of 1.1 mm 
emission superimposed.  

The chain of  bright 1.1 mm  clumps extending from Sgr B2 (G0.68,--0.03)
to G0.26+0.03  and the  surrounding lower level diffuse emission is seen in 
silhouette  against  the bright background in the infrared image.  The 20$'$
halo of 1.1 mm emission  surrounding Sgr B2 can also be seen in 
silhouette against the infrared background.  Thus, most of BGPS emission
between  $l$ = 1.0$\arcdeg$ and 0.25$\arcdeg$ must originate in front of 
most of the stars and  mid-infrared light in the central bulge of the Galaxy.  
This is consistent with the  suspected orientation of the central bar whose 
positive longitude part is closer.   

Several features of the dust  continuum emission from the CMZ are apparent.    
First, most of  the 1.1 mm emission  with a flux density greater than about  1 
Jy / beam  originates  from a few dozen clumps (in the Bolocat clump 
catalog, many of these objects are subdivided further).   The two brightest 
peaks  correspond to Sgr B2  and  the Sgr A*.  At lower levels, the CMZ can 
be decomposed into  hundreds  of individual clumps and filaments.  
Bolocat contains 1428 entries in the field covered in Figure \ref{fig1}. 
Second,  most  of the  1.1 mm  emission in the CMZ  closely follows  the  
distribution of  dense molecular  gas traced  by high-dipole moment   
molecules  and the  large line-width  CO emission 
\citep{bally87, bally88,oka98,oka01}.    In the  inner square  degree,  both 
the  1.1 mm  dust  continuum and the molecular  tracers  of  the  CMZ exhibit 
a smaller  scale-height than emission  from the  Galactic  disk measured  
beyond $|l| > 1\arcdeg$.   Near Sgr A ($l \approx$ 0\arcdeg ), the vertical 
extent (orthogonal to the Galactic plane)  of the brightest BGPS emission 
and broad-line  molecular gas is only about 2\arcmin\  to 3\arcmin\  (5 to 8  pc); 
near Sgr B2 around $l \approx$ 0.7\arcdeg\,  this layer has a thickness of
about 10\arcmin\  to 15\arcmin\ (25 to 40 pc).   Beyond  $|l| \sim$ 1\arcdeg ,  
the scale-height  of  1.1 mm clumps  associated with the CMZ   increases  
dramatically.    Around $l$ = 1.3\arcdeg\ and  3.2\arcdeg\, the gas layer 
is nearly 60\arcmin\ (150 pc) thick.  Away from the Galactic center, between
$l$ = 5\arcdeg\  and 35\arcdeg\ the dust layer has an average thickness of 
about  0.8\arcdeg\  or 125 pc.    The spatially averaged 1.1 mm flux from the 
Galactic center  features is about 3 to 5 times  higher  than the average 
flux from the roughly $\pm$  0.4\arcdeg\  scale-height 
clump  population  of the  Molecular Ring which is at least  3 kpc  in front of 
(or behind)  the Galactic center.     Third,   many 1.1 mm features can be 
seen in  silhouette  in the  8 $\mu$m   IRAC and  24 $\mu$m 
MIPS images.    Fourth,   at levels up to several hundred mJy/beam, a complex 
network of filaments forms a frothy background of emission which surrounds 
voids.  The filaments may either trace the walls of stellar-wind 
bubbles or SNR, the warm edges of molecular clouds, cavities produced
by HII regions,  or ``fossil" cavities whose energy sources are extinct.

Comparison with the 20 cm radio maps (Figure \ref{fig11}) 
shows that only a small subset of these 1.1 mm cavities are associated
with radio continuum emission.
Figure \ref{fig11} shows the 20 cm emission imaged with 30\arcsec\
resolution \citep{yusef-zadeh2004} superimposed on the 1.1 mm
image.  The Sgr A, B1, B2, C, and D regions that contain young massive
stars and star clusters stand out in both the radio continuum and at
1.1 mm.   The non-thermal radio continuum filaments  tend to 
be located adjacent to 1.1 mm clumps.  For example, the largest and brightest  
group of filaments near $l$ = 0.18\arcdeg\  crosses the Galactic plane at 
the positive latitude end of the ridge of clumps associated with the 50 \kms\ 
cloud  that stretches from $l$ = 0\arcdeg\  to 0.17\arcdeg .   The fainter 
non-thermal filament located directly above (in latitude) Sgr A known 
as N5 \citep{yusef-zadeh2004} passes through the cluster of clumps 
associated with the HII region G359.96,+0.17.    The brightest part of
filament N8 abuts a diffuse 200 mJy BGPS clump at [359.78,0.17] (not marked).

Some of the diffuse 1.1 mm emission between $l$ = 0.24\arcdeg\ and 
--359.6$\arcdeg$ located at positive latitudes is visible in emission 
at 8 $\mu$m.    At positive longitudes,   this region contains the thermal  
``arched  filaments"  near the  Arches cluster.  The 1.1 mm 
emission may  either trace warm dust or  free-free emission from 
dense, photo-ionized  plasma illuminated by  the Arches Cluster.   
At negative longitudes,   1.1 mm clumps associated with the high-latitude, 
positive velocity  portion of the  ``expanding molecular ring" ($b$ = 0.0\arcdeg\ 
to about 0.2\arcdeg ; $V_{LSR}$ = 100 to 200 km~s$^{-1}$) are also associated 
with 8 $\mu$m emission.   At negative longitudes, the 20 km~s$^{-1}$ 
cloud  located south of  Sgr A*  is  visible in silhouette and must therefore 
be in  front of the central cluster.  A pair of oppositely facing comet-shaped 
clouds  near the right edge of Figures \ref{fig9} through \ref{fig11} 
(associated with Sgr C and  G359.62--0.24) are both seen prominently 
in extinction.  Except for  these clouds, few of  the 1.1 mm sources at 
negative longitudes are  visible in extinction in the IRAC data.

The fainter 1.1 mm continuum is  structured and organized into a  filamentary  
network of arcs and circular  features.     This is  most  evident in 
left half of  Figure \ref{fig9}.  Some rings and  partial rings 
surround  radio continuum features and therefore probably mark dust heated 
in PDRs and the edges of cavities created by massive stars and 
star clusters.  However, most  rings, arcs, and filaments are not 
associated with known HII regions or supernova remnants.  
Massive  stars that have shed their natal cocoons are hard 
to separate from the multitude of intrinsically fainter foreground stars,  because in
the IR  distant massive stars and closer lower-mass stars have similar colors
and spectral shapes. These cavities may also be the remnants 
of older  generations of  stars and clusters whose most massive members 
have either exploded as supernovae more than a million years ago 
so that all radio continuum  has long since faded, or whose stars 
were ejected by dynamical  processes that create run-away, high-velocity
stars.   Run-away stars are most common among spectral type O
\citep{Gies_Bolton1986,Gies1987}; many of these stars were ejected by
dynamic interactions in dense clusters \citep{Gualandris2004}.
The shells and filaments  may be tracers of fossil star  and cluster 
formation in the CMZ.   However, it is also  possible that randomly 
superimposed filaments create the impression of cavities.

The large kidney-bean-shaped clump at  [0.26,0.03] (see Figures \ref{fig9}
and \ref{fig12}) is remarkable for being 
one of the brightest features at 1.1 mm and for being the most prominent 
infrared dark cloud (IRDC) in the entire Galactic center region \citep{Liszt09}.  
This feature is  seen in absorption at 24 $\mu$ (Figure \ref{fig14}). 
Although  Bolocat lists two entries for this object (Nos. 96 and 99), dense gas 
tracers such as CO, CS,  and HCN indicate that the entire structure is at 
approximately the same radial velocity (V$_{LSR} \approx$ 40  km~s$^{-1}$
with a line-width extending from 20 to 60 km~s$^{-1}$).  
Liszt (2009) use the MSX data to  set a lower bound on the column density 
of N(H$_2$) $> ~1.0 \times 10^{23}$ cm$^{-2}$ (A$_V  > $ 53 magnitudes),
consistent with the 1.1 mm-based column density estimate in a 40\arcsec\
aperture.   The higher resolution 350 $\mu$m observations (Figure \ref{fig12})
indicate that it has  considerable sub-structure with consequently larger 
column densities.

Figure \ref{fig13} shows the ratio map formed from the 350 $\mu$m and
1.1 mm data in the vicinity of Sgr A.   In regions with significant emission
at both wavelengths, there is a general trend that the clump centers where
the fluxes peak have the smallest flux ratios.  The flux ratios tend to increase
towards the clump edges.   Ratios at a number of  locations in Figure \ref{fig13}
are indicated by the blue arrows.   In Figure \ref{fig13}, the flux in the 1.1 mm
BGPS maps have been scaled-up by a factor of 1.5.  Without this scaling, 
a large fraction of the pixels have ratios larger than 100.   Figure \ref{fig13}
shows that clumps in the CMZ have warmer dust than either foreground or
background clumps in the Molecular Ring.

Sgr B2 is the brightest  clump in the entire BGPS survey.   
Figure \ref{fig15} shows the 350 $\mu$m image.  As shown in
Table~2, it  has the largest total mass ($> 5 \times 10^5$ M$_{\odot}$),
highest volume-averaged density ($n(H_2) > 10^4$ cm$^{-3}$), and
greatest column density  ($N(H_2) > 2 \times 10^{24}$ cm$^{-2}$ in 33\arcsec\
beams centered on Sgr B2 Main and North) of any BGPS clump.   Thus, the
average extinction towards Sgr B2 M and N corresponds to  at least 
1,000 magnitudes at visual wavelengths.    Molecular line measurements
indicate that the Sgr B2 complex  has a mass of at least $5 \times 10^6$ M$_{\odot}$,
at least a factor of 10 greater than implied by the total 1.1 mm flux detected 
by BGPS and listed in Table 2  \citep{Jones2008}.   Over 50 compact HII regions 
and powerful masers indicate that a major burst of star formation is occurring 
here  \citep{Gaume1995,DePree1998,DePree2005}. The molecular line
data provide evidence that the intense burst of star formation occurring
in  Sgr B2 may have been caused by the supersonic collisions between 
two giant Galactic center molecular  clouds \citep{Jones2008,Liszt09}.
Sgr B2 may be  located near the high-longitude  end of `x2' family of orbits 
in a bared potential \citep{Contopoulos1980}
where it may encounter gas moving along the `x1' family of orbits.
Sgr B2  is located at the high-longitude
end of a chain of 1.1 mm clouds, mid-IR extinction features, embedded 24 $\mu$m
sources, and HII regions starting near the less-extincted and older Sgr B1 complex
\citep{Yusef-Zadeh2009}.  Sgr B2  may represent the most recent event in a chain
of sequential star formation that started near Sgr B1.   
The Sgr  B2 complex may be giving birth to a massive star cluster or even a 
super star cluster comparable to the somewhat older Arches and Quintuplet 
clusters located in the CMZ.    

Figures \ref{fig17} and \ref{fig18} show the BGPS 1.1 mm contours superimposed on
the Spitzer 8 and 24 $\mu$m images respectively.    These images clearly show that
most of the dust associated with the Sgr B2 region is in front of the bulk of the 
8 and 24 $\mu$m emission.  Furthermore,  several of the roughly circular cavities
in the BGPS images are translucent in the IR data.  The large Sgr B1 complex of HII regions
and infrared sources has created a giant cavity on the low-longitude and low-latitude
side of Sgr B2.  Highly obscured nebulosity and IR sources abut the the Sgr B2
sub-mm/mm peaks on the low-latitude side.  However, no IR sources can be seen
at the locations of Sgr B2N and Sgr B2M. 
More  discussion of the Sgr B2 region is given in the Appendix.

Figure \ref{fig16} shows the 350 $\mu$m / 1.1 mm flux ratio map derived from
the 350 $\mu$m image matched in spatial frequency response to BGPS divided
by a 1.1 mm map scaled-up by a factor of 1.5.  No
correction for line contamination has been applied.    Sgr B2 is likely
to be most affected by such contamination; \citet{LisGoldsmith91} found
about 30\% of the 350 GHz flux is due to lines and contamination.  The two
brightest peaks, corresponding to Sgr B2 North and Sgr B2 Middle (or Main) 
have the smallest flux ratios (25 and 34, respectively) which for $\beta$ = 2 dust, 
implies temperatures
of 14 -- 17 K.     It is likely that the dust is internally heated 
by forming massive stars to at least 20 K, and possibly 40 K \citep{lis91}.     As 
these authors proposed, it is possible that the low flux ratio is the result of an 
unusually low $\beta$;   for $\beta$ = 1.5, ratios of 25 and 34 imply 21 K and 33 K.

\subsection{Sgr A$^*$ and its Environment}

The second most prominent peak  in the 1.1 mm data (Sgr B2 is the first) is 
centered on Sgr A* which is thought to mark the location of the  
$4.5 \pm 0.4  \times 10^6$~M$_{\odot}$ black hole at the Galactic 
center \citep{Genzel2003,Ghez2008}.  The extended radio source
known as Sgr A consists of  Sgr A East, a several arc-minute diameter oval 
of non-thermal radio continuum emission thought to trace a supernova remnant,  
and Sgr A West, an HII region which consists of  a  ``mini-spiral"  of dense, thermal 
plasma with the central black hole,  Sgr A* at its center.    The ``mini-spiral" is 
surrounded by the  $\sim$ 2 pc radius ``circum-nuclear-ring" (also
referred to as the ``circum-nuclear disk'' -- CND) of dense 
molecular gas and dust  \citep{MorrisSerabyn1996}.  The CND is clearly
seen in the 350 $\mu$m image shown in Figure \ref{fig12}.

Emission from Sgr A* peaks 
near a wavelength of 1 mm \citep{serabyn97}.  Between 1993 and 1996 
the peak flux near this wavelength was around 3.5 Jy (Figure 2 
in \citet{serabyn97}).  In July 2005, Sgr A* was about 1.6 times 
brighter at 1.1 mm with a flux density of 5.7 Jy.   This may be an 
underestimate because the BGPS pipeline may not fully recover the flux
of bright sources or those surrounded by an extended envelope 
\citep{Aguirre2010,Rosolowsky2010}.   Diffuse emission produced 
by dust or hot plasma along the line of sight contributes at most 1 Jy 
per beam.  Thus, the increased brightness of Sgr A* in the 1.1 mm maps is 
most likely due to variability which is consistent with a non-thermal 
origin for most of the 1.1 mm continuum emission from that source.

Figure \ref{fig12} shows the 350 $\mu$m SHARC-II image of the  30\arcmin\  
field surrounding Sgr A.  Figure \ref{fig14} shows contours of  
350 $\mu$m emission superimposed on the Spitzer 24 $\mu$m image.    
In contrast to 1.1 mm, there is very little 
350 $\mu$m emission from Sgr A*  itself.  At 350 $\mu$m,  Sgr A* has  
a flux density between 0.5 and 1.5 Jy  depending on how the large
background flux is subtracted.    While emission from the 
CND and Sgr A* are blended at  1.1 mm by  the 33\arcsec\ effective
resolution,    in the 350  $\mu$m image, the CND is resolved 
(Figure \ref{fig12})  and appears as a nearly complete oval. 
Several dozen discrete clumps and filaments can be identified in the region 
surrounding the ring in addition to the prominent 20 and 50 km~s$^{-1}$ 
clouds.    Interferometric  observations
of molecules such as HCN resolve the CND into a collection of  dense
clumps  \citep{Christopher05}.     The detection of maser emission and
hyper-compact sources of radio continuum emission indicates 
on-going massive star formation in the CND \citep{Yusef-Zadeh2008}.

The central 30\arcmin\ region  contains several chains of 1.1 mm
clumps and  filaments elongated towards  Sgr A (turquoise lines in Figures \ref{fig9}
through \ref{fig11}  red lines in  Figures \ref{fig12} and \ref{fig14}).     
These chains are  narrower at the end 
facing Sgr A and may therefore be cometary clouds sculpted by this source.
The two most prominent chains are located about 1\arcmin\  south of 
the Arches cluster and 2\arcmin\ south of the Quintuplet cluster and Pistol Star;
these chains are each about 6\arcmin\  to 8\arcmin\  long.  A shorter,  fainter chain is
located between the Arches and Quintuplet clusters, and a nearly 10\arcmin\ long
chain is located about 3\arcmin\  north of the Arches cluster.   These features 
are also seen in the 350 $\mu$m image (Figures \ref{fig12} and \ref{fig14}) where 
they break-up into complicated substructures with an overall elongation
towards or away from Sgr A.   

About a half dozen compact clumps located between the IRDC at [0.26,0.03] and
the Quintuplet cluster near  $l \sim$ 0.22\arcdeg\  are also elongated along
the Galactic plane and may be smaller cometary clouds less than 1\arcmin\  
to 2\arcmin\ in  length.     These are best seen in Figure \ref{fig9}.   
Figure \ref{fig11},  shows that these clumps are located at the 
high-longitude side of the prominent non-thermal filaments that cross the 
Galactic plane at $l ~\approx $ 0.18\arcdeg .

Mid-infrared images from the MSX and ISO satellites reveal a very bright,
limb-brightenned bubble of infrared emission surrounding the Quintuplet and
Arches clusters   \citep{Moneti2001}.   Inspection of the 24 $\mu$m images 
obtained by the Spitzer Space Telescope (program P20414; PI -  F. Yusef-Zadeh)
also shows the prominent elliptical bubble with a major axis
bounded towards positive latitudes by the Arched Filaments around 
[$l,b$]  $\approx$ [+0.11\arcdeg ,+0.10\arcdeg ] and toward negative 
latitudes by a rim at [$l,b$]  $\approx$ [+0.12\arcdeg ,--0.18\arcdeg ] and
extending from $l \approx$ 0.0 to 0.18.    The oval infrared shell  is clearly
seen in Figure \ref{fig14}.
The location of the 24 $\mu$m ovoid is indicated by the large ellipse in 
Figures \ref{fig9} through  \ref{fig14}.

The chains facing Sgr A are in the projected interior of the infrared ``bubble"
surrounding the Arches and Quintuplet clusters.   It is likely that this bubble
traces the walls of a cavity  that extends from the non-thermal filaments
near $l \approx$ 0.18\arcdeg\  to at least  $l \approx$.0.0\arcdeg\  near the 
50 km~s$^{-1}$ cloud and
contains both the Arches and Quintuplet clusters and possibly the Sgr A complex.    
Figure \ref{fig11} shows that
this region is filled with diffuse 20 cm continuum emission.   Figure \ref{fig12} shows
this region at  350 $\mu$m.  

The dashed line in Figure \ref{fig2} outlines the boundary of a large region relatively
devoid of clouds as traced by 8 or 24 $\mu$m emission. 
This feature may trace an extension of the bubble
associated with the Arches and Quintuplet clusters  that opens-up above and below 
the Galactic plane.  While this feature can be traced to the edge of the observed
field above Sgr A, below the Galactic plane it becomes confused with the foreground 
bubble centered on [$l,b$] = [+0.41\arcdeg ,--0.50\arcdeg ].   An extended bubble 
blowing-out of the Galactic plane may have been created by energy release
from Sgr A,  Arches, and the Quintuplet regions.

\subsection{The Inner Galactic Plane: Bania's Clump 2}

The Galactic center contains several remarkable cloud complexes
which exhibit line-widths   $\Delta V ~>$ 100 \kms\ over a 
region typically about 0.5\arcdeg\ in diameter 
\citep{Bania_Clump1,Bania_Clump2,oka98,Liszt06,Liszt08}.    
Although Bania's  Clump 1 at [$l,b$] $\approx$ [--355\arcdeg ,+0.4\arcdeg ]
and Clump 2 [$l,b$] $\approx$  [+3.2\arcdeg ,+0.2\arcdeg ]  were the first to be 
noted, similar but less distinct, large line-width
complexes  exist near    [+1.3\arcdeg ,+0.2\arcdeg ] and near
[+5.0\arcdeg ,+0.2\arcdeg ]  \citep{oka98,oka2001b,dame2001,bally87,bally88}.     
Bania's Clump 2, located at a projected distance of about 450 pc from the 
Galactic center,  is the most prominent of these features with a line width of
$\Delta V \approx$ 200 \kms .  The feature extends from the mid-plane to 
about  $b$ = +0.8\arcdeg\ and consists of about 16 CS-emitting clumps
each with masses of order $5 \times 10^5$ \msol\ and H$_2$ densities 
in excess of $2 \times 10^4$~cm$^{-3}$ \citep{Bania_Clump2} surrounded
by an envelope of CO emitting material.  Such dense and massive clumps
are usually associated with on-going  massive star formation.    

Figure \ref{fig19} shows the BGPS 1.1 mm map of the region around 
$l$ = 3\arcdeg  .   Some BGPS data were obtained up to  $b =  \pm$ 1.5\arcdeg\ 
that shows several additional features associated with Bania's Clump 2.   
These data are noisier than the
BGPS data along the plane and are not shown here. 
The 1.1 mm clumps coincide with the CS and CO clouds in  Bania's Clump 2.   
Figure  \ref{fig20} shows the Spitzer Space Telescope
8 $\mu$m image from GLIMPSE  \citep{GLIMPSE} with superimposed 1.1 mm
contours.   The most striking result is the absence of extended infrared sources
or bubbles associated with most Clump 2 features.   In contrast to Clump 2, 
1.1 mm clumps located along the Galactic plane tend to contain  
infrared sources at 4.5, and 5.8 $\mu$m, extended 8 $\mu$m 
emission, and prominent bubbles, indicating massive star and star 
cluster formation.  Inspection of the 24 $\mu$m MIPS scan-maps available from
the Spitzer Science Center archives also 
shows a general absence of mid-IR sources in Bania's Clump 2.
This result is consistent with the lack of 20 cm radio continuum emission
\citep{yusef-zadeh2004}.

Comparison of the 1.1 mm and Spitzer images shows that there is a slight
depression in the spatially-averaged 8 and 24 $\mu$m emission from the bulge of
the Galaxy at the location of the 1.1 mm features in Clump 2.  The Clump 2 
dust features are weak infrared dark clouds (IRDCs) indicating 
that most of the dust in this region must be somewhat in the foreground of 
the bulge. This is consistent with the suspected location of the Clump 2 
material in the  near-side of the central bar.

Two clumps located in Bania's Clumps to appear to be foreground objects.
The 1.1 mm clump located at [$l,b$] = [--3.405,+0.877] is associated with
IRAS 17470-2533 at the high-latitude  end of Clump 2. 
Located several arcminutes southwest of the clump, there is an arc-minute 
diameter  bubble of 8 $\mu$m emission  which  contains a small cluster.  
The BGPS clump is seen as an IRDC
projected in front of the background star field and diffuse emission.     
Its high latitude and appearance as an IRDC is consistent with this clump 
and adjacent HII region being located in the foreground in the Galactic disk.  
The bright BGPS clump at [$l,b$] = [+3.34,+0.44] is associated with
IRAS 17484-2550 and the clump  CO003.34+00.44 in the 
Galactic center survey of \citet{oka98,oka01}.   The radial velocity
ranges  from 0 to 40 \kms .  A low-contrast IRDC is associated with this feature.

\subsection{Foreground Features and the  Galactic Disk}

Several prominent 1.1 mm clumps are located at relatively high 
Galactic latitudes $|b| ~>$ 0.25$\arcdeg $.    Some of these
clumps are associated with prominent HII regions or bubbles visible in the IRAC
images.  The Spitzer 8.0 $\mu$m images (Figure \ref{fig10}) reveal a prominent 
bubble centered on the 1.1 mm clump located at 
[$l,b$] = [0.41\arcdeg ,--.50\arcdeg ]  (Figure \ref{fig9}).
In the 8.0 $\mu$m Spitzer image, the bubble appears as a nearly
circular ring of emission with a radius of about 7\arcmin\ ($\sim$ 17 pc)
with a dark region centered on the 1.1 mm clump.  The 1.1 mm clump is 
located at the apex of a cometary cloud seen at 8 $\mu$m that projects 
from more negative latitude toward the center of the bubble (Figures \ref{fig9} 
and \ref{fig10}).      The bubble is rimmed  by several 1.1 mm clumps, 
the brightest of which is located at  [$l,b$] = [0.28\arcdeg , --0.48\arcdeg ].    
Several additional 1.1 mm clumps are located  at larger projected radii 
from the center of the bubble along its rim.   Figure \ref{fig11} shows that 
the bubble interior is lit-up with faint 20 cm  radio continuum emission.    
This region is reminiscent of the IC 1396 bubble in Cepheus with its 
intruding tongue of molecular  gas.  

Several other relatively low-latitude 1.1 mm clumps such as [$l,b$] = [359.71,--0.37], 
[$l,b$] = [359.91,--0.31], and  [$l,b$] = [359.97,--0.46] form a chain below the 
bulk of dust associated with the Galactic center.   This chain of clouds and 
those associated  with the Spitzer ring have been recently noted by 
\citet{Nagayama08} who identified them in near-IR extinction maps as well as 
in CO data.   These authors estimate distances using the cumulative numbers of stars
and reddening to determine distances and column densities.   The distance to this chain of 
clouds located  about 0.2\arcdeg\ to 0.4\arcdeg\ below the Galactic plane ranges 
from  3.6 to 4.2 kpc. K-band extinctions range from 0.4 to 1.0 magnitudes.
In Table~2, we assume a uniform distance of 3.9 kpc for mass estimation.

Another high-latitude region is associated with the compact cluster of
1.1 mm clumps listed in Table~1 and in the Figures as [$l,b$] = [359.96,+0.17].
The Spitzer image (Figure 10) shows bright 8.0 $\mu$m emission
at the high-longitude (northeast) side of the cluster of BGPS clumps.   
This region may be a foreground HII region complex
exhibiting a ``champagne flow" towards high Galactic 
longitudes from a molecular cloud containing several dense 
clumps.

The clumps at  [$l,b$] = [0.53,+0.18]  and [$l,b$] = [0.84,+0.18] 
are located well away from  the dense-gas layer of the CMZ 
which suggests that they may be foreground clumps in the 
Galactic plane far from the Galactic center region.   Thus, for these 
objects, as well as the cluster of clumps near [$l,b$] = [359.96,+0.17],
we assume a distance of 3.9 kpc.

\section{Discussion}

\subsection{A Cavity Extending 30 pc from Sgr A?}

Millimeter and sub-mm imaging of the CMZ reveals a network of dust
filaments, shells, and dense clumps.    The $\sim$ 30 pc diameter infrared
bubble between Sgr A and  $l \approx$ 0.2\arcdeg\ is associated with a 1.1
mm and 350 $\mu$m cavity that contains the Quintuplet and Arches clusters
and Sgr A.  Several chains of clumps in the bubble interior point towards 
Sgr A.   The high combined luminosity of the Arches and 
Quintuplet clusters and of circum-nuclear cluster surrounding Sgr A* 
can plausibly produce the cavity.   Below,  the physical parameters of
this feature are estimated.  

The central few hundred parsecs of the Galaxy contain at least 4 different
types of gas reservoir;   cold (T $< ~10^2$ K) and dense molecular 
clouds  (n(H$_2$) $>$ $10^4$ cm$^{-3}$) 
that comprise the CMZ \citep{MorrisSerabyn1996,oka98},  
a warm inter-cloud medium  (T $\approx$  $10^2$  to $10^3$ K)  traced by 
species such as H$_3^+$  in absorption  \citep{oka98,oka05,Goto2008} 
and diffuse  $^{12}$CO and CI   in emission  \citep{Martin04},  
a warm ionized  component (T $\approx$ $10^4$  K)  
traced by  recombination and fine structure lines,  radio scattering 
\citep{LazioCordes98},  and free-free emission and absorption,  and an 
ultra-hot component  (T $> ~10^6$ K) traced by  X-rays \citep{wang02}. 

The volume filling factor of the dense, gravitationally bound cold molecular 
phase is in the range 0.01 \citep{Cotera00} to 0.1\citep{MorrisSerabyn1996}.
The remaining volume is either filled with gravitationally unbound (to individual
clouds or star clusters) warm neutral atomic and molecular, photo-ionized, 
or X-ray emitting gas \citep{Martin04,
oka05,Goto2008}.    Abundant  3 to 4 $\mu$m H$_3^+$ absorption towards 
massive stars in the central  0.2\arcdeg\  of the Galaxy indicates that warm, but 
mostly neutral gas unbound to any particular cloud  may occupy a large fraction 
of the volume in the CMZ.  However, the large angular-diameters of background 
extra-galactic radio sources  produced by scattering of their radio continuum 
emission by  a foreground screen of electrons \citep{LazioCordes98} observed 
towards  the CMZ indicates the  presence of relatively dense HII regions towards 
most lines-of-sight through the CMZ.
\citet{YusefZadeh07}  show that  6.4 keV  Fe K$\alpha$  emission fills 
the projected area of the cavity delineated by the Spitzer 24 $\mu$m
bubble and the 1.1 mm cometary clouds  (see their Figure 13).  
However, the strongest  6.4 keV emission appears
to be associated with the molecular clouds traced by the 1.1 mm continuum.
These authors suggest that the X-ray emission is produced by the interaction
of low-energy cosmic rays with dense clouds.

We consider a scenario in which UV radiation and  winds emerging
from the Arches, Quintuplet, and Sgr A* clusters and the central black 
hole are responsible for carving out the cavity.    The massive  Quintuplet 
and Arches clusters may have contributed to formation of the individual 
cells in which these clusters are embedded.   Towards positive longitudes
the outer-boundary  of the cavity is near $l \approx$~0.20\arcdeg\   
at a projected distance of  about 30 pc from  Sgr A* (Figure \ref{fig9}).    
As shown in Figure \ref{fig12},   the high longitude
end of the cavity is marked by a chain of small dust clouds located around
$l$  = 0.22\arcdeg\  (northeast of the Quintuplet cluster marked with a ``Q" and
below the bright 1.1 mm cloud at [0.26\arcdeg ,+0.03\arcdeg ]).   
The cell associated with  the  Quintuplet cluster may extend to 
$l$ = 0.28\arcdeg , or  about 40 pc from Sgr A.
Below (to the right of) Sgr A, the intense emission from the 
20 and 50  km~s$^{-1}$  clouds masks the cavity if it extends this far.   
The 20 and 50  km~s$^{-1}$ clouds are physically close to Sgr A; 
they may block the penetration of  UV irradiation and winds. 
The Spitzer 24 $\mu$m image shows that the low-longitude end of the 
cavity is near $l$ = 0\arcdeg .   

An order-of-magnitude estimate of the parameters of an ionized cavity can
be obtained from its size and the 20 cm radio continuum flux, which imply
that the average electron density in the cavity can not be much above 
100 cm$^{-3}$.   Assuming that the cavity interior is ionized and has a 
uniform hydrogen  density,  the  Lyman continuum luminosity (in units of 
the number of  ionizing photons emitted per second) required to ionize
the cavity is   $L(LyC) = 3 \times 10^{50}  n^2_{100} r^3_{10}$
where $n_{100}$ is the average ionized hydrogen density in 
units of 100 cm$^{-3}$ and $r_{10}$ is the mean radius of the cavity
in units of 10 pc.     The above parameters imply 
an ionized mass of about  $M_{HII} \approx 1.4 \times 10^4  
n_{100} r^3_{10}$ Solar masses around the Arches, Quintuplet, and
Sgr A.  

The Lyman continuum luminosity of the Sgr A region  has been estimated 
from the flux of free-free radio continuum to  be around 
$L(LyC) \sim few \times 10^{51}$
ionizing photons per second.  The entire CMZ has a Lyman continuum
luminosity of around $1 - 3 \times 10^{52}$ ionizing photons per second
\citep{MorrisSerabyn1996}. This radiation 
field is produced mostly by  young massive stars with a possible contribution 
from the central black hole in the immediate vicinity of Sgr A.  Dust within
the cavity and the CMZ may absorb a substantial fraction of the Lyman
continuum radiation.  However, even if nearly 90\% is absorbed,  the 
UV radiation field is sufficient to maintain the central cavity in an ionized state
for a mean density of order $10^2 - 10^3$ cm$^{-3}$. 

The orbital time scale in the center of the Galaxy sets a constraint on
the age of the cavity and its cometary clouds.   Using a typical orbit 
velocity $V_o$ = 120 km~s$^{-1}$, a test particle in a circular orbit 
would have an orbital period of 
$t_o = 2 \pi r / V_o ~\sim$ 0.5 $r_{10} V^{-1}_{120}$ Myr where  
$V_{120}$ is the orbit velocity in units of 
120 km~s$^{-1}$ and $r_{10}$ is the distance from the Galactic center
in units of 10 pc.  Portions of clouds at different radii that are gravitationally
unbound from each other would be sheared on a time-scale comparable
to the orbit time.   The presence of recognizable cometary clouds implies
that they have been ionized on a time-scale shorter than the orbit time.
On the other hand, dense clouds are expected to orbit on largely radial
trajectories.  Thus, the observed cometary features may be objects that
have only recently fallen into the interior of cavity with a speed comparable 
to the orbit speed.  In this case, the elongation of these features away from 
Sgr A may be a result of the survival of dense gas and dust in the region
shielded by the dense head facing Sgr A.   Alternatively,  this cavity may
be only recently ionized by the Arches and Quintuplet clusters.    If these 
clusters have high space velocities, they may have completed several orbits 
around the  nucleus during their lifetimes \citep{Stolte2008} and the cometary 
clouds in their vicinity may be a consequence of this motion.

\subsection{Fueling Star Formation in Sgr A}

The formation mechanism of  massive stars within the central parsec of the Galaxy
has been a long-standing mystery.     \citet{Wardle08}  present a model in 
which a giant molecular cloud passes through the center.   Gravitational 
focusing of trajectories  by the central black hole and the central stellar cluster of 
older stars causes gas to collide and dissipate angular momentum and kinetic 
energy in the wake.   This results in the  trapping  of some gas  in  eccentric,  
circum-nuclear  orbits, forming a disk of  compressed  material.   Gravitational 
instabilities  in the  disk can result in the formation of  massive stars.   

It has been long suspected that the two most prominent molecular clouds
near  Sgr A, the 20 and 50 km~s$^{-1}$ clouds,  are located within
the central 30 pc  \citep{MorrisSerabyn1996,CoilHo1999,Wright2001,
McGary2001,Christopher05,Lee08}.    
High-dipole moment molecules such as HCN and CS show that they are
among the densest and warmest clouds in the CMZ \citep{MiyazakiTsuboi2000}.  
Together, these two clouds give the CMZ its apparently small (few arcminute)
scale-height near Sgr A.  
 
The 50 km~s$^{-1}$  cloud is located about 10 pc in projection from the 
CND towards higher  longitudes.    Several 350 $\mu$m   filaments connect the  
50 km~s$^{-1}$ cloud  to the CND (Figure \ref{fig12});    these features  correspond 
to ammonia streamers seen by \citet{McGary2001}.     \citet{Lee08} present a 
3D model based   on near-infrared observations of  H$_2$ and argue that the 
2 pc radius CND  was  recently fueled by the passage of  the 50 km~s$^{-1}$ 
cloud within a few pc of the  nucleus.     Assuming that the 50 km~s$^{-1}$ cloud 
is on an  x2 orbit,   the projected separation of about 10 pc from Sgr A* implies that for a 
proper motion of  100 km~s$^{-1}$,  it passed by the nucleus  about $10^5$ years 
ago.   The passage  of the 50 km~s$^{-1}$  cloud may have formed the CND, 
leading to  the current  episode of massive  star formation activity within its clumps. 
     
The 20 km~s$^{-1}$  cloud  is the most prominent object at 350  $\mu$m  in the
inner square degree of the Galactic center (Figures \ref{fig12} and \ref{fig14}); 
it is located about 7 to 20 pc from the nucleus in projection.   It is elongated with a 
length-to-width ratio of about 5 to 10:1 with a major axis that points about  to 2\arcmin\ 
below the CND (Figure \ref{fig12}).  Table 2 shows that  for an assumed grain 
temperature of 20 K,  the 20 km~s$^{-1}$  cloud is  the second most massive 
and densest  cloud in  the CMZ (Sgr B2 is the most  massive and dense).   A 
bright filament of 350 $\mu$m and 1.1 mm dust emission  connects  the 20 km~s$^{-1}$ 
cloud to the CND;  this feature is  likely to be  the same as the one detected in 
ammonia by \citet{CoilHo1999}.    This structure  provides  support  for  the model  
in which CND continues to be actively fueled, currently by flow from  the 
20  km~s$^{-1}$ cloud  \citep{CoilHo1999}.

If  the center of the 20 km~s$^{-1}$ cloud is currently located about 20 pc 
from the nucleus, and  on a near collision course with the nuclear cluster, 
it may provide fuel for further star formation in the nucleus in the near future.    
Assuming that this cloud  is on an x2 orbit  \citep{binney1991,Marshall07,
RodriguezFernandez08} whose velocity is mostly orthogonal to our line-of-sight, 
and has speed of about 100  km~s$^{-1}$,  it will come closest to the nuclear region 
in a few hundred thousand years.

The IRS 16  cluster of massive stars surrounding Sgr A* could not have formed from
any currently visible cloud.   The IRS 16 cluster has an age of at least several 
Myr;   \citet{Tamblyn1993} give an age of 7 to 8 Myr.   The formation of a previous
circum-nuclear ring would have required the passage of a cloud about 5 to 10 Myr ago.
Any such cloud would have moved too far from the nucleus to have a clear connection to it
and may have been disrupted by a combination of tidal forces, UV radiation and winds.

\subsection{Little Star Formation in the Large Velocity Dispersion Complex,
Bania's Clump 2}

Figure \ref{fig19} shows the 1.1 mm image of the region containing Bania's Clump 2.
Figure \ref{fig20} shows this image superimposed on the Spitzer Space Telescope
IRAC image.   These images show that  despite the presence of dozens of clumps
emitting brightly at 1.1 mm, Clump 2 lacks significant mid-infrared emission from 
either extended or point sources.  
It is  not forming stars at a rate comparable to similar clumps elsewhere in the 
Galaxy.    In the Galactic plane, HII regions and nebulae excited by young, massive
stars are frequently seen at 3.6 to 5.8 $\mu$m.   The 8 $\mu$m Spitzer images trace 
O, B, and A stars by the presence of  bubbles rimmed by bright emission from 
UV-excited PAH molecules and nano-particles.   Few of these signatures of massive 
stars are observed towards Bania's Clump 2.  The longer wavelength Spitzer images 
(e.g. 24 $\mu$m) trace highly embedded young stars and clusters.    Inspection
of the 24 $\mu$m Spitzer images  shows that Bania's Clump 2  also lacks a
substantial population  of embedded sources visible at 24 $\mu$m.   Perhaps 
the clumps are embedded in a medium having an unusually high pressure or large 
non-thermal motions which suppress  gravitational instabilities on the mass-scale of 
stars or even clusters. Many of the  Clump 2 features  are seen in silhouette against 
background star light and diffuse  emission from the  ISM in the 3.6 to 8 $\mu$m 
Spitzer images, indicating that they  are in front of much of the  central bulge.

The absence of emission in the Spitzer images is not the result of high extinction.
Regions having similar amounts of 1.1 mm emission closer to the mid-plane of the 
Galaxy show a factor of 2 to 4 times more emission between 6  and 24 $\mu$m.
Furthermore, the brightest 1.1 mm emission maxima towards Bania's Clumps 2 imply 
localized extinctions of A$_V < $10 magnitudes averaged over a 40\arcsec\ aperture.    
However, the low-spatial frequency emission resolved-out by BPGS may amount 
to as much as A$_V$ = 30 magnitudes based on the line-of-sight  column density of CO.   
But, this would be insufficient to completely hide  all Spitzer mid-IR emission.

The exceptionally large line-widths and velocity extent of this complex may 
be caused by three processes:  
First, the abrupt change in the orientation 
of the orbital velocity where a spur or inner dust lane 
encounters the ridge of dense gas at the leading edge of the Galactic bar.   
Second, at  the ends of the so-called ``x1" orbits in a tri-axial potential where
they become self-intersecting 
\citep{Contopoulos1980,binney1991,Marshall07,RodriguezFernandez08}.
In either mechanism cloud-cloud and cloud-ISM collisions result in high ram-pressures,
shocks, and turbulence that can produce large line-widths and non-thermal motions.
Third, broad-line features such as Bania's Clump 2 may consist of unrelated 
clouds along an x1 orbit which has a leg aligned with our line-of-sight.    In this
scenario, Bania's Clump 2 traces gas in a dust lane at the leading edge of the
bar which is elongated along our line-of-sight.
 
Figure \ref{fig21} shows a cartoon of the possible configuration of the central 
regions of the Milky Way as seen from the north Galactic pole.  
The x1 orbits are elongated along  the major  axis of the Galactic bar in a frame 
of reference rotating with the bar.   As a family of orbits with  decreasing 
semi-major  axes (corresponding to lower angular momenta about the nucleus)  
are considered,    the ends first  become more ``pointed",   then they  become 
self-intersecting.     Clouds whose orbits 
decay eventually enter the regions of self-intersection where they may  encounter 
other clouds or the lower density ISM on the same trajectory with supersonic velocities.  
The resulting shock  waves and consequent dissipation of orbital energy causes the 
clouds to rapidly migrate onto the smaller ``x2" orbits whose major axes are 
{\it orthogonal}  to the bar.    Most of the observed  dense gas in the CMZ is 
thought to occupy the near-side of the X2 orbits at positive longitudes where 
they are seen as IRDCs.

Models of gas flows in a tri-axial potential can reproduce the
major features observed in the longitude-velocity diagrams of molecular
tracers of the inner Galaxy \citep{Bissantz03, RodriguezFernandez06,
RodriguezFernandez08,RodriguezFernandez09,Pohl08,Liszt09}.   
It is suspected that the major axis of the central bar is currently oriented within 
15\arcdeg\ to 45\arcdeg\ of our line-of-sight.   The conventional interpretation 
of the Galactic center  molecular line emission places the high-velocity 
rhombus evident in $l$-V diagrams  in the  ``180 pc molecular ring"  
(sometimes called the ``Expanding  Molecular Ring" or EMR - see  Figure 4 in 
Morris and Serabyn 1996).   This feature is tilted with respect to the Galactic 
plane so that the positive latitude part is seen at positive velocities.   
In Figure \ref{fig4}, this feature is labeled as the ``Leading Edge of the Bar"  
since in most models, the gas that is transiting from x1 to x2 orbits is found 
in such a dust lane  (see Figure 5 in Morris and Serabyn 1996). 

Bania's Clump 2 is located at higher longitudes than the  180-pc molecular
ring. Thus, it  is either  located  where the x1 orbits become
self-intersecting, or where a spur encounters the dust lane at the leading-edge
of the bar, or traces gas in the dust lane which is oriented nearly parallel to
our line-of-sight.  \citet{RodriguezFernandez06} shows the likely  location of 
Bania's Clump 2 in a hypothetical  face-on view of the Galaxy (see their 
Figure 2).  The superposition  along the line-of sight of  clouds moving 
near the apex of  the innermost,  self-intersecting x1 orbits can 
exhibit a  large range of radial velocities within a very restricted spatial region.  
This phenomenon is caused by the  near co-location of clouds moving towards 
and away from the Galactic center. Some of these clouds will be colliding, 
resulting in compression, heating, and turbulence generation.   While 
low-velocity cloud-cloud collisions can trigger star formation, high-velocity 
collisions are likely to be disruptive;  sufficiently fast collisions dissociate 
molecules and raise post-shock temperatures.    
 
The presence of 1.1 mm clumps and associated gas seen in high-dipole 
moment molecules such as CS and HCN suggests that the smallest mass that can 
become gravitationally unstable is comparable to the observed clump mass
of 10$^3$ to $10^4$ M$_{\odot}$.  The low-rate of star formation in these clumps
implies that fragmentation is suppressed in this environment, possibly
by shear or large-fluxes of energy and momentum from large-scales to small.

If cloud collisions and supersonic internal motions suppress star formation,
the dissipation of even a small fraction of this energy must emerge as radiation,
resulting in the excitation of  fine structure cooling lines such as the 63 $\mu$m [OI],
157 $\mu$m [CII], and 205 $\mu$m [NII], or high-J lines of CO and OH.  
A substantial portion of the internal energy may also go into heating grains 
which would emerge as sub-mm continuum radiation.
The Herschel Space Observatory may be able detect the resulting fluxes.

While such shocks may promote star formation in other environments such as in
Sgr B2, they have failed to do so here.  Perhaps the mm sources in Bania's Clump 2
are in an earlier, pre-stellar stage of evolution.   Clumps could evolve to denser 
configurations capable of star formation in a few crossing times.   Using a typical 
Galactic center cloud line-width of $\Delta V$ = 20 km~s$^{-1}$ for each clump,  a 
clump with a  radius of $r$ = 1\arcmin\  ($\sim$ 2.5 pc) will evolve on a time-scale
$t \sim 2 R /  \Delta V f \sim 2.5 \times 10^5 f^{-1}$ years where $f$ (= 0.1 to 1)
is the area filling factor of gas in the clump.  Thus, it is possible that we are seeing 
the Clump 2 objects in an early state of evolution and that they will become active star 
formers in the  next few million years.   

The  Clump 2 complex has a diameter of about 0.5 \arcdeg (70 pc), implying 
R$_2 \sim 2.3 \times  10^{20}$ cm (25 pc), and a velocity extent of about $2 V_2$ 
= 200 km~s$^{-1}$.     The time-scale on  which the entire Clump 2 region should 
evolve is about  $t_2 \sim 2 R_2 /  \Delta V_2 f _2 \sim 3.6 \times 10^5 f^{-1}$ years.
Figure \ref{fig19} implies  an area filling factor of about  $f_2 \sim$ 0.1 for the 1.1 
mm clumps, so the evolutionary time-scale for the entire complex is a few Myr, 
comparable to evolutionary time-scales of individual clumps.   Both of these time-scales
are short compared to the $\sim$ 15 Myr orbital time-scale at the projected speration
between Sgr A$^*$ and Clump 2.

\section{Conclusions}

The first wide-field $\lambda$ = 1.1 mm continuum map
of the inner six degrees of the Galaxy is presented.   This emission
traces dozens of bright star forming clumps and an extended, nearly continuous
filamentary network.  Filaments tend to outline a lattice of voids.  Some
contain bright, diffuse nebulosity in infrared images that may indicate the
presence of HII regions.  A few may contain supernova remnants.  However,
many cavities do not contain obvious energy sources.   They may trace fossil 
cavities carved in the Galactic center ISM by the action of massive stars
that have died in the recent past.   The large velocity dispersion of gas
in the Galactic center region combined with the shear of differential
rotation would tend to erase unsupported cavities on a relatively short 
time-scale.   Thus, the cavities require energy injection within the last few Myrs.
The four main results of this study are:

First, the 1.1 mm Bolocam Galactic Plane Survey (BGPS) reveals over a thousand
individual clumps within a few degrees of the Galactic center that are associated
with the Central Molecular Zone (CMZ).    Their small scale height compared to the 
rest of the Galactic plane and association with the dense molecular gas 
in the CMZ indicates that over  80\% of these clumps are likely
to be within the central few hundred parsecs of the Galaxy.     The 20 and 50 
km~s$^{-1}$ clouds near Sgr A, and the Sgr B2 complex,  host the brightest 1.1 
mm sources in the field of study.  Comparison of the 1.1 mm emission with  350
$\mu$m maps of the regions around Sgr A and Sgr B2 is used to constrain
the dust emissivity,   grain temperatures, and optical depths.    These measurements 
indicate peak 1.1 mm optical depths of around 0.1, and grain temperatures around
20 to 70 K for an emissivity power-law index of 2.0.   Sgr B2 appears to have
an atypically low emissivity index for reasonable dust temperatures, indicating that
substantial grain growth has occurred.    Both the 350 $\mu$m and 1.1 mm emission
provide evidence for about a half dozen elongated, cometary dust clouds with 
dense heads pointing towards the Sgr A complex.  Some of these features 
delineate the walls of a multi-celled cavity having a mean radius of at least 30 pc.   
The most prominent cometary dust clouds delineate cells located near the
massive Arches and Quintuplet clusters.    These dust features point towards Sgr A, 
and indicate that UV radiation from the central cluster of massive stars has
sculpted the Galactic center environment within the last 1 Myr.  The massive
20 and 50 km~s$^{-1}$ clouds appear to be located in the interior of this cavity
and may shield the nuclear ISM at low longitudes and latitudes from ionization.

Second, the collection of clouds located between $l$ = 3\arcdeg\  and 3.5\arcdeg\ 
and $V_{LSR}$ = 0 to 200 km~s$^{-1}$  known as ``Bania's Clump 2" are
highly deficient in active star formation given their 1.1 mm dust continuum 
fluxes.  It is possible that the clumps in Clump 2 are in a pre- star forming 
state.   This observation provides evidence for models in which Clump 2 
consists of gas and dust that has recently been shocked and rendered highly 
turbulent.  This feature may either trace gas that has recently fallen in from the
innermost x1 orbit (elongated along the bar)  onto the outermost x2 orbit
(elongated orthogonal to the bar), or gas that is entering the shock on the leading
edge of the bar from a spur, or a dust lane at the leading-edge of the bar that
is elongated along our line-of-sight.

Third, under the assumption of a constant dust temperature in the CMZ, 
the mass spectrum of clumps  M $>$ 70  M$_{\odot}$
in the Galactic center is characterized by a steep power-law  
$dN / dM = k M^{-2.14 (-0.4, + 0.1)}$.   This index is significantly steeper than
found from CS observations of the Galactic center and similar to the index
found for star forming cores in the Solar vicinity.     While such a steep index
may be a real feature of the Galactic clump population, it may in  part be an 
observational artifact produced by the attenuation of low spatial-frequency 
flux, source confusion, and  the watershed algorithm used to construct Bolocat 
which tends to subdivide  large objects into clusters of smaller ones.  

Fourth, comparison of the 350 and 1100 $\mu$m images places new constrains
on dust properties.  Values of $\beta$ between 1.5 and 2.0 give the most reasonable
dust temperatures with values ranging from 15 K to as high as 80K.  However, 
Sgr B2 is anomalous.  The low flux ratios here require lower values of $\beta$,
providing evidence for substantial grain growth.

\acknowledgments{
The BGPS project is Supported in part by the National Science Foundation
through NSF grant AST-0708403.   J.A. was supported by a Jansky Fellowship from 
the National Radio Astronomy Observatory (NRAO).    The first observing runs for BGPS were
supported by travel funds provided by NRAO.   Support for the development
of Bolocam was provided by NSF grants AST-9980846 and AST-0206158.   
NJE and MKD were supported by NSF grant AST-0607793.  
This research was performed at the Caltech Submillimeter Observatory 9CS), supported by
NSF grants AST-05-40882 and AST-0838261.
CB and CC were supported by a National Science Foundation Graduate Research 
Fellowship.   We thank Farhad Yusef-Zadeh for providing his 20 cm radio image.
We thank an anonymous referee for very helpful comments that improved the 
manuscript.}

\clearpage

\section{Appendix I:  Comments on Individual Regions}

In this section,  individual regions  Listed in Table~2 
are discussed in order of increasing 
Galactic longitude.    Fluxes and mass  estimates for the marked features 
(based on visual inspection of the images rather than Bolocat) 
are given in Table 2. 

{\it  The Great Annihilator (1E1749.7-2942)}; [$l,b$] = 359.116,--0.106):
Long suspected to be a major source of 511 keV emission from decaying
positronium, this hard X-ray and $\gamma$-ray source was initially
suspected to be stellar-mass black hole accreting from a molecular
cloud \citep{BallyLeventhal91,Mirabel91}.   However, recent observations
indicate that this object may be a low-mass X-ray binary 
\citep{Main99}.  The molecular cloud located adjacent to 
1E1749.7-2942 was recently observed at millimeter wavelengths
with the CARMA interferometer by \citet{HodgesKluck09} who
provides some mass estimates.   The molecular cloud is detected
in our 1.1 mm  image  (Figure \ref{fig22}).    Thus, we can compare
the dust-based mass estimate with the gas-based estimate.    Interestingly,
while the clump near 1E1749.7-2942 is clearly seen, the brightest 
HCN and HCO$^+$ clump in the maps of \citet{HodgesKluck09}
is not evident in the 1.1 mm dust continuum.   The 1.1 mm flux at the location of
their HCO$^+$ and HCN peak 2b is only about 30 mJy, comparable to the
noise in our maps.

The masses estimated for the cloud adjacent to the Great Annihilator
from the 1.1 mm emission in  40\arcsec\ and  300\arcsec\ apertures are 
$2.7 \times 10^2$  and $7.2 \times 10^3$ M$_{\odot}$ (for fluxes scales-up
by a factor of 1.5).   The mass in the smaller  aperture
was measured at the location of the HCO$^+$ peak found by  \citet{HodgesKluck09}.
On the other hand, the mass in the larger aperture was determined in an 
aperture centered on the 1.1 mm peak at [$l,b$] = 359.135, --0.100. 
 \citet{HodgesKluck09} use the Virial theorem to estimate the mass from the clump-size and 
line-width for their clump 1 (at the location shown by the small circle
immediately next the the Great Annihilator in Figure \ref{fig22}).
They find M = $3.9 \times 10^3$ M$_{\odot}$,  in between the two 
BGPS estimates in  40\arcsec\  and 300\arcsec\ apertures.

{\it  Sgr C} ([$l,b$] = 359.47,--0.11):
There are relatively few major star forming complexes at negative
Galactic longitudes.  The exception is the Sgr C complex of HII regions.
The BGPS images reveal a prominent comet shaped cloud clump 
complex facing west  in the Sgr C region.  

[$l,b$] = 359.47,--0.03:
This feature is a 1.1 mm clump located at the low-longitude end of a chain of
clumps that can be trace back to just above Sgr A.   This chain is part of the
so-called ``Expanding Molecular Ring"  that may mark gas and dust 
at the leading edge of the Galactic center bar.   In CO spatial-velocity diagrams,
this feature is part of the positive latitude,  positive velocity feature that 
defines the ``rhombus" thought to mark the innermost x1 orbit (marked in Figure
4).  A 20 cm non-thermal filament crosses the Galactic plane on the low-longitude
side of the marked clump in Figure \ref{fig11}. 

\begin{figure}
\epsscale{1.0}
\center{\includegraphics[width=1.0\textwidth,angle=0]{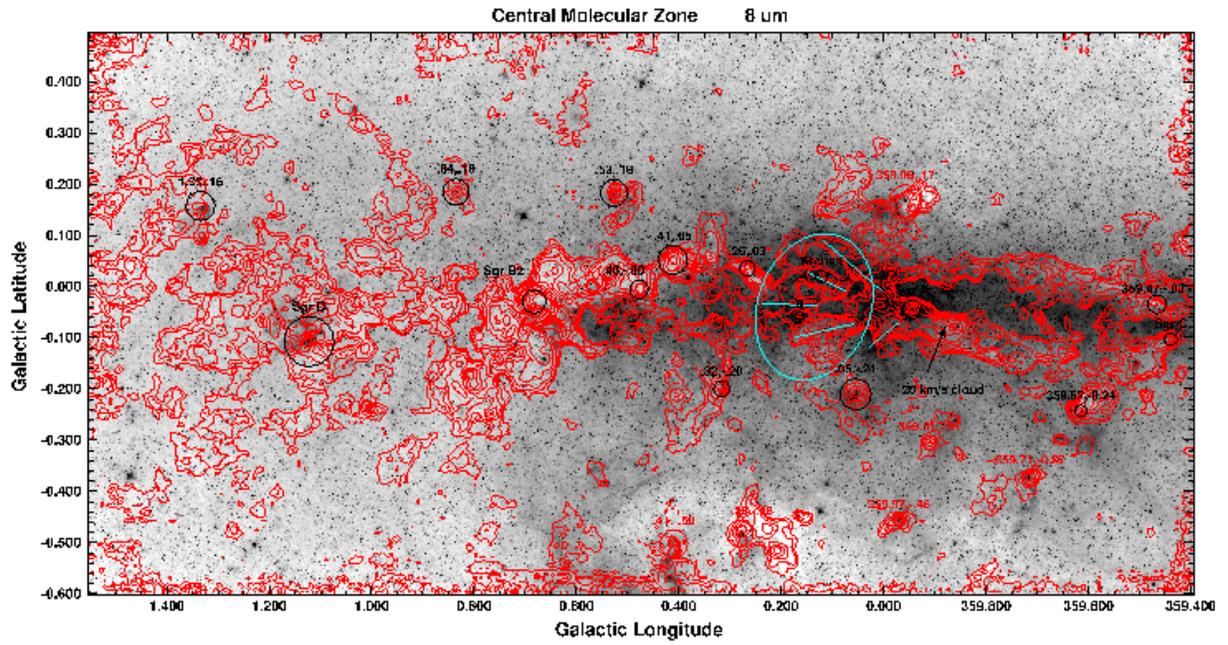}}
\caption{Contours of 1.1 mm dust continuum emission from the CMZ superimposed on
the Spitzer IRAC 8 $\mu$m image  \citep{Arendt2008} show with a logarithmic greyscale
presented at the same spatial scale as Figure 9.  
The 1.1 mm contour levels are 0.3, 0.45, 0.75, 1.2, 2.0, 3.0, 5.0, 7.5, 12.0, 20, and 30 Jy/beam.}
\label{fig10}
\end{figure}

\begin{figure}
\epsscale{1.0} 
\center{\includegraphics[width=1.0\textwidth,angle=0]{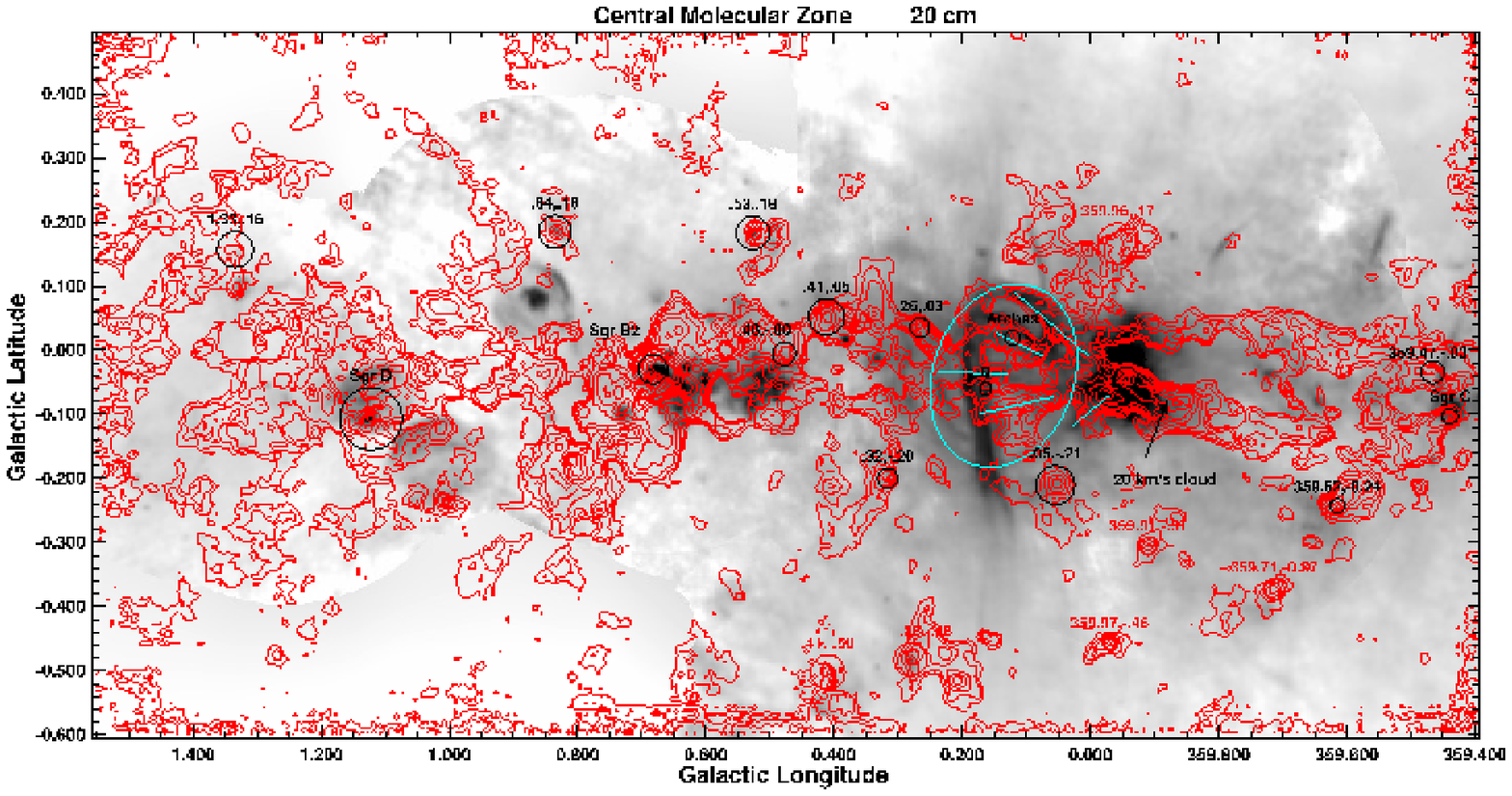}}
\caption{Contours of 1.1 mm dust continuum emission from the CMZ superimposed on
 the 20 cm  radio continuum image from  \citet{yusef-zadeh2004} shown on a
 logarithmic greyscale
 and presented at the same spatial scale as Figure 9.
 The 1.1 mm contour levels are 0.3, 0.45, 0.75, 1.2, 2.0, 3.0, 5.0, 7.5, 12.0, 20, and 30 Jy/beam.}
\label{fig11}
\end{figure}

\begin{figure}
\epsscale{1.0} 
\center{\includegraphics[width=1.0\textwidth,angle=0]{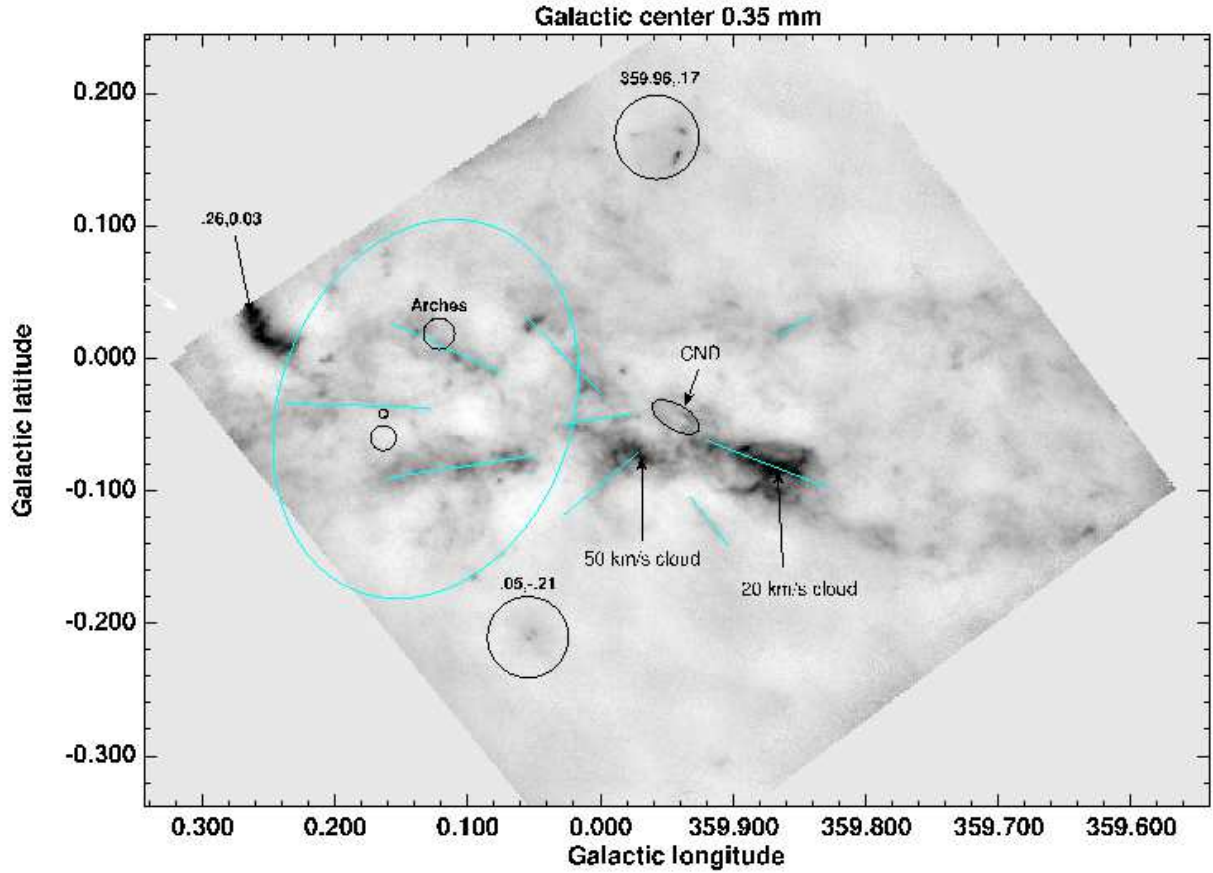}}
\caption{A SHARC-II 350 $\mu$m  image showing warm dust continuum
emission from the vicinity of Sgr A in the center of the Galaxy with 9\arcsec\ resolution.
This image has been "unsharp-masked" to suppress large scale structure and
to match the spatial frequency response of the Bolocam 1.1 mm images. }
\label{fig12}
\end{figure}

\begin{figure}
\epsscale{1.0} 
\center{\includegraphics[width=1.0\textwidth,angle=0]{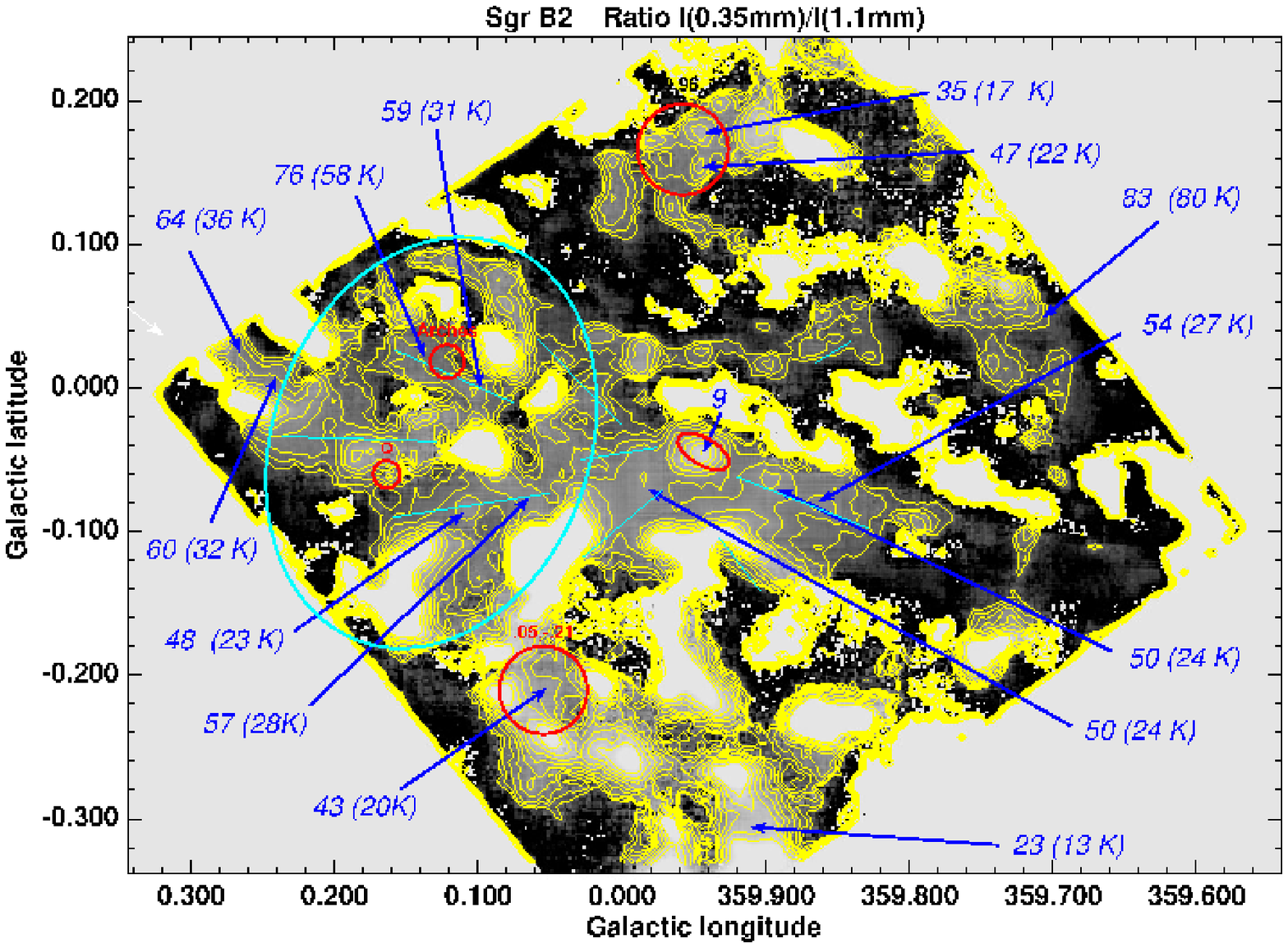}}
\caption{A map showing the ratio of the surface brightness  of 350 $\mu$m emission 
divided by 1.1 mm emission in the vicinity of Sgr A.    The display range goes from 
0 (white) to 100 (black).     Regions with ratios larger than 100, which corresponds to the
Rayleigh-Jeans limit for $\beta$ = 2.0 dust have been masked and are shown as white.    
The ratios at the edge of the SHARC-II field are  spurious.    As discussed in the text,
the 350 $\mu$m map has been convolved with a gaussian kernel to have the
same effective beam size as the 1.1 mm map (33\arcsec\  effective beam).    The italic 
numbers indicate the flux ratio at the locations indicated by the associated 
arrows. The  numbers in parentheses give the derived dust temperatures for an 
emissivity power-law index of $\beta$ = 2.0.    Emission from the central black hole 
dominates the 1.1 mm flux originating 
in the center of the CND.  Because the source of emission is unlikely to dust, 
no dust temperature is given.  }
\label{fig13}
\end{figure}

\begin{figure}
\epsscale{1.0} 
\center{\includegraphics[width=1.0\textwidth,angle=0]{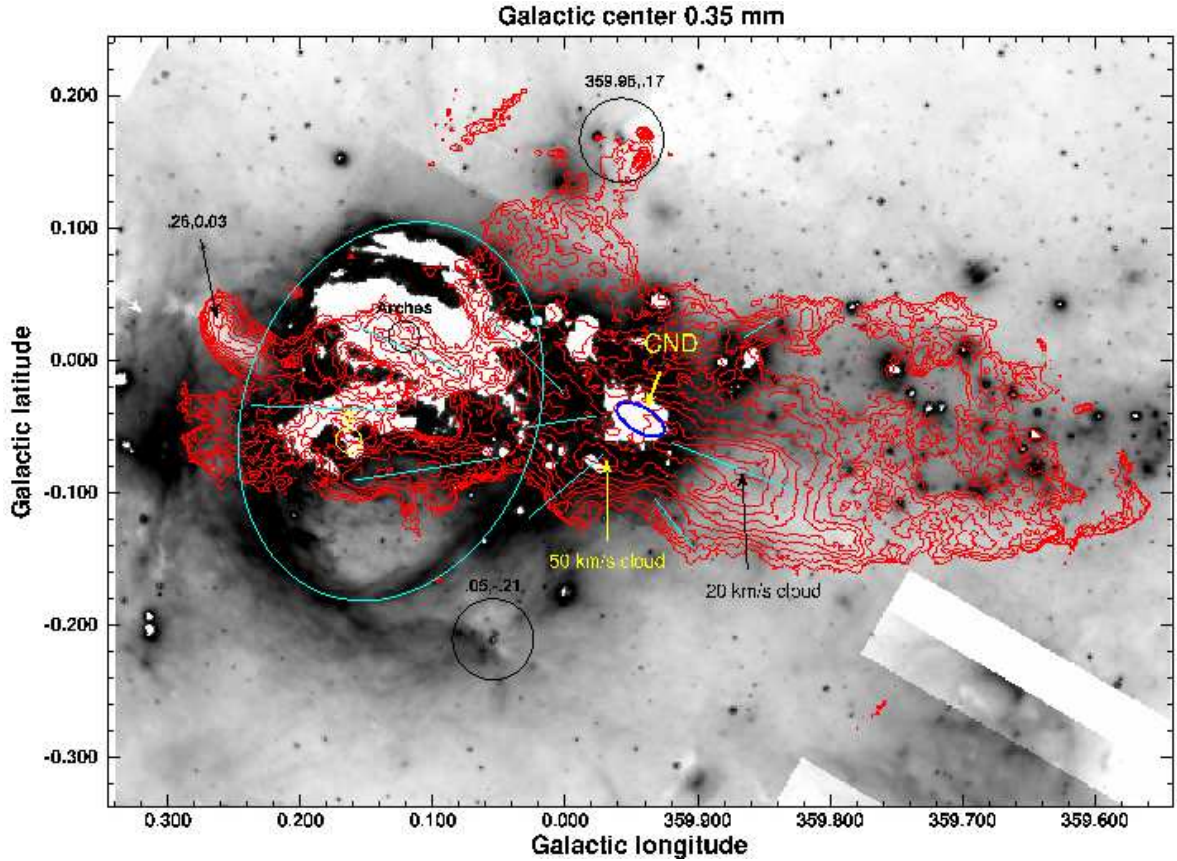}}
\caption{Contours of  SHARC-II 350 $\mu$m  emission superimposed 
on a greyscale rendition of the 24 $\mu$m Spitzer image showing the
vicinity of Sgr A
\citep{Yusef-Zadeh2009} displayed on a logarithmic intensity scale. 
The sharp-edged white regions indicate locations where the Spitzer
image is saturated.     Contour levels are:  4.5, 7, 9.0, 13, 18, 26, 
37, 53, 75, 106, and 150 Jy/beam.}
\label{fig14}
\end{figure}

\begin{figure}
\epsscale{1.0} 
\center{\includegraphics[width=0.9\textwidth,angle=0]{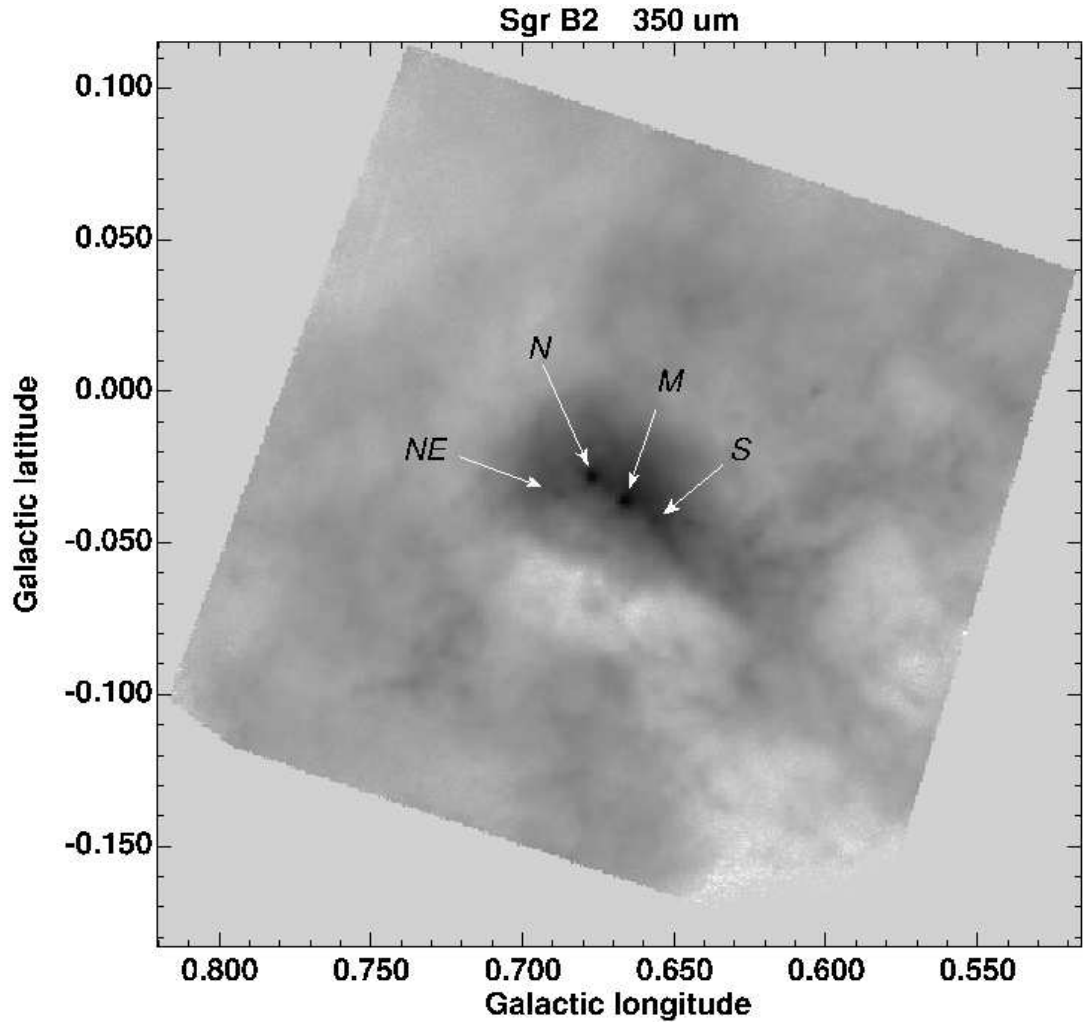}}
\caption{A SHARC-II  0.35  mm image showing the Sgr B2 complex on 
a logarithmic intensity scale.  The bright emission is contoured with levels
shown at 100, 130, 170, 220, 285, 371, 482, 627, 814, 1060, and 1375 Jy/beam.
The two peaks in the center correspond to Sgr B2N and SgrB2 Main. }
\label{fig15}
\end{figure}

[$l,b$] = 359.62,--0.24:
Located 15\arcmin\  east of Sgr C, there is a second, east-facing cometary 
cloud in the 1.1 mm continuum image.  The dense head of this cometary feature
abuts the HII region G359.62-0.25 which consists of a series of 20 cm  filaments
east of the 1.1 mm clump.   Extended emission east of the clump is evident in the
GLIMPSE 8 $\mu$m image (Figure \ref{fig10}).  The
1.1 mm clump is  the brightest portion of a roughly east-west
ridge that extends  west towards  Sgr C.   Both the Sgr C and G359.62-0.24 
cometary clouds are  clearly seen in silhouette against background 8 $\mu$m emission.

[$l,b$] = 359.71,--0.37:
A compact 1.1 mm knot associated with a cometary feature in the GLIMPSE images.
This is one of the clumps associated with the band of foreground clouds located
at about 3.9 kpc from the Sun \citep{Nagayama08}.

[$l,b$] = 359.91,--0.31:
This is another of  the clumps associated with the band of foreground 
clouds located at about 3.9 kpc from the Sun \citep{Nagayama08}.  
It is a cometary cloud facing the Galactic plane that is visible in silhouette 
against the 8 $\mu$m background in the Spitzer GLIMPSE images.

{\it The 20  km~s$^{-1}$ cloud.} ([$l,b$] = 359.88,--0.08):  
The central twenty parsecs of the Galaxy contains the Sgr A radio
complex and the 20 and 50 km~s$^{-1}$ clouds;  both clouds are seen in 
silhouette in the Spitzer 8 images (Figure \ref{fig10}).  
The 50 km~s$^{-1}$ cloud located east-northeast of Sgr A*
contains bright MSX point sources while the 20 km~s$^{-1}$ cloud
lacks bright MSX counterparts.  Thus,  the former is
more evolved and is actively forming stars while the latter is more
similar to IRDCs and likely to be either in a pre-star forming stage
of evolution, or its star formation may be inhibited by its proximity
to the Galactic nucleus.

The 20  km~s$^{-1}$ cloud is the third brightest source of 1.1 mm 
emission in the Central Molecular 
Zone (next only to Sgr B2 and and Sgr A*) and  has a mass of at least   
$1.6 \times 10^5$~M$_{\odot}$.   The cloud is elongated along
a direction pointing towards  Sgr A*.   The 350 $\mu$m image (Figure
\ref{fig12}) shows complex structure consisting of several
bright arcs of dust protruding above and below the cloud.
Although there is abundant extended  20 cm  continuum emission in 
the general vicinity, no radio features can be directly associated 
with the cavities outlined by these arcs.    Filaments and clumps of dust 
connect this cloud to the Circum-Nuclear Disk  (CND) surrounding 
Sgr A* and to the 50 km~s$^{-1}$ cloud discussed below.    As discussed 
in Section 4.2  this cloud may be plunging towards the central
parsecs of the Galaxy on an x2 orbit.  If it is on a radial trajectory towards
Sgr A*, it is the most likely cloud to inject new gas into the central
few parsecs of the Galaxy.  

{\it The G359.94,+0.17  Complex}:
Located 0.2\arcdeg\ directly above (in Galactic latitude) the
Galactic nucleus, this complex is associated with several compact 
8 $\mu$m  bubbles.   It contains IRAS 17417-2851,  a compact cluster 
of at least four prominent 1.1 mm clumps adjacent to an HII region complex that
wraps around  the clumps with PDRs and ionization fronts on the north, 
east and south sides.   The non-thermal 
filament, N5 in Figure \ref{fig11} \citep{yusef-zadeh2004} is
located along the line-of-sight to this clump.  However, because of
its high-latitude location and visibility in the Spitzer IRAC data
as both an IRDC and as a bright region of emission at 8 $\mu$m, 
we assume that it is a foreground complex at an approximate distance 
of about  3.9 kpc (Table~2).

{\it The 50 km~s$^{-1}$ cloud.} ([$l,b$] = 359.98,--0.08):
Located only a few arc-minutes east of Sgr A*, the passage of
this cloud near the Galactic center may have fueled the injection
of gas into the circum-nuclear ring about $1$ to $2 \times 10^5$
years ago.   The cloud contains several compact radio sources
\citep{goss1985,yusef-zadeh2004,Yusef-Zadeh2008} 
and evidence for star formation.
Tendrils of dust continuum emission appear to link this
cloud to both the CND and to the 20 km~s$^{-1}$ cloud
discussed above.   

[$l,b$] = 359.97,--0.46:
An isolated cometary cloud associated with an
8 $\mu$m bubble, presumably an HII region.  This
clump appears to be part of band of foreground 
clouds located at about 3.9 kpc from the 
Sun \citep{Nagayama08}. 

[$l,b$] = 0.05,--0.21:
A 1.1 mm clump located directly below Sgr A* and the 50 km~s$^{-1}$ 
cloud with a prominent Spitzer nebulosity on its low-longitude side. 

{\it The Quintuplet and Arches Clusters}   ([$l,b$] = 0.163,--0.060 and 
 [$l,b$] = 0.121,+0.018):
One of the most striking features of the  20 cm image are series of
concentric shells centered at $l,b$ = [0.15,--0.06] near the location
of the ``Quintuplet" cluster of massive stars \citep{figer1999}.   
The innermost ring may trace the ionization front and PDR associated 
with this cluster; the northeast part of this ring is known as the  
``Sickle" due to its radio morphology  \citep{lang2005}.  
The outer, northwest portion of these concentric rings comprise the
``Arched"  filaments  thought to be illuminated by the massive Arches cluster
\citep{figer02,stolte05}.   

While there is no 1.1 mm clump at the location of the Quintuplet cluster,
an extensive network of clumps is located adjacent to and to the northeast. 
The Quintuplet appears to be in a cavity which is consistent with its 
relatively evolved status.  
The 4 to 5 Myr old Quintuplet cluster  contains one of the most luminous 
stars yet discovered: the Pistol Star.    The VLA 20 cm 
continuum map  shows a prominent ionization front  north and east of
this cluster that is  associated with a ridge of 1.1 mm dust continuum
emission along the Galactic plane.  

The Arches Cluster is another of the  most massive young star clusters 
in the Galaxy \citep{figer02,stolte05}  and is associated with a  
diffuse 1.1 mm source.   The thermal
radio continuum filaments known as the `Arches'  trace ionization
fronts and photon-dominated regions (PDRs) illuminated by this
cluster.   Several clumps  of 1.1 mm  emission are embedded
within the ridges of radio continuum emission.    Fainter filaments of
1.1 mm continuum follow the Arches about 
10\arcsec\ to 30\arcsec\ due east of the radio continuum ridges.  As
shown by \citet{serabyn99}, the radio emission traces the western
edges of a chain of molecular clouds seen in CS.  The 1.1 mm emission
is located between the radio filaments and the CS clouds, indicating a
layer of warm dust on the side of these clouds facing the Arches
cluster.  This layer is most likely associated with a
PDR. \citep{yusef-zadeh03}.    Some of the
1.1 mm emission may be free-free emission from the plasma.  \citet{Wang2006}
proposed that the Arches cluster is colliding with the molecular clouds
in this region.  

{\it [$l,b$] = 0.26,+0.03}:
The  ``lima bean'" shaped molecular cloud extending from [$l,b$] = 0.23,+0.01
to 0.26,+0.03 is remarkable for being a bright 1.1 mm source not associated
with radio continuum sources or any other indicators of on-going star
formation such as bright sources in MSX or Spitzer images.  In
the 2MASS,  Spitzer, and MSX images,  
GCM 0.25+0.01 is seen as  the most  prominent  absorption feature in the
CMZ  (Figure \ref{fig10}).   The cloud qualifies as an
infrared-dark-cloud (IRDC) which does not yet contain evidence of 
star formation.  In the SHARC II 350 $\mu$m
images GCM 0.25+0.01 is resolved into a dozen individual
clumps which could be precursors to star-forming clumps. 
\citet{lis98} find broad molecular line profiles typical of Galactic 
center Giant Molecular Clouds (GMCs) and a line center velocity of 
20 to 40 \kms\ similar to the GMCs known to be interacting with the
non-thermal filaments that cross the Galactic plane at l = 0.18 near
the Pistol and Quintuplet Clusters.  GCM 0.25+0.01  may  be the best
example of a high column density GMC in a pre-star forming
state in the inner molecular zone of the Galaxy.

[$l,b$] = 0.28,--0.48:
This is the brightest clump in a cluster of clumps  associated with 
the  band of foreground  clouds located at about 3.9 kpc from the 
Sun \citep{Nagayama08}.  This clump is located on the rim of 
the large Spitzer 8 $\mu$m bubble centered on the clump at 
[$l,b$] = 0.41,--0.50.   [$l,b$] = 0.28,--0.48 is associated with its
own bright Spitzer nebulosity which may indicate the presence of a
compact HII region whose birth may have been triggered by the
expansion of the Spitzer bubble.

[$l,b$] = 0.32,--0.20:
A clump of 1.1 mm emission associated with a bright but compact
cluster of extended Spitzer nebulosity.  The clump lies at the
low-longitude end of fan-shaped group of at least three filaments 
that converge on this object and which can be traced for at least 
0.1\arcdeg\  towards higher longitudes in both 1.1 mm emission and
8 $\mu$m extinction. 

[$l,b$] = 0.41,--0.50:
The easternmost 1.1 mm clump associated with the band of 
foreground  clouds located at about 3.9 kpc from the 
Sun \citep{Nagayama08}.   This particular clump  is located
at the high-latitude tip of a large pillar of dust and is centered
on the 0.15\arcdeg\  (10 pc) radius ring of 8 $\mu$m Spitzer emission. 

[$l,b$] = 0.41,+0.05:
The low-longitude end of the chain of clouds that extend towards and
includes Sgr B2.  

[$l,b$] = 0.48,--0.00:
A clump in this chain located directly above the Sgr B1 complex of HII
regions..

{\it Sgr B1}  ([$l,b$] = 0.506,--0.055  - not marked on the images):
Sgr B1  is  a giant HII region to the southwest of  Sgr B2  and associated with 
extensive recent star formation traced by radio continuum
emission, masers, and IR sources. In the BGPS map, it appears as a
large cavity, marked only by a few
faint filaments and  bright clumps located to the northwest.   Thus,
this region must be more evolved than Sgr B2 because gas and
dust has been expelled.

[$l,b$] = 0.53,+0.18:
The brightest member of a cluster of 1.1 mm clumps located above the
Galactic plane and Sgr B1.  At least three extended 8 $\mu$m Spitzer
emission nebulae are associated with this complex. 

{\it Sgr B2}  ([$l,b$] = 0.687,--0.030): 
The Sgr B2 complex consists of a 5\arcmin\ long north-to-south chain
of  bright cloud clumps, surrounded by a `halo' of  fainter ones,
filaments, and shells extending over nearly 20\arcmin\ in R.A. and
10\arcmin\  in Declination.  Sgr B2 clump is  the hottest
and currently most active star forming region in a roughly 25 by 50
parsec complex of emission.  Sgr B2  is associated with molecular
gas between V$_{LSR}$ = 30 to 80 km~s$^{-1}$ as seen in 
tracers such as CS, HCN, and HCO$^+$ \citep{bally87, bally88, jackson96}.
Figure \ref{fig15}  shows the Sgr B2 region at  0.35 mm; 
Figure \ref{fig17} shows the BGPS 1.1 mm contours superimposed on the 
Spitzer 8 $\mu$m image;
Figure \ref{fig18} shows the BGPS 1.1 mm contours superimposed on the 
Spitzer 24 $\mu$m image.

The brightest emission at 1.1 mm originates from the northern component 
of the Sgr B2 complex (Sgr B2N) which has  $S_{1.1} \sim 100$~Jy / beam
and $S_{0.35} \sim 1,303$~Jy / beam.    Sgr B2M is a bit fainter at 1.1 mm
with $S_{1.1} \sim 80$~Jy / beam.   However, it is the dominant source
at 350 $\mu$m with  $S_{0.35} \sim 1,375$~Jy / beam.  (Note that the 1.1 mm
beam has a diameter of  33\arcsec\ while the 0.35 mm beam has
a diameter of about  9\arcsec ).

Figures  \ref{fig17} and \ref{fig18} show that the large cavities located at low longitudes
and latitudes with respect to Sgr B2 are filled with bright IR emission.
Furthermore, there is a close correlation
between the mm and sub-mm emission and dark clouds seen in silhouette
against background stars and diffuse infrared light.

Sgr B2  exhibits a rich millimeter-wave spectrum consisting of thousands 
of individual spectral lines tracing transitions of over 100 molecular species.
Combined with the 10 to 50 km~s$^{-1}$ Doppler widths of individual lines,
this results in a nearly continuous forest of spectral features.  In Sgr B2, 
the combined flux between  218 to 236 GHz from mostly heavy organic 
molecules provides a  frequency variable background spectrum with 
peak temperatures ranging from 0.1 to over 10 K \citep{nummelin98} 
that results  in a pass-band  averaged flux in this spectral range of 
about 1 K.    \cite{nummelin98}   argue that  about 22\% of the total flux 
from  Sgr B2N, and 14\% of  the flux from the clump known as M is produced 
by spectral lines.    Thus,  in the densest 
clumps,   up to tens of percent  of the  flux in the  
1.1 mm BGPS pass-band is produced by spectral lines rather than dust.   
Away from these few  bright regions, the contribution of 
spectral lines is likely to be smaller than  10\% and less than the 
flux calibration error.

The average 1.1 mm flux  of  the Sgr B2 complex in a 150\arcsec\  radius 
circle centered on Sgr B2N is 10.7 Jy.    Assuming a dust temperature 
of 20 K, the mass enclosed in this region is about $8.3 \times 10^5$  M$_{\odot}$
This  region is surrounded by a 0.2\arcdeg\ by 0.25\arcdeg\
diameter envelope of filamentary emission with an average 1.1 mm flux of 
about 0.9 Jy that implies an additional mass of at least 
$1.3 \times 10^6$  M$_{\odot}$. 
Thus, the total mass of the Sgr B2 complex is at least
$1.0 \times 10^6$ M$_{\odot}$.  This estimate is  a severe lower
bound because the spatial frequency transfer function of the BGPS
observations and pipeline attenuates emission that is smooth on
scales larger than about 3\arcmin\ and completely filters out emission
uniform on scales larger than 7.5\arcmin .

The extended Sgr B2 complex
is the densest and most massive giant molecular cloud complex in the
Galaxy \citet{lis91}.   About a dozen cavities are apparent in the 1.1 mm image.
The large 0.15\arcdeg\  by 0.08\arcdeg\ cavity at located at low Galactic longitudes 
(centered roughly near $l$ = 0.50\arcdeg , $b$ = --0.06\arcdeg\ 
(Figures \ref{fig10} and \ref{fig18}) contains bright extended 8 and 24 $\mu$m  
emission    in the Spitzer IRAC images that consists of diffuse emission, filaments,
and compact sources.  This region apparently contains mature HII regions
and massive stars formed in a previous episode of star formation.  A similar but
smaller cavity, centered at $l,b$ = 0.680\arcdeg , --0.600\arcdeg\ is also
filled with  8 $\mu$m  emission.   These cavities  contain an earlier
generations of massive stars and mature  HII regions.

[$l,b$] = 0.84,+0.18:
An isolated clump associated with extended Spitzer emission.

{\it Sgr D}  ([$l,b$] = 1.12,--0.11):
The brightest 1.1 mm clump within a 0.5\arcdeg\ region.  It is associated with bright
Spitzer emission, and two bubbles of 20 cm continuum located north and south of the
clump (Figures \ref{fig9} through \ref{fig11}). 

[$l,b$] = 1.33,+0.16:
A typical 1.1 mm clump in this portion of the survey.  However, this one is associated
with a blister HII region that extends to the south of the mm clump.  

[$l,b$] = 1.60,+0.02:
One of the brightest  clumps in the $l$ = 1.3\arcdeg\ complex of molecular clouds.  
The surrounding  complex consists of several prominent CO-emitting  filaments that extend
vertically out of the Galactic plane for at least 0.5\arcdeg\ ($\sim$ 75 pc).   This 
one degree diameter complex
has the largest scale-height in the Central Molecular Zone.  The radial 
velocities of CO emitting clouds extend over at least 120 km~s$^{-1}$. 
This region may be similar in nature to Bania's Clump 2.  

{\it IRAS 17481-2738} ([$l,b$] = 1.74,--1.41):
Located at the low-latitude end of the $l$ = 1.5\arcdeg\ complex of molecular clouds,
this is the second brightest clump in the region.

{\it IRAS 17469-2649} ([$l,b$] = 2.30,+0.26):
A typical example of one of the faint, diffuse clumps in this portion of the Galactic
plane.  It was singled out because it is associated with bright and extended
Spitzer emission as well as a 20 cm radio continuum complex. 

[$l,b$] = 2.62,+0.13:
A prominent 1.1 mm clump associated with IRDCs, a complex of Spitzer nebulosities,
and a bright, compact cometary 20 cm continuum feature facing east.

[$l,b$] = 2.89,+0.03:  The low-latitude end of a complex of 1.1 mm clumps which 
contains a compact Spitzer source and filament of bright emission.   A small and
bright 20 cm radio source may trace a compact HII region.  A low-surface
brightness and extended bubble of 20 cm continuum emission extends about 
0.1\arcdeg\ towards low-longitudes.  This is one of the few 1.1 mm clumps in the
Bania's Clump 2 region that exhibits signs of recent massive star formation.  
The mm clump is seen as a dim IRDC.  It may be located in the foreground portion
of the Galactic plane.

{\it IRAS 17496-2624} ([$l,b$] = 2.96,--0.06):
This 1.1 mm clump is associated with the brightest 20 cm radio continuum source
in the Bania's Clump 2 region.   Spitzer shows a compact limb-brightened bubble
opening towards high latitudes.  This feature may trace a compact HII region 
less than 1\arcmin\  in diameter. 

[$l,b$] = 3.09,+0.16:
One of the brightest 1.1 mm clumps in Bania's Clump 2.  It is not associated with
either a 20 cm or a Spitzer nebula.  Located at the high-longitude end of a complex
of clumps that may be faintly seen in silhouette at 8 $\mu$m. 

[$l,b$] = 3.14,+0.41:
A relatively bright 1.1 mm clump in Bania's Clump 2.  It is not associated with
either a 20 cm or a Spitzer nebula.

[$l,b$] = 3.31,--0.40:
A bright 1.1 mm clump associated with a compact IRDC and with no bright Spitzer 
sources. 

[$l,b$] = 3.34,+0.42:
The largest and brightest  1.1 mm clump in Bania's Clump 2.  It does not contain any
Spitzer sources.  However, it is seen faintly in silhouette at 8 $\mu$m and
may be associated with very dim, extended 20 cm continuum (0.04 Jy/beam).

[$l,b$] = 3.35,--0.08:
A very bright 1.1 mm and Spitzer source which is associated with a bright cometary 
20 cm continuum source located at the high-longitude end of a large HII region 
complex.

[$l,b$] = 3.44,--0.35:
The brightest 1.1 clump in this part of the survey which is associated with
a compact and prominent IRDC.  A dim 8 $\mu$m stellar object is located
near the center of the cloud.

{\it IRAS 17470-2533} ([$l,b$] = 3.41,+0.88):
The highest latitude object near Bania's Clump 2, this 1.1 mm clump
is  surrounded by two tails, one extending towards the
Galactic plane, and another towards higher longitudes.  The clump contains a bright
Spitzer and IRAS source and extended emission .  It may be a foreground
complex. 

{\it IRAS 17504-2519} ([$l,b$] = 4.00,+0.34):
An isolated 1.1 mm clump seen as an IRDC that contains a bright and compact
8 $\mu$m Spitzer source, likely to be a compact HII region.

[$l,b$] = 3.39,+0.08:
A collection of 1.1 mm clumps  associated with a  complex of Spitzer  emission 
nebulae.

[$l,b$] = 4.43,+0.13:
The brightest 1.1 mm clump associated with the above complex.  The Spitzer
8 $\mu$m emission shows that this clump is a IRDC superimposed on the
extended bright emission.  The IRDC consists of a compact arc of absorption
no more than 8\arcsec\ wide, indicating that the feature is much thinner in
one dimension than the Bolocam beam.

\clearpage

\section{Appendix II:  Bolocat Clump Masses listed in Table 3}

The electronic version of this paper contains a table of masses, column densities,
and densities computed using the methods outlined in Section 2.1.  These
masses are based on the Bolocam V1.0 fluxes scaled-up by the empirically
determined factor of 1.5.   The various  entries in Table 3 are described here.

\noindent
Column 1:  The clump number in Bolocat V1.0 released in June via the IPAC website
at 

    http://irsa.ipac.caltech.edu/data/BOLOCAM$\_$GPS/  .
 
\noindent
Column 2:  Galactic longitude in degrees.

\noindent
Column 3:  Galactic latitude in degrees.

\noindent
Column 4:  J(2000) Right Ascension  in degrees.

\noindent
Column 5:  J(2000) Declination in degrees.

\noindent 
Column 6:  Mass estimate for each clump measured in a 40\arcsec\ aperture
  assuming T = 20 K as discussed in Section 2.1.

\noindent
Column 7: H$_2$ column density  estimate for each clump measured in a 
40\arcsec\ aperture   assuming T = 20 K as discussed in Section 2.1.

\noindent  
Column 8: H$_2$ volume  number density  estimate for each clump measured in a 
40\arcsec\ aperture   assuming T = 20 K as discussed in Section 2.1.

\noindent
Column 9: The mass derived from the Bolocat flux in the effective beam-deconvolved 
  area of each clump tabulated  in column 19 of the Bolocat catalog available from
  the IPAC website.
  
 \noindent  
Column 10: The Bolocat effective clump radius tabulated  in column 12 of the 
  Bolocat catalog available from the IPAC website.

\noindent  
Column 11:  H$_2$ column density  estimate for each clump measured in the
  beam-deconvolved effective area and based on the mass in column 10
  assuming T = 20 K as discussed in Section 2.1.

\noindent  
Column 12: H$_2$ volume  number density  estimate for each clump measured 
in the   beam-deconvolved effective area and based on the mass in column 10
assuming T = 20 K as discussed in Section 2.1.

\noindent
Column 13: Mass estimate for each clump measured in a 120\arcsec\ aperture
  assuming T = 20 K as discussed in Section 2.1.

\noindent
Column 14: Mass estimate for each clump measured in a 40\arcsec\ aperture
  assuming a temperature gradient that declines as a power-law with distance
  from Sgr A*.  The power-law has a slope $\gamma _g$ = 0.2 and is normalized
  to have T = 50 K at a distance of 10 pc from Sgr A*.  This power-law implies
  T = 27.5 K at $d$ = 200 pc from Sgr A*, 15.5 K at d = $3$ kpc, and 
  12.5 K at $d$ = 8.5 kpc.   See discussion in  Section 3.1.

\clearpage

\bibliography{ms}

\begin{deluxetable}{lll}
\tablewidth{0pt}
\tablecaption{  Observing Epochs for the Galactic center data }
\tablehead{\colhead{Begin  } & \colhead{End }  & Instrument}
\startdata
\label{tab:Runs}
2 Jul 2005  &  9 Jul 2005  & Bolocam \\
2 Jun  2006   & 29 Jun  2006  & Bolocam + SHARC-II \\
7 Jul  2007   & 14 Jul  2007 & Bolocam  + SHARC-II  \\
13 Apr  2008   & 14 Apr  2008 & SHARC-II \\
\enddata
\end{deluxetable}

\clearpage

\begin{deluxetable}{ccccccccccl}
\tabletypesize{\scriptsize}
\tablecaption{ 1.1 mm Peak Fluxes and Masses of Selected Regions 
      \label{table2}}
    
\tablehead{
    \colhead{$l$} & 
    \colhead{$b$} &
    \colhead{ $S_{40}^1$}   & 
    \colhead{ $S_{300}^1$} &
    \colhead{M$_{40}^2$}   & 
    \colhead{M$_{300}^2$} & 
    \colhead{N$_{40}^3$}  &
    \colhead{N$_{300}^3$}  &
    \colhead{n$_{40}^4$ }  &
    \colhead{n$_{300}^4$}  &
    \colhead{Comments}      \\

    \colhead{deg} & 
    \colhead{deg} & 
    \colhead{ Jy} & 
    \colhead{ Jy} & 
    \colhead{$10^2 M_{\odot}$}  &    
    \colhead{$10^2 M_{\odot}$}   &
    \colhead{$10^{21} cm^{-2}$}  &
    \colhead{$10^{21} cm^{-2}$}  &
    \colhead{$10^2      cm^{-3}$}  &
    \colhead{$cm^{-3}$}  
      }
      
\startdata
\label{tab:Masses}
359.13      &--0.10     & 0.45  & 0.14 &       4     	&       68    	&     9 	&   3		&  27 	&   111 	& Great Annihilator \\
359.44 	&--0.10	&  6.9   & 1.3   &     66     	&      561   	& 138 	&  23 	& 413 	& 1017	& Sgr C  \\
359.46 	&--0.03	&  1.4   & 1.3   &     13      	&      561   &   27 	&  23 	&  81 	& 1017 	& In Exp. Mol. Ring  \\
359.62	&--0.24	&  8.1   & 0.60 &     77      	&     300    & 162 	&   12 	& 485 	&    495	&  E. facing comet	\\
359.71	&--0.37	&  2.1  & 0.36  &       4       	&       38    &    42 	&     8 	& 273 	&    647 	&  	D=3.9 kpc \\
359.86	&--0.08	&  9.6  & 3.36  &     90      	&   1680   	& 192	&  66 	& 475	&  2769	& 20 km~s$^{-1}$ cloud \\
359.91      &--0.31     &  1.5  & 0.32   &       3     	&       33    	&   30 	&   6  	& 195 	&    566 	&      	D=3.9 kpc \\
359.94      &--0.04     & 11.7 & 2.90   &      --       	&   1425    & 234	&  57	         &  701 	& 2348      & Sgr A*, 50 km~s$^{-1}$ cloud \\
359.96	& +0.17	& 3.3  & 0.77    &      7      	&        81    &   65 	&  15 	& 431 	& 1374	& IRAS 17417-2851, D=3.9 kpc \\
359.97	&--0.46	& 3.3  & 0.41    &      7       	&       42	&    66 	&    8 	& 431  	&   728	 & D=3.9 kpc \\
0.05   	&--0.21	& 1.8  & 0.41    &    17       	&      195	&    36 	&    8 	& 108 	&   335	 & \\
0.12	         & +0.02	& 1.4  & 0.78    &     13       	&      390	&    27 	&  15 	&    81 	&   642	 & Near Arches cluster	\\
0.16	          &--0.06	& 0.3  & 0.83    &      3      	&      413	&      6 	&  17 	&   18 	&   680 	& Near Quintuplet cluster	\\
0.26   	&--0.03	& 5.7  & 1.50    &     54       	&      750	&  114	&  30 	&   341	&  1235	& Lima bean	\\
0.28  	&--0.48	& 2.3  & 0.80    &       5       	&        84	&    40 	&  17 	&   396	&  1428	& 	for D=3.9 kpc  \\
0.32	          &--0.20	& 3.9  & 0.59    &    38      	&      293	&    78	&   12 	&   233	&    482     & IRAS 17439-2845 \\
0.41  	&--0.05	& 4.2  & 1.2      &    39      	&       600	&    84	&   24 	&   251	&    989	& \\
0.41   	&--0.50	& 1.8  & 0.33    &      4       	&         35	&    36	&     6 	&   236	&    593	& 					for D=3.9 kpc \\
0.48           &--0.00	& 7.7   & 2.0     &     72    	&       975	& 153 	&   39 	&   458	&   1607    & \\
0.53  	& +0.18	& 2.4   & 0.50   &     23      	&       240  &    48 	&   11 	&   144	&    408	& \\
0.68           &--0.03	&140   & 16.1  & 1320    	&    8010  &  2796      & 323	&  8351     & 13224    & Sgr B2	\\
0.84  	& +0.18	& 2.0 & 0.32  &      18    	&       158	&  39		&   6 		&  117 	&  260 	& 	\\
1.12  	& -0.11	& 5.1 & 0.92  &      48    	&       458	&  68 	&  12 	& 204 	&  503 	& Sgr D	\\

1.33  	& +0.16	& 0.6 & 0.36  &        6     	&       180	&   12	&   8		&    36  	&  297	& IRAS 17450-2742 \\
1.60  	& +0.02     & 2.1 & 0.48  &      20    	&       240  &   42	&   9 		&  126 	&  396 	& \\	
1.74  	&--0.41	& 2.3 & 0.47  &      21    	&       233	&   45	&   9 		&   135 	&  383	& IRAS 17481=2738 \\	
2.30  	& +0.26	& 0.5 & 0.05  &       4     	&        23	&     9 	&   2 		&    27	&    38	& IRAS 17469-2649 \\	
2.62  	& +0.13	& 1.5 & 0.12  &     14    	&        60	&   30	&   3 		&    90	&    99	&  \\	
2.89           & +0.03	& 0.9 & 0.30  &       9    	&       150	&   18	&   6 		&    54	&  248 	& \\		
2.96           &--0.06	& 0.6 & 0.14  &       6     	&         68   &   12	&   3 		&    36	&   111	& IRAS 17496-2624 \\		
3.09           & +0.16	& 0.9 & 0.15  &       9     	&         75	&    18	&   3 		&    54	&   123	& \\		
3.14           & +0.41	& 0.9 & 0.15  &       9     	&         75	&    18	&   3 		&    54	&   123	& \\		
3.31           &--0.40	& 1.5 & 0.17  &     14      	&         83   &   30	&   3 		&    90	&   137	& \\		
3.34           & +0.42	& 1.2 & 0.30  &      11      	&       150	&    24	&   6 		&    72	&   248	& \\		
3.35           &--0.08	& 2.3 & 0.12  &       14    	&          60 	&    45	&   3 		&  135	&     99      & \\			
3.41           & +0.88	& 0.8 & 0.06  &      7      	&          30	&    15	&   2 		&    45	&     50	& IRAS 17470-2533 \\	
3.44           &--0.35	& 4.8 & 0.26  &     45   	&        128	&    96	&   5 		&  288	&    210	& \\		
4.00           & +0.34	& 0.9 & 0.06  &       9    	&          30	&    18	&   3 		&    54	&      50	& IRAS 17504-2519 \\		
4.39           & +0.08	& 0.8 & 0.15  &       7    	&          75	&    15	&   3 		&    45	&    123     & \\		
4.43           & +0.13	& 2.1 & 0.23  &      20  	&         113	&    42	&   5 		&  126	&  186       & \\		
				
\enddata
Notes:  
{\bf{[1]   Flux densities:}}
$S_{40} $ is the flux (in Jy per beam)  measured in a 33\arcsec\  
diameter effective CSO beam (equivalent to a 40\arcsec\ diameter 
top-hat beam with a radius of 19.82\arcsec ).
$S_{300}$ is the average flux measured (in Jy per beam) in a solid 
angle corresponding to a  300\arcsec\  diameter "top-hat" beam.
{\bf{[2]  Masses:}}   
$M_{40}$ and $M_{300}$ are the masses estimated for an assumed 
dust temperature of 20 K and a distance of D = 8.5 kpc 
(unless otherwise noted) in 40\arcsec\ and 280\arcsec\ apertures.
The tabulated values are in Solar masses ($M_{\odot}$) divided by $10^2$.
{\bf{[3]:    Column densities:}}
$N_{40}$ and $N_{300}$ are the equivalent average column 
densities estimated in the 40\arcsec\ and 300\arcsec\ apertures, 
assuming that the emission is uniformly spread over the
aperture.  These are lower bounds on the actual column densities
as discussed in the text.  The units are in $cm^{-2}$ divided by
$1.0 \times 10^{21}$ which is about $A_V \approx 1 $ magnitude.
{\bf{[4]  H$_2$ volume densities:} }  
$n_{40}$ and $n_{300}$ are the equivalent average volume densities
of $H_2$ estimated in the 40\arcsec\ and 300\arcsec\ apertures
under the assumption that the clumps are spherical with a radius
equal to the projected radius of the measurement aperture in which
the emitting material is spread uniformly.  These are a lower 
bounds on the actual densities as discussed on the text. 
The units for $n_{40}$ are in $cm^{-3}$   divided by $10^2$.
The units for $n_{300}$ are in $cm^{-3}$.
\end{deluxetable}

\clearpage

\clearpage

\clearpage

\clearpage

\clearpage

\clearpage

\clearpage

\clearpage

\clearpage

\clearpage

\clearpage

\clearpage

\clearpage

\clearpage

\clearpage

\clearpage

\begin{figure}
\epsscale{1.0} 
\center{\includegraphics[width=1.0\textwidth,angle=0]{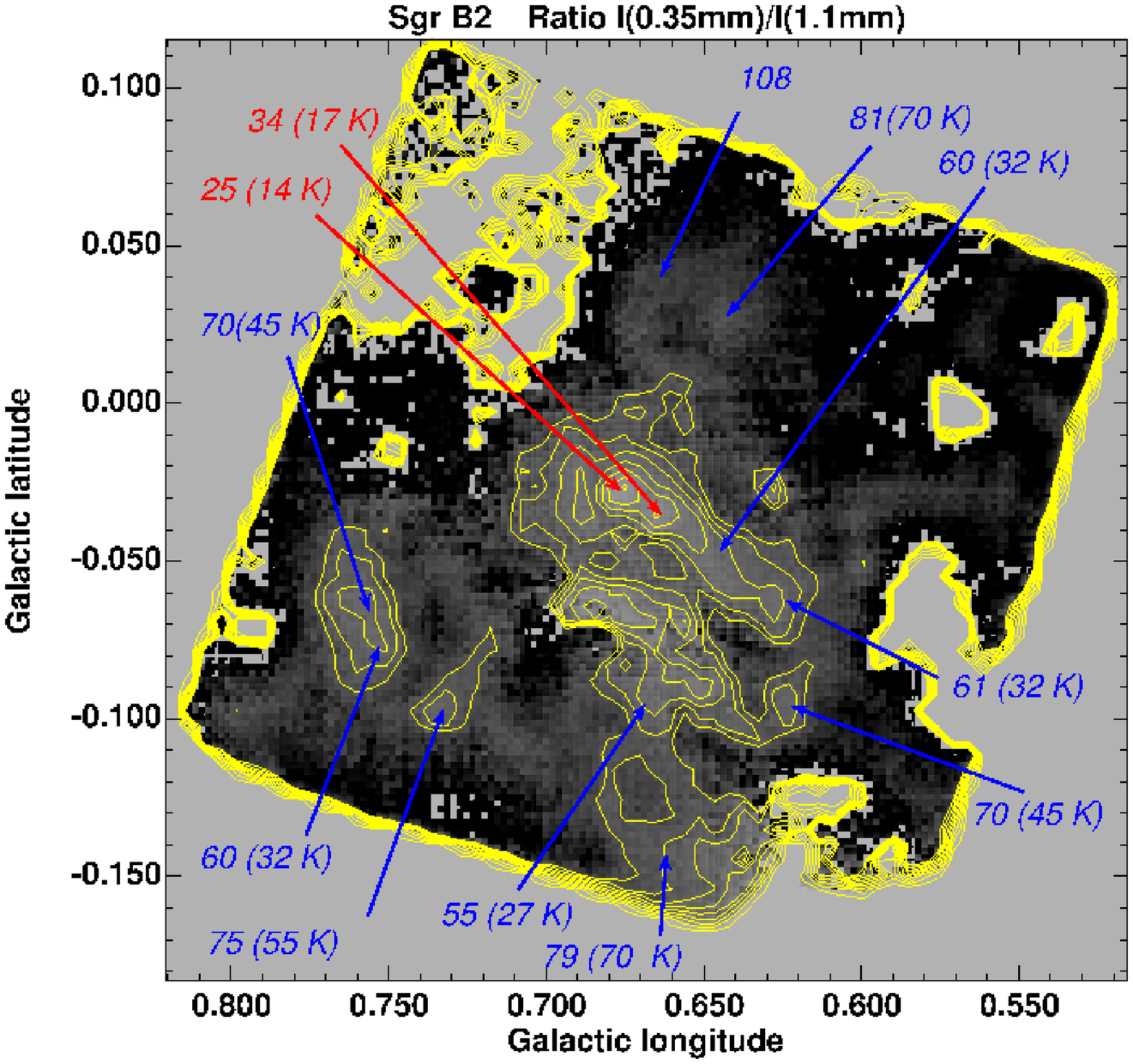}}
\caption{A map showing the ratio of the surface brightness  of 350 $\mu$m emission 
divided by 1.1 mm emission in the vicinity of Sgr B2.    The display range goes from 
0 (white) to 100 (black).     Regions with ratios larger than 100, which corresponds to the
Rayleigh-Jeans limit for $\beta$ = 2.0 dust have been masked and are shown as white.    
The ratios at the edge of the SHARC-II field are  spurious.    As discussed in the text,
the 350 $\mu$m map has been convolved with a gaussian kernel to have the
same effective beam size as the 1.1 mm map (33\arcsec\  effective beam).     
The italic numbers indicate the flux ratio at the locations indicated by the associated 
arrows. The numbers in parentheses give the derived
dust temperatures for an emissivity power-law index of $\beta$ = 2. }
\label{fig16}
\end{figure}

\clearpage

\begin{figure}
\epsscale{1.0} 
\center{\includegraphics[width=1.2\textwidth,angle=0]{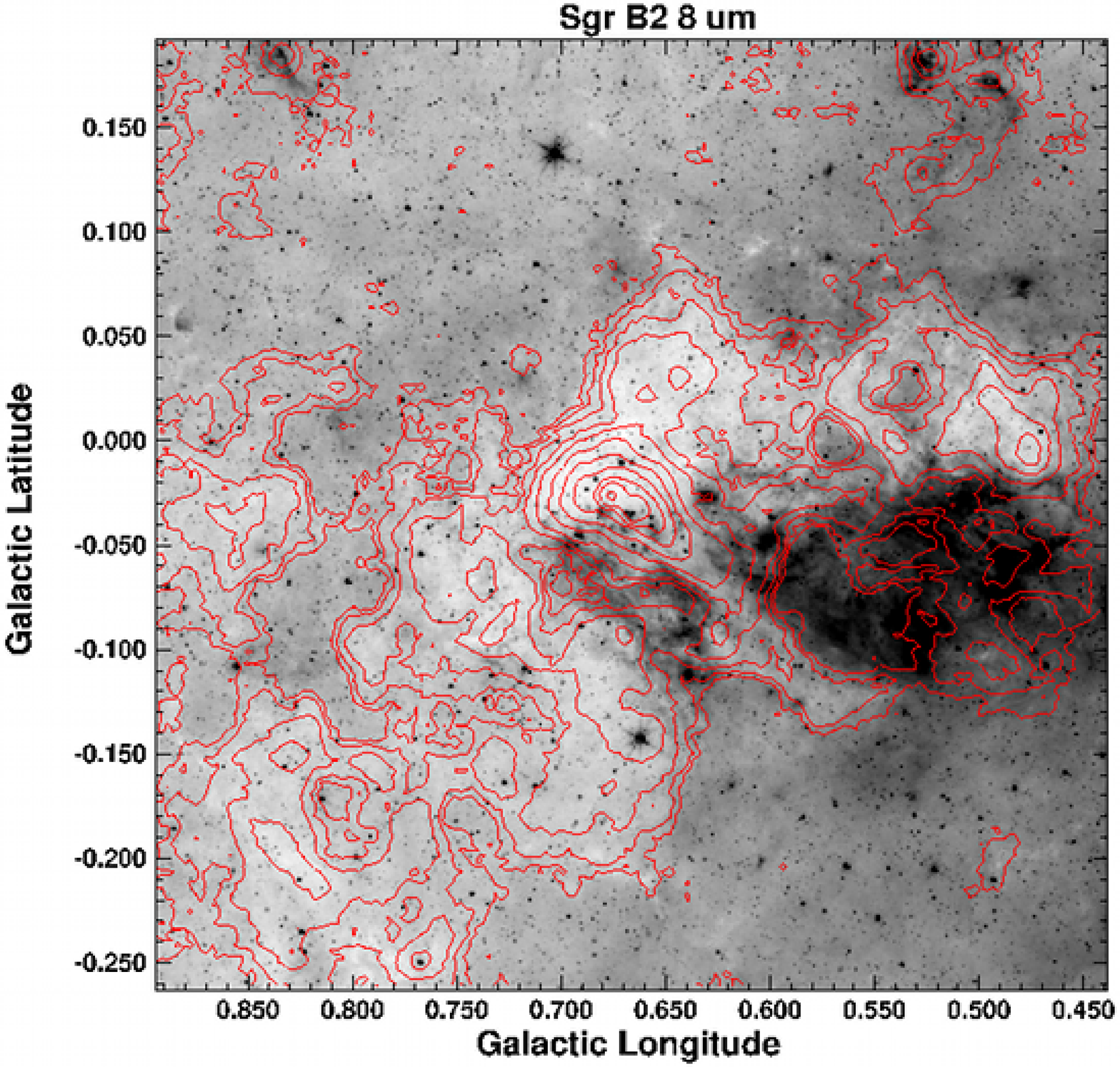}}
\caption{The Bolocam 1.1 mm contours superimposed on a Spitzer
8 $\mu$m image displayed in logarithmic intenstiy.  Contour levels are at
0.30, 0.54, 1.0, 1.8, 3.3, 6.0, 11, 20, 36, 66, and 120.0 Jy/beam.} 
\label{fig17}
\end{figure}

\clearpage

\begin{figure}
\epsscale{1.0} 
\center{\includegraphics[width=1.2\textwidth,angle=0]{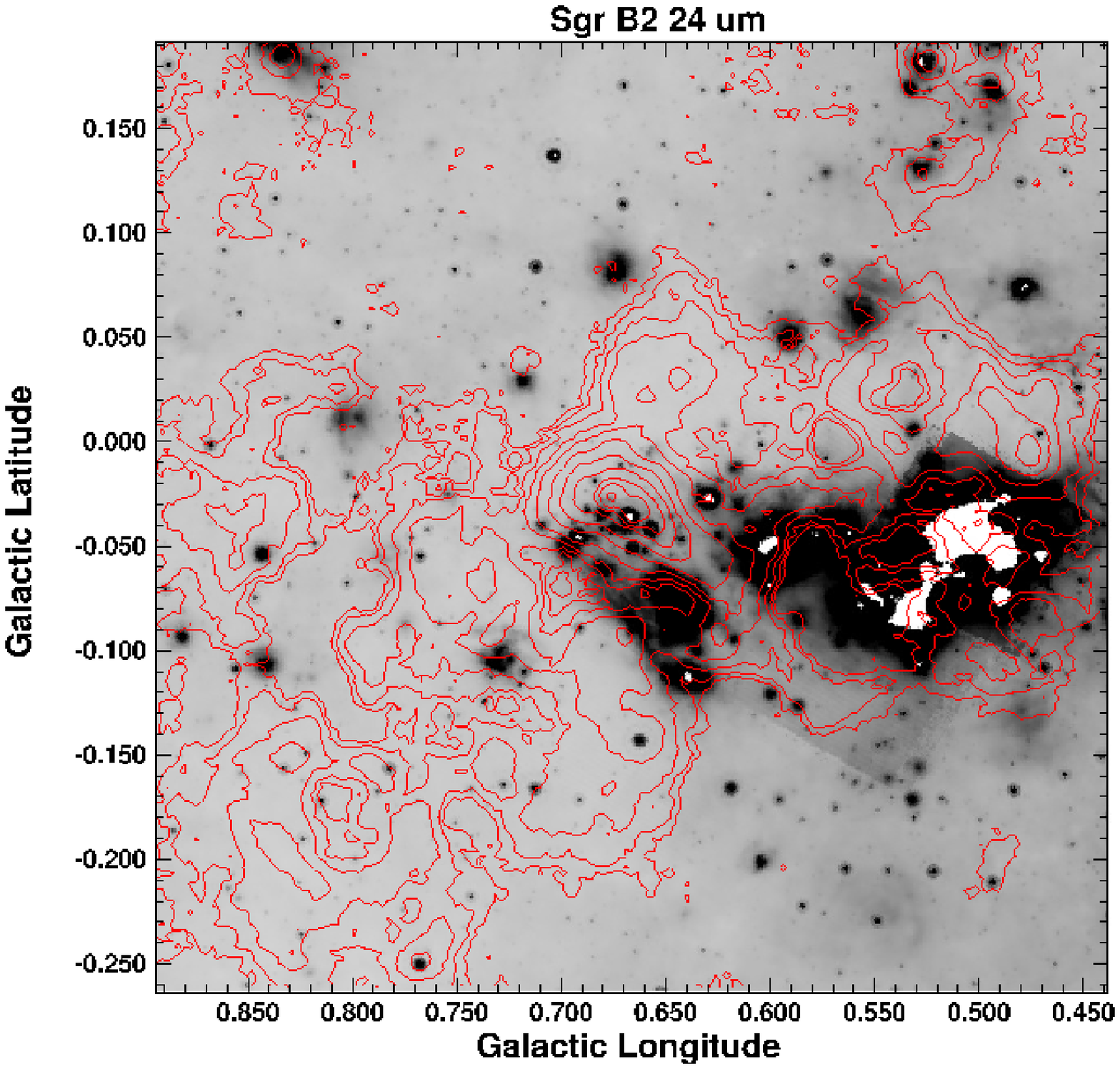}}
\caption{The Bolocam 1.1 mm contours superimposed on a Spitzer
24 $\mu$m image displayed in logarithmic intenstiy.  Contour levels are at
0.30, 0.54, 1.0, 1.8, 3.3, 6.0, 11, 20, 36, 66, and 120.0 Jy/beam.} 
\label{fig18}
\end{figure}

\clearpage

\begin{figure}
\epsscale{1.0} 
\center{\includegraphics[width=1.0\textwidth,angle=0]{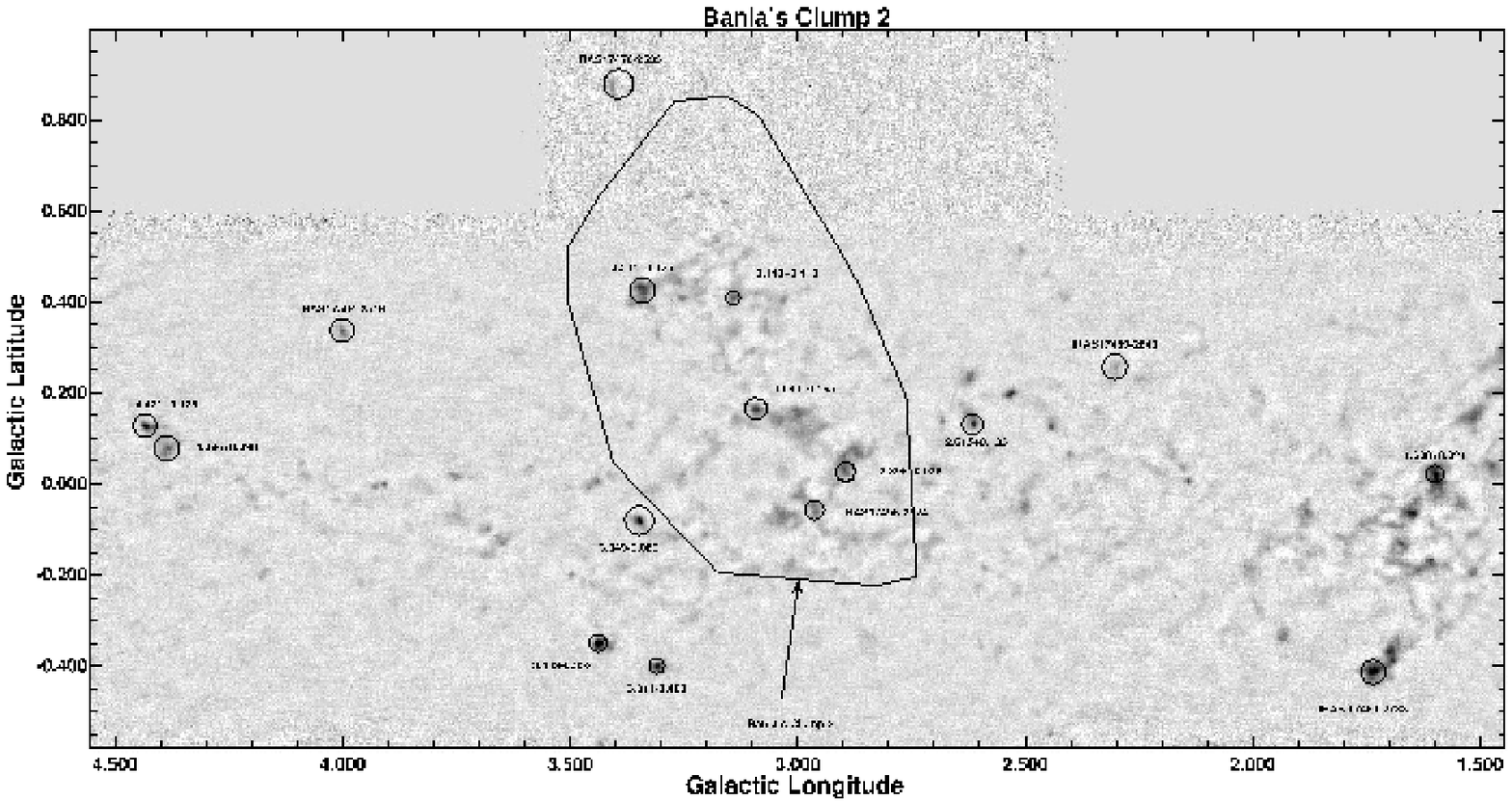}}
\caption{A 1.1 mm image centered at $\sc l$ = 3\arcdeg , the region that
contains Bania's Clump 2 with carious clumps marked.}
\label{fig19}
\end{figure}

\clearpage

\begin{figure}
\epsscale{1.0} 
\center{\includegraphics[width=1.0\textwidth,angle=0]{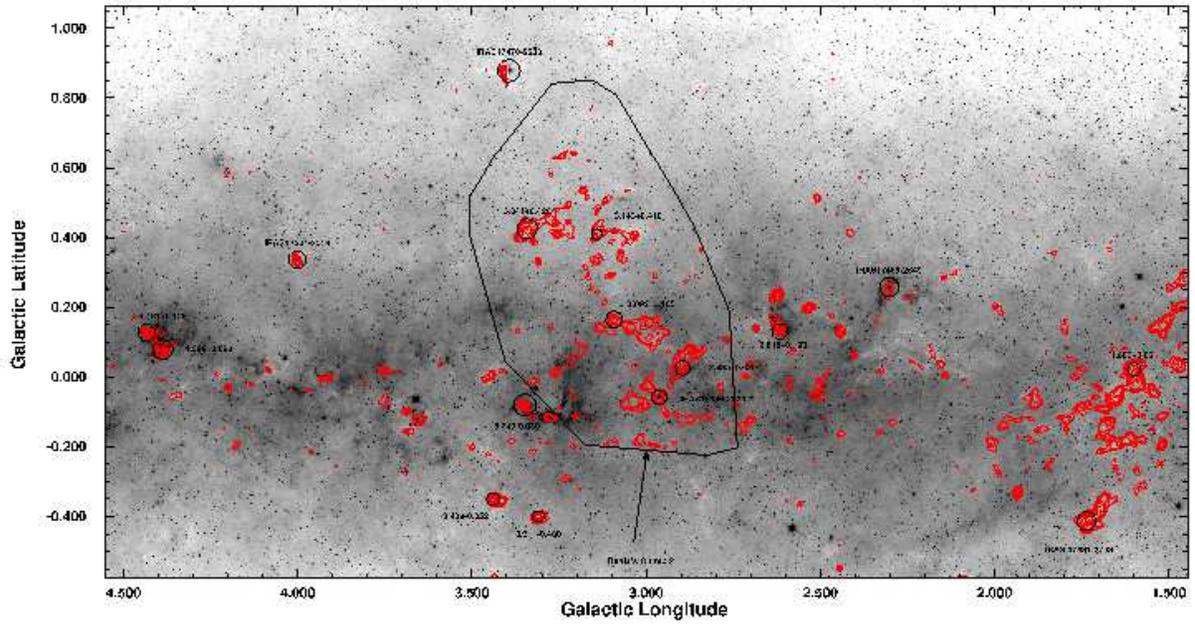}}
\caption{Contours of 1.1 mm continuum emission centered at $\sc l$ = 3\arcdeg ,
the region that contains Bania's Clump 2 (red), superimposed on the 
GLIMPSE2 8  $\mu$m image.   Contour levels are at 
0.15, 0.24, 0.38, 0.6, 1.0, 1.5, 2.4, 3.8, 6.0, 10, and 15 Jy/beam.}
\label{fig20}
\end{figure}

\clearpage

\begin{figure}
\epsscale{1.0} 
\center{\includegraphics[width=0.6\textwidth,angle=0]{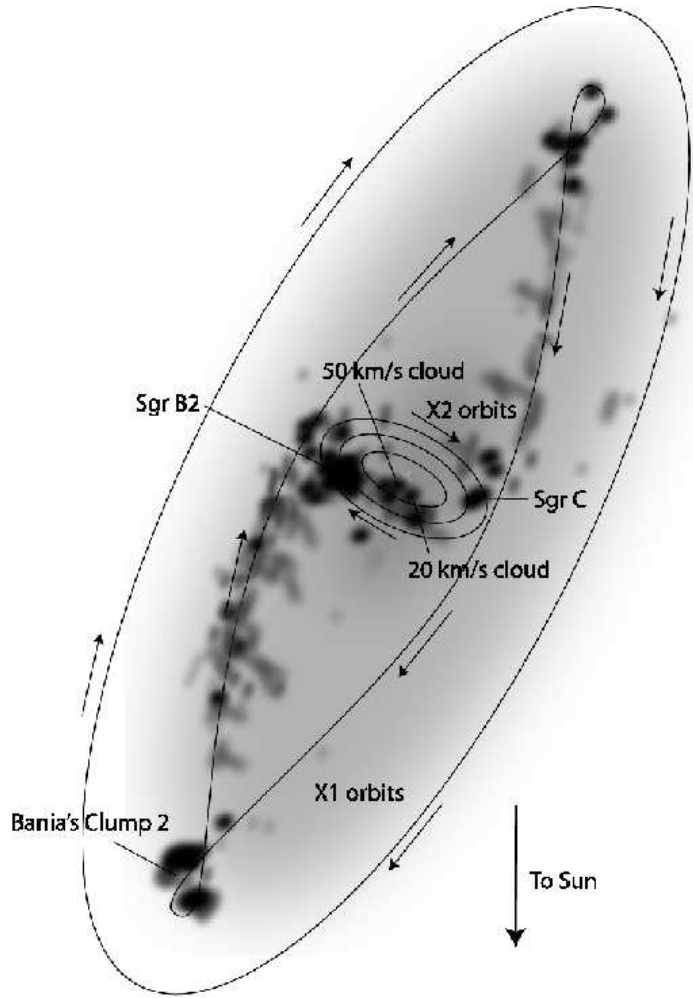}}
\caption{A cartoon showing a face-on view of the central 500 pc region of the Milky Way as
viewed from the northern Galactic pole.  The Sun is located below the figure and
positive longitudes to the left.   The diffuse grey-scale shows the current orientation of the
stellar bar, thought to have its major axis tilted between 20 to 45\arcdeg\ with respect to our
line-of sight.  The large oval show a non intersecting x1 orbit;  Lower angular momentum x1 orbits
become self-intersecting.  The "rhombus" of molecular emission in $l$-$V$ diagrams probably 
occupies the receding and approaching portions of the innermost x1 orbits.  The 3 smaller 
inscribed ellipses show x2 orbits.  Bania's Clump 2 is thought to be located on the receding 
portion the last stable x1-orbit where it is self-intersecting.
Approaching gas presumably enters as an atomic phase;  shock-copresison and subsequent
cooling results in its conversion to the molecular phase.   The receding gas above Bania's clump 2
traces the leading edge of the bar and populates extended network of positive velocity clouds in
$l$-$V$ diagrams such as Figure 4.   The CMZ emission is associated with gas mostly located on the
near-side of the Galactic center on the x2 orbits.  The possible locations of the Sgr C, 20 km~$^{-1}$,
50 km~$^{-1}$, and Sgr B2 complexes are indicated.   } 
\label{fig21}
\end{figure}

\clearpage

\begin{figure}
\epsscale{1.0} 
\center{\includegraphics[width=1.0\textwidth,angle=0]{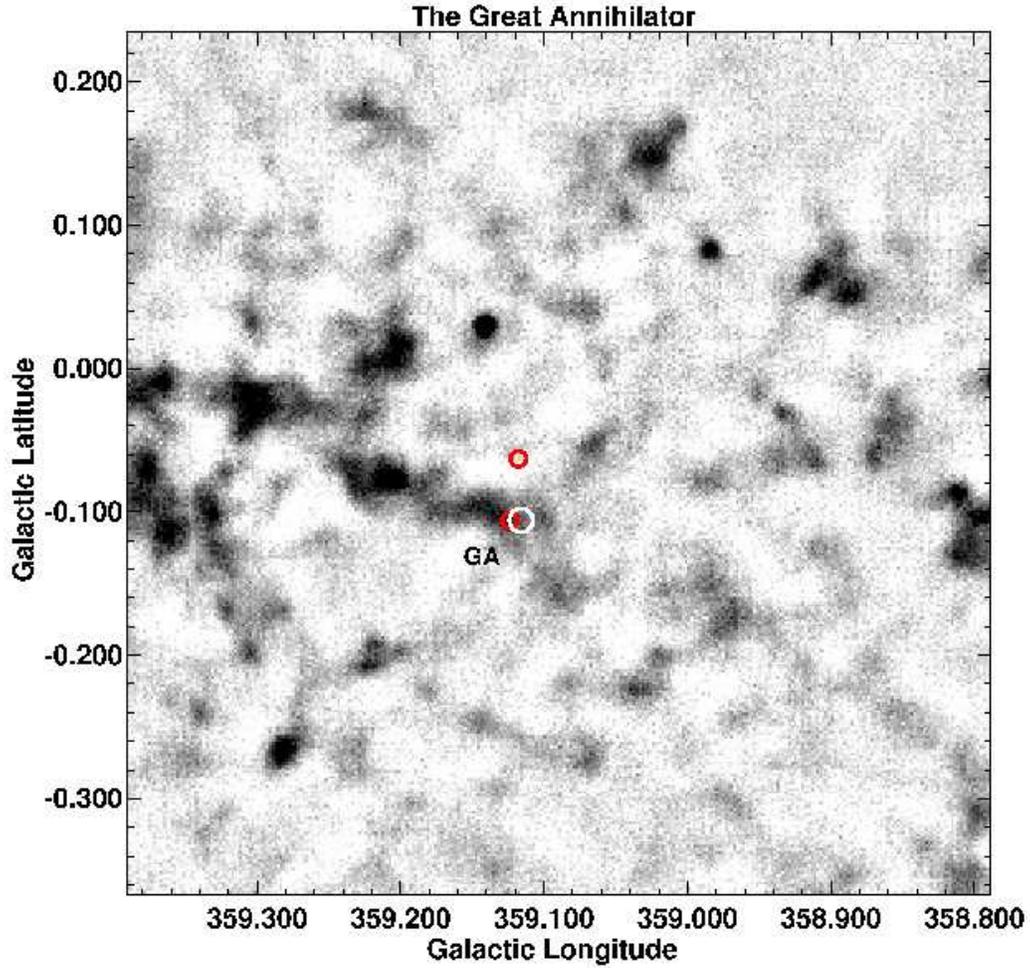}}
\caption{A 1.1 mm image showing the field of view surrounding the
"Great Annihilator"  (GA) thought to be a major source of 511 keV positronium
emission.  The large white circle is 40\arcsec\ in diameter and shows the
location of the X-ray source 1E1740.7-2942.  The small circles 
mark the position centroids of two HCO$^+$ clumps detected by 
\citet{HodgesKluck09}.   The circle next to the GA is their clump 1; the
circle above is their clump 2b which does not appear to be associated
with 1.1 mm emission.  The display is linear intensity.
}
\label{fig22}
\end{figure}

\clearpage


\begin{deluxetable}{lccccccccccccc}
\tabletypesize{\scriptsize}
\setlength{\tabcolsep}{0.02in}

\tablecaption{ 
      1.1 mm Masses, Column Densities, and Densities for 1428 Bolocat Clumps 
      \label{table3}}
    
\tablehead{
    \colhead{BGPS}    &
    \colhead{$l$} & 
    \colhead{$b$} &
    \colhead{$R.A.$}             &
    \colhead{$Dec.$}              &
    \colhead{M(40\arcsec )$^1$}   & 
    \colhead{N$_{40}(H_2)^1$} &
    \colhead{n$_{40}(H_2)^1$}   & 
    \colhead{M(r)$^1$} & 
    \colhead{r)}  &
    \colhead{N$_{r}(H_2)^1$} &
    \colhead{n$_{r}(H_2)^1$}   &
    \colhead{M(120\arcsec )$^1$}   & 
    \colhead{M(gradient  T)}      \\

    \colhead{(\#)} & 
    \colhead{deg} & 
    \colhead{ deg} & 
    \colhead{ J(2000)} & 
    \colhead{ J(2000)}  &    
    \colhead{ M$_{\odot}$}   &
    \colhead{ cm$^{-2}$}  &
    \colhead{ cm$^{-3}$}  &
    \colhead{ M$_{\odot}$}   &
    \colhead{(\arcsec )} &
    \colhead{ cm$^{-2}$}  &    
    \colhead{ cm$^{-3}$}  &
    \colhead{ M$_{\odot}$}   &
    \colhead{ M$_{\odot}$}   
    }
      
\startdata

     1 &   0.000 &   0.057 & 266.349 & -28.906 &   269.6 & 5.7E+21 &  1660.5 &   946.4 &    38.4 & 2.0E+22 &   826.8 &  1434.5 &    99.3  \\
     2 &   0.004 &   0.277 & 266.138 & -28.788 &   292.3 & 6.2E+21 &  1800.3 &  1983.6 &    79.6 & 4.2E+22 &   193.7 &  1410.4 &   137.0  \\
     3 &   0.006 &  -0.135 & 266.540 & -29.001 &   312.2 & 6.6E+21 &  1922.7 &  2002.0 &    82.1 & 4.2E+22 &   178.3 &  1975.1 &   112.9  \\
     4 &   0.010 &   0.157 & 266.258 & -28.846 &   861.3 & 1.8E+22 &  5304.9 &  5037.0 &    81.4 & 1.1E+23 &   459.8 &  3310.2 &   364.7  \\
     5 &   0.016 &  -0.017 & 266.431 & -28.931 &  1755.1 & 3.7E+22 & 10810.8 & 15522.4 &    93.2 & 3.3E+23 &   945.7 &  8845.2 &   588.0  \\
     6 &   0.018 &  -0.431 & 266.837 & -29.144 &   114.9 & 2.4E+21 &   707.9 &   217.1 &    33.0 & 4.6E+21 &   297.7 &   271.0 &    56.2  \\
     7 &   0.020 &   0.033 & 266.385 & -28.902 &  1172.0 & 2.5E+22 &  7218.8 &  8606.9 &    77.2 & 1.8E+23 &   922.9 &  6892.9 &   425.0  \\
     8 &   0.020 &  -0.051 & 266.467 & -28.946 &  1620.4 & 3.4E+22 &  9980.5 &  9038.2 &    62.3 & 1.9E+23 &  1841.8 &  8159.9 &   540.3  \\
     9 &   0.022 &   0.251 & 266.174 & -28.786 &   140.5 & 3.0E+21 &   865.2 &   367.5 &    33.0 & 7.8E+21 &   503.9 &   614.4 &    64.8  \\
    10 &   0.034 &  -0.437 & 266.852 & -29.134 &    89.4 & 1.9E+21 &   550.6 &   157.5 &    33.0 & 3.3E+21 &   216.0 &   254.0 &    43.9  \\

\enddata

\tablecomments{
{\bf[1]}   Assuming a dust temperature of 20 K.
{\bf[2]}   This is only a sample table containing 10 entries.  
               The electronic version of this paper contains 1428 entries.
               }
\end{deluxetable}

\clearpage

\end{document}